\documentclass[preprint,12pt,authoryear,3p,twocolumn]{elsarticle}

\usepackage{amssymb}
\usepackage{amsmath}

\usepackage{graphicx}
\usepackage{pdflscape}
\usepackage{multirow}
\usepackage{rotating}
\usepackage{xcolor}
\usepackage{soul}
\usepackage[caption=false]{subfig}
\usepackage{natbib}
\usepackage{txfonts}

\usepackage{hyperref}

\defcitealias{Paiano2018_TXS}{[1]}
\defcitealias{ZBLL}{[2]} 
\defcitealias{Shaw2012}{[3]}

\journal{Astroparticle Physics}

\begin{document}

\begin{frontmatter}



\title{TXS 0506+056-like blazar sources and their role as possible neutrino emitters}

\author[label1]{I. Viale}
\ead{ilaria.viale@unipd.it}
\affiliation[label1]{organization={University of Padova and INFN},
            postcode={I-35131},
            city={Padova},
            country={Italy}}

\author[label2, label3]{G. Principe} 
\ead{giacomo.principe@ts.infn.it}
\affiliation[label2]{organization={University of Trieste and INFN},
            postcode={I-34127}, 
            city={Trieste},
            country={Italy}}
\affiliation[label3]{organization={INAF - Istituto di Radioastronomia},
            postcode={I-40129}, 
            city={Bologna},
            country={Italy}}

\author[label4]{C. Righi} 
\affiliation[label4]{organization={INAF, Osservatorio Astronomico di Brera},
            postcode={I—23807}, 
            city={Merate},
            country={Italy}}

\author[label5]{M. Cerruti} 
\affiliation[label5]{organization={Université Paris Cité, CNRS, Astroparticule et Cosmologie},
            postcode={F-75013}, 
            city={Paris},
            country={France}}

\author[label4]{F. Tavecchio} 
\author[label1]{E. Bernardini} 

\begin{abstract}
The interest in blazars as candidate neutrino emitters grew after the 3$\sigma$ evidence for a contemporaneous joint photon and neutrino emission from the flaring blazar TXS~0506+056 in 2017. Blazars, a class of extragalactic sources with relativistic jets pointing toward Earth, present a broadband emission interpretable via leptonic and hadronic processes, the latter relevant for proton acceleration and neutrino production. Several emission models have been developed to explain this multi-messenger observation, but the details of the neutrino production and the nature of TXS~0506+056 are not yet fully understood.
In this work we investigate the properties of sources similar to TXS~0506+056.
We select a sample of 
blazars from the Fermi 4LAC-DR2 catalog by constraining a number of key parameters in ranges centered on TXS~0506+056 values. We estimate their disk accretion efficiency and model their spectral energy distribution (SED) in terms of lepto-hadronic emission, gaining information respectively on 
the potential similarity of their environment with that of TXS~0506+056
and on their neutrino flux and detectability prospects at TeV energies.
Our study shows the candidates’ high energy emission to be dominated by leptonic processes. Part of them also show a high accretion rate, characteristic of FSRQs. 
For these sources, the very high energy (VHE) and neutrino fluxes appear undetectable by current and future instruments in an average emission state. 

\end{abstract}




\end{frontmatter}



\section{Introduction}

The origin of cosmic-rays is still an open issue.
In this context the detection of high-energy neutrinos can provide a unique signature of cosmic ray interactions and pinpoint genuine cosmic-ray accelerators, given that neutrinos are expected to originate from the interactions of hadrons with ambient matter or radiation fields.
Many advancements have been made in the last decade, from the detection of a diffuse neutrino flux \citep{IceCube:2013low}, to the evidence of 
the first potential neutrino sources \citep{TXS_2017, IceCube:2022der, nu_from_GalPlane} by the IceCube Neutrino Observatory \citep{IC2017_system}.
For example, the multi-messenger observation of a high-energy neutrino in coincidence with the flaring blazar TXS~0506+056, in 2017 \citep{icecube2017-txs}, strengthened the interest in blazars as possible neutrino-emitting candidates \citep{icecube2017-txs}, as already suggested by e.~g. \citet{Mannheim1993Feb, Atoyan2001Nov, Bottcher2013Apr}.

Blazars are a subclass of active galactic nuclei (AGNs) hosting a relativistic jet pointing in the direction of Earth and behaving as a natural accelerator of particles.
These sources are thus expected to provide a favorable environment for proton acceleration 
to very-high energies, and consequent neutrino production via photo-meson interactions inside the jet.
They can reach a total luminosity ranging from $10^{44}$ erg/s to $10^{48}$ erg/s.
The spectral energy distributions (SEDs) of blazars are characterized by two, broad non-thermal components peaking in the infrared-to-soft X$-$ray band and 
$\gamma-$ray band, respectively.
The former is unanimously explained as originating from synchrotron emission from a population of relativistic electrons in the jet, while the origin of the latter is still debated.
It can be interpreted as inverse Compton (IC) scattering of relativistic electrons with soft photons.
The target photons involved can be the ones emitted through the synchrotron radiation at lower energies (in this case we talk about synchrotron self Compton, SSC) or can originate in external radiation fields coming from e. g. the accretion disk, the broad line region (BLR) or the torus (in this case we talk about external Compton, EC).
Another interpretation invokes hadronic emission processes, as synchrotron emission from relativistic protons or radiation from 
secondary particles produced via 
proton-photon or proton-proton interactions.
Mixed lepto-hadronic models assuming the high-energy component to be produced by both leptonic and hadronic emission processes are also possible.
Among the sources interpreted with such scenarios (see e. g. \citet{Rodrigues2021, Oikonomou2021, Acharyya2023})
there is TXS~0506+056, where the high-energy emission was explained as dominated by IC scattering, with a minor contribution from hadronic processes in X$-$rays and VHE $\gamma-$rays \citep{Ansoldi2018, Keivani2018_txs, Cerruti19, Gao2019_txs}.

Blazars are divided into two sub-classes: BL Lac objects and Flat-Spectrum Radio-Quasars (FSRQ).
They were initially classified based on the intensity of the emission lines in their optical spectra \citep{Urry1995_bll-fsrq-ew}, considering as FSRQs the sources showing prominent emission lines, and as BL Lacs those with faint or absent emission lines.
This classification can be related to the efficiency of the accretion flow onto the supermassive black hole \citep{Ghisellini2009, division_bll-fsrq}: FSRQs tend to be more powerful and are thought to have a radiatively efficient accretion flow and a BLR rich in gas, while BL Lacs are thought to have an inefficient disk and a weak BLR.
Differences between these two sub-classes can be observed also in their SEDs.
FSRQs usually present larger bolometric luminosity, lower frequency of the peaks,
larger Compton dominance\footnote{The Compton dominance is defined as the ratio between luminosity of the SED at the two peak frequencies (higher-energy peak divided by the lower-energy one).} and softer $\gamma-$ray spectra than BL Lacs. 
%
This behaviour is well highlighted by the so-called blazar sequence, an anticorrelation between the SED luminosity and the peaks frequency firstly observed by \citet{Fossati1998}. Here, the authors started from radio and X-ray samples of blazars and divided them into radio luminosity bin, finding the above mentioned anticorrelation together with an increase of the Compton Dominance with frequency. 
The same behaviour was later confirmed from \citet{sequence2.0}, who used a larger sample comprising $\gamma-$ray sources and divided it into $\gamma-$ray luminosity bins. 
The 
However, since its first proposal in 1998, the blazar sequence has been subject of active debate, with different authors arguing it to be the result of selection effects and observational biases \citep[see e.~g.][]{Giommi2012Mar, Keenan2021Aug}. 
%
However, the division between the two sub-classes is not sharp and the classification of a single blazar is often not straightforward.
For example, TXS~0506+056 is usually referred to as a BL Lac object, but in the last years a potential nature as FSRQ was suggested \citep{Padovani2019_TXS}, making it a so-called masquerading BL Lac. 
These kinds of sources are intrinsically FSRQ with an efficient accretion disk and prominent optical emission lines which appear faint due to a highly boosted non-thermal continuum.
After TXS~0506+056, other blazars have been reclassified as potential masquerading BL Lacs (see e.~g.  \citet{pks1424+240_masq-bll, pks0735+17_masq-bll, Padovani2022_SIN2, Paiano2023_SIN3}).

In this work we present the selection of a sample of blazars similar to TXS~0506+056 aiming to investigate their accretion efficiency and their potential multi-messenger role as possible neutrino-emitting candidates.
In Sect.~\ref{sec:selection} the details of the candidates selection are explained, with Sect.~\ref{sec:acc_rate} showing the estimation of the accretion rate of the sources.
Sect.~\ref{sec:analysis} reports information on the data and the analysis methods used.
Analysis results for each single candidate are reported in Sect.~\ref{sec:analysis_results}.
The modeling of the SEDs of the candidates in the context of a lepto-hadronic emission is reported in Sect.~\ref{sec:model} for each object.
Finally in Sect.~\ref{sec:discussion} we present the interpretation and discussion of our results.

\section{Sample of candidates}
\label{sec:selection}
This Section is dedicated to the selection of candidates with similar features to those of TXS~0506+056.

\subsection{Selection from \textit{Fermi} 4LAC-DR2 catalog}

The sources have been selected from the second release of the Fourth Catalog of Active Galactic Nuclei \citep[4LAC-DR2,][]{4LAC, 4LAC-DR2} from the \textit{Fermi} Large Area Telescope (LAT), based on the second release of the  Fourth \textit{Fermi}-LAT catalog \citep[4FGL-DR2,][]{4FGL, 4FGL-DR2}\footnote{At the time of this analysis, the 4LAC-DR2 catalog was the latest available resource. Following the recent release of the 4LAC-DR3 catalog \citep{2022ApJS..263...24A}, we compared the spectral parameters between the 4LAC-DR2 and 4LAC-DR3 versions. This comparison confirmed that no significant changes were observed for the selected sources.}.
The 4FGL-DR2 contains $\gamma$-ray sources detected in the energy range from 50 MeV to
1 TeV in the first 10 years of \textit{Fermi} activity, from 2008, August 4
to 2018, August 2.
The 4LAC catalog was developed to identify AGN counterparts to 4FGL $\gamma$-ray sources. 
The new catalogue contains 285 additional AGNs with respect to the first release, the majority of which are blazars.
In particular, the 4LAC-DR2 contains 689 FSRQs, 1111 BL Lacs, 1277 blazar candidates of uncertain type (bcu) and 70 non-blazar AGNs.

\begin{figure}
	\includegraphics[width=\columnwidth]{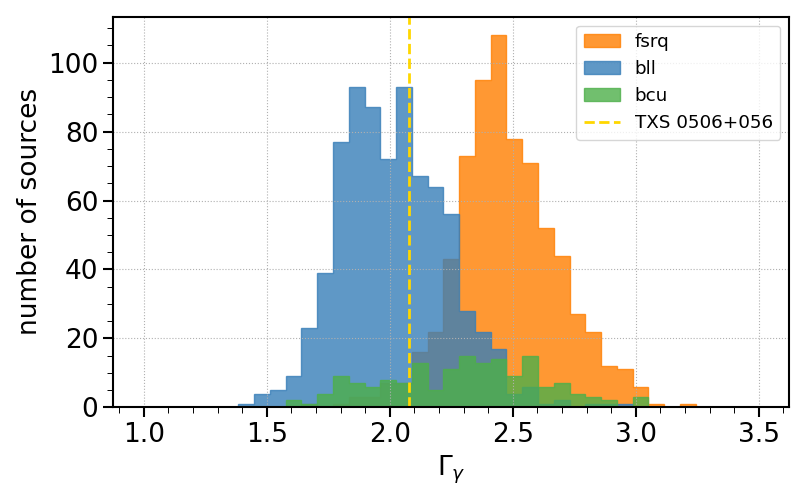}
	\caption{Histogram of the photon indexes of all sources in 4LAC-DR2 with known redshift. The legend reports: BL Lacs (bll), FSRQs (fsrq), blazar candidate of unknown type (bcu) and TXS~0506+056.}
	\label{fig:PL_hist}
\end{figure}

\begin{figure}
	\includegraphics[width=\columnwidth]{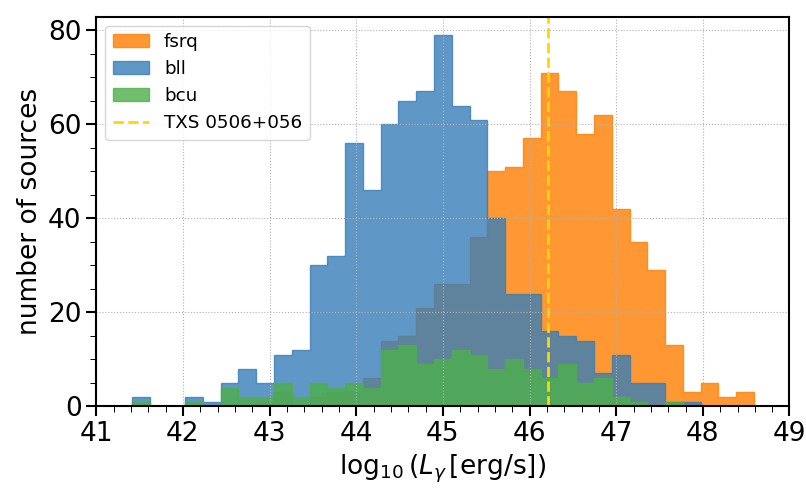}
	\caption{Luminosity histogram of all sources in 4LAC-DR2 catalog with known redshift. 
    The legend is the same as in Fig.~\ref{fig:PL_hist}.
    }
	\label{fig:L_histo}
\end{figure}

\begin{figure}
	\includegraphics[width=\columnwidth]{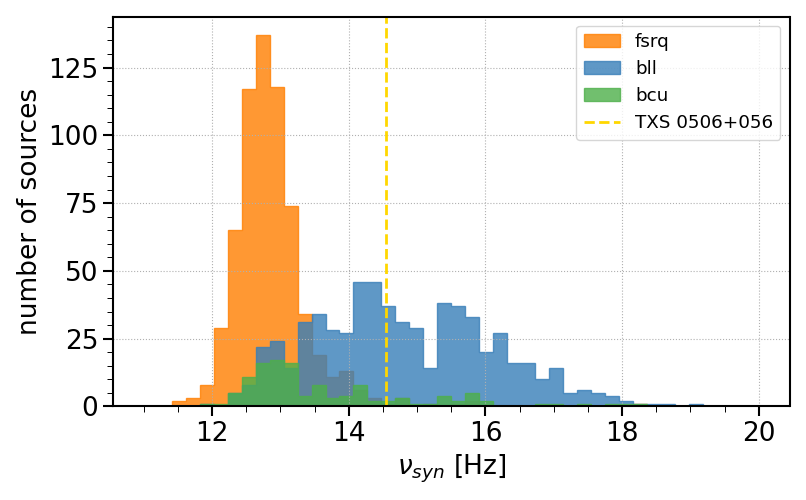}
	\caption{Histogram of the synchrotron peak frequencies of all sources in 4LAC-DR2 catalog with known redshift. 
    The legend is the same as in Fig.~\ref{fig:PL_hist}.
    }
	\label{fig:nusyn_histo}
\end{figure}

Assuming the existence of the blazar sequence and taking its results as starting point for our selection, we first investigated the distribution of three source parameters linked to the SED shape of blazars. 
The chosen parameters are the $\gamma-$ray photon index, $\Gamma_\gamma$, the $\gamma-$ray luminosity (integrated in the $1-100$ GeV range), $L_\gamma$, and the synchrotron peak frequency, $\nu_{syn}$. In order to compute the luminosity, we took into account all blazars in the catalog with a known redshift. 
The distributions of the two blazar subclasses are shown in Figures \ref{fig:PL_hist}, \ref{fig:L_histo}, and \ref{fig:nusyn_histo}.
Although overlapping, the distributions of the photon index, $\Gamma_\mathrm{\gamma}$, and $\gamma-$ray luminosity, $\mathrm{L}_\gamma$, show clearly distinct values with the FSRQs tending to have higher values of both $\Gamma_\mathrm{\gamma}$ and $\mathrm{L}_\gamma$ with respect to BL Lacs.
For the synchrotron peak frequency, FSRQs show smaller values than BL Lacs, with a very narrow distribution peaked around $\sim10^{13}$ Hz, while the distribution of BL Lacs spans a wider range (from $\sim10^{13}$ Hz to $\sim10^{18}$ Hz).
These distributions are in line with the behaviour highlighted by the blazar sequence.
Looking at these distributions, TXS~0506+056 seems not to hold the properties of neither a pure FSRQ, nor a pure BL Lac, being it more likely to be sampled from the BL Lac distribution in Figures \ref{fig:PL_hist} and \ref{fig:nusyn_histo}, and from the FSRQ distribution in Fig. \ref{fig:L_histo}. 
Its classification may therefore be non-straightforward.
Indeed, given the discrepancy between its usual classification as a BL Lac and the interpretation from \citep{Padovani2019_TXS},  
the nature of TXS~0506+056 is not firmly established and it is possible that the source has transitional properties between the two subclasses of blazars.


Motivated by these arguments, we performed a selection based on the above-mentioned parameters, which are related to the  SED characterization and the emission properties of these
sources and, consequently, on their internal environment and possible classification into FSRQs and BL Lacs.
We note that observational biases might be introduced with the used criteria, as well as with other criteria that might be adopted.
The photon index and the synchrotron peak frequency were extracted directly from the catalog, while the luminosity was derived according to \cite{Ghisellini2009}.
Since we chose the luminosity as a selection parameter, all sources without a known redshift are excluded a priori from the selection. 
The chosen parameters were constrained to be close to those of TXS~0506+056, 
within reasonable uncertainties in their values,
in order to plausibly select sources 
with the same properties.
The defined ranges are: 
\begin{gather*}
    \Gamma_{\gamma,0506}-0.5<\Gamma_\mathrm{\gamma}<\Gamma_{\gamma,0506}+0.5, \\ \log_{10}(\mathrm{L}_{\gamma,0506})-0.5<\log_{10}(\mathrm{L}_{\gamma})<\log_{10}(\mathrm{L}_{\gamma,0506})+0.5,\\ \log_{10}(\nu_{S,0506})-0.7<\log_{10}(\nu_\mathrm{S})<\log_{10}(\nu_{S,0506})+0.7,
\end{gather*}
where $\Gamma_{\gamma,0506}=~2.079\pm~0.014$, $\mathrm{L}_{\gamma,0506}=(1.62\pm0.04)\times10^{46}\, \mathrm{erg/s}$, and $\nu_{S,0506}=3.55\times10^{14} \, \mathrm{Hz}$, are referred to TXS~0506+056\footnote{Note that the 4LAC-DR2 catalog does not report an error on the synchrotron peak frequency.}. 
With this choice we expect to select blazars belonging to the same $\gamma-$ray luminosity bin of the blazar sequence \citep{sequence2.0} and source class as TXS~0506+056. Note that this is not the only possible choice for the selection but it represents a reasonable and reliable possibility since the chosen parameters act as effective proxies for the SED shape.

The sample resulting from this selection contains 27 sources,
of which 25 are blazars, and, in particular, 23 are classified as BL Lac objects. 
The full list of the selected sources can be found in Table \ref{tab:full_selection}, together with TXS~0506+056 for comparison.

\begin{sidewaystable*}
	\centering
 	\caption{Parameters of the 27 selected sources. All of them except the luminosity are directly extracted from the 4LAC-DR2 catalog. The columns report: name of the source from the catalog, photon flux, photon index, source type (bll = BL Lac object, fsrq = flat spectrum radio quasar, bcu = blazar candidate unknown, css = Compact Steep Spectrum radio source), redshift, SED class, frequency of the synchrotron peak, luminosity in $1 - 100$ GeV energy range, information on spectral analysis in the literature (z ll = redshift lower limit, z only = only redshift value provided, yes = both redshift and information on spectral lines provided), and related references.
	The sources highlighted in bold represent those we chose after the redshift inspection. 
	}
	\label{tab:full_selection}
	\scriptsize
	\begin{tabular}{lccccccccc}
	\hline
	\multirow{2}{*}{4LAC-DR2 name} & 	Photon Flux  & \multirow{2}{*}{Photon index} & \multirow{2}{*}{Type} &  \multirow{2}{*}{Redshift} & \multirow{2}{*}{SED class} & $\nu_\mathrm{S}$ & $\mathrm{L}_\gamma$ & Spectral & Refs. \\
	& [10$^{-10}$ cm$^{-2}$s$^{-1}$] &  & & & & [10$^{14}$ Hz] & [10$^{46}$ erg/s] & analysis &  \\
	\hline
    \textbf{4FGLJ0509.4+0542}  &   \boldmath{$80\pm2$} & \boldmath{$2.08\pm0.01$} & \textbf{bll}   & \boldmath{$0.3365$} & \textbf{ISP}  & \boldmath{$3.55$} & \boldmath{$2.00\pm0.05$} & \textbf{yes} &  \cite{Paiano2018_TXS, ZBLL} \\
    \hline
    \hline
	4FGLJ0004.0+0840  &   $1.9\pm0.4$  &  $2.0\pm0.1$ & bll  & 2.057 & ISP & 3.02  &   $4.3\pm0.8$ & $z$ ll & \cite{ZBLL} \\
	4FGLJ0035.2+1514  &  $11.6\pm0.7$ & $1.9\pm0.04$  &  bll     & 1.090  & ISP  & 9.71  &  $5.9\pm0.4$ & $z$ ll & \cite{TeVbll} \\
	\textbf{4FGLJ0050.7-0929}  &   \boldmath{$39\pm1$}  & \boldmath{$2.02\pm0.02$}  & \textbf{bll}   &\textbf{0.635}  & \textbf{ISP}  & \textbf{1.32}  &  \boldmath{$4.8\pm0.2$} & \textbf{yes} & \cite{ZBLL}  \\
	\textbf{4FGLJ0114.8+1326}  & \boldmath{$11.1\pm0.6$} & \boldmath{$2.11\pm0.04$} &  \textbf{bll}     & \boldmath{$0.583$}  & \textbf{ISP}  & \boldmath{$6.31$} & \boldmath{$1.02\pm0.06$} & \textbf{yes} & \cite{Shaw2012} \\
	4FGLJ0153.9+0823  &   $16.2\pm0.8$ & $1.95\pm0.03$ & bll      & 0.681  & ISP  & $8.91$ & $2.5\pm0.1$ & $z$ only & \cite{Shaw2013} \\
	4FGLJ0253.2-0124  &  $2.9\pm0.4$  & $2.1\pm0.1$ & bll     & 1.480  & ISP  & $2.24$ &  $2.8\pm0.4$ & no & ---  \\
	4FGLJ0434.1-2014  &  $3.4\pm0.4$  & $2.27\pm0.08$ & bll    & 0.928  & ISP  & $1.12$ & $0.9\pm0.1$ & $z$ only & \cite{Shaw2013}  \\
	4FGLJ0528.7-5920  &   $2.8\pm0.3$  & $2.06\pm0.09$ & bll & 1.13 & HSP  & $15.85$ &  $1.4\pm0.2 $ & no & --- \\
	4FGLJ0849.5+0456  &  $5.4\pm0.5$  & $2.08\pm0.07$ & bll  & 1.06959 & ISP & $2.45$ & $2.3\pm0.2$ & no & --- \\
	4FGLJ0908.9+2311  &   $4.8\pm0.5$ & $1.82\pm0.07$ & bll   & 1.184  & ISP  & $5.07$ & $3.1\pm0.3$ &  $z$ only & \cite{Shaw2013}  \\ 
	4FGLJ0921.7+2336  &  $3.6\pm0.4$ & $2.06\pm0.09$ & bll & 1.062 & ISP  & $2.00$ & $1.5\pm0.2$ & no & --- \\
	4FGLJ1003.4+0205  &    $1.5\pm0.3$ & $2.0\pm0.2$ & bcu & 2.075 & ISP & $5.72$ &  $3.5\pm0.7$  & no & ---  \\
	4FGLJ1103.0+1157  &   $12.2\pm0.6$ & $2.41\pm0.03$ & fsrq  & 0.914  & ISP  & $1.27$  &  $3.0\pm0.2$ & no & ---   \\
	4FGLJ1107.8+1501  &   $5.8\pm0.5$ & $1.95\pm0.06$ & bll    & 0.6    & ISP  & $3.63$  &  $0.65\pm0.06$ & $z$ only & \cite{Shaw2013}   \\
	4FGLJ1155.8+6137  &  $0.9\pm0.2$ & $1.9\pm0.2$ & bll  & 1.525 & ISP  & $2.66$ & $1.0\pm0.3$ & no & --- \\
	4FGLJ1213.6+1306  &  $6.1\pm0.5$  & $2.47\pm0.05$  & fsrq  & 1.14 & LSP & $0.87$  & $2.7\pm0.2$ &  $z$ only  & \cite{Hewitt1987}  \\
	4FGLJ1215.1+5002  &  $3.9\pm0.4$  & $1.88\pm0.07$  & bll   & 1.545  & ISP & $7.94$ & $4.5\pm0.4$ & no & --- \\ 
	4FGLJ1245.1+5709  &  $3.6\pm0.3$  & $2.13\pm0.07$  & bll  & 1.545  & ISP  & $7.94$  & $3.8\pm0.3$ & no & ---    \\
	4FGLJ1253.8+6242  &  $4.1\pm0.3$ & $1.99\pm0.06$  & bll  & 0.867  & LSP  & $0.09$  &  $1.1\pm0.1$ & $z$ only & \cite{Shaw2013}   \\
	\textbf{4FGLJ1309.4+4305}  &  \boldmath{$16.4\pm0.7$} & \boldmath{$1.90\pm0.03$} & \textbf{bll}  & \boldmath{$0.691$} & \textbf{ISP} & \boldmath{$5.19$} & \boldmath{$2.7\pm0.1$} & \textbf{yes} & \cite{ZBLL} \\
	4FGLJ1459.0+7140  &  $2.9\pm0.3$ & $2.50\pm0.06$ & css  & 0.91  & ISP  &  $3.83$  &  $7.12\pm0.07$ &  $z$ only & \cite{Hewitt1987}  \\
	4FGLJ1509.7+5556  &  $3.4\pm0.3$ & $1.88\pm0.07$  & bll  & 1.68 & ISP  & $1.41$  & $4.9\pm0.5$ & no  & ---   \\
	4FGLJ1530.9+5736  &  $2.2\pm0.3$ & $2.06\pm0.09$ & bll  & 1.1 & ISP & $2.00$ & $1.0\pm0.1$ & no & --- \\
	4FGLJ2116.2+3339  &  $34\pm 1$ & $1.90\pm 0.02$ & bll  & 0.35 & ISP  & $7.94$ & $1.13\pm0.04$ & $z$ ll & \cite{Paiano2017}  \\
	4FGLJ2145.5+1006  &  $1.0\pm0.3$ & $1.8\pm0.2$ & bll  & 2.826 & ISP & $1.82$  &  $5\pm2$ & no &  ---   \\
	4FGLJ2156.9-0854  &  $2.7\pm0.4$ & $2.1\pm0.1$ & bll  & 1.017 & ISP & $1.74$ & $1.0\pm0.2$ & no & --- \\
	\hline
	\end{tabular}
\end{sidewaystable*}

\subsection{Selected sources}

Since the aim of this work is to find a sample of sources with similar physical properties to TXS~0506+056, we intend to give an estimate of the accretion rate of the candidates.
For the computation of this quantity we need not only the redshift value of the candidates but also information on the emission lines in their optical spectra, in particular the luminosity of the lines.
For this reason we will keep in the sample only those sources satisfying these requirements, although this reduce significantly the number of selected candidates.

Starting from the sources in Table \ref{tab:full_selection}, we applied a further selection in order to only keep objects having a detailed spectral analysis in the optical and/or ultraviolet band.
For each object we performed a search in the literature to inspect the reliability of the redshift value reported in the 4LAC-DR2 catalog and the presence of information on the emission lines.
Details are reported in the last column of Table \ref{tab:full_selection} and listed in the following.
We found that almost half of the sources in the sample (13 objects)
do not report any optical/UV spectral analysis in the literature; 
7 sources  
have a reliable redshift 
but no detailed information on their spectral lines is reported. 
The redshift of some of these objects has been reviewed by \cite{foschini2022} through photometric measurements. However, the absence of a detailed spectral analysis limits their suitability for our purposes, since they do not include information on the single spectral lines. In addition, within these sources, one went out
of the selection after the redshift inspection (4FGL J0908.9+2311): the redshift reported in the catalog for this source is $z= 1.184$, while the one provided in the reference \cite{Shaw2013} is $z=0.431$.
With this new value, the luminosity of the source does not lie anymore in the range defined for the selection.
Within the remaining seven sources, 
three only have a redshift lower limit given by 
the absorption lines of the host galaxy. 
Finally, there are three sources, besides TXS~0506+056, that have a detailed optical/UV spectral analysis.
All of them are classified as BL Lac objects.
Our work is then based on the comparison of these 3 sources, namely 4FGL J0050.7-0929, 4FGL J1309.4+4305, and 4FGL J0114.8+1326, with TXS~0506+056 (4FGL J0509.4+0542). 
In Table \ref{tab:info_lines}, for each of them, we report the selection parameters (photon index $\Gamma_\mathrm{\gamma}$, synchrotron peak frequency $\nu_\mathrm{S}$, luminosity $\mathrm{L}_\gamma$) as derived from the 4LAC-DR2 catalog, together with the information on their redshift and spectral lines and the related references.
We also report the associated counterpart and include TXS~0506+056 for comparison.
From this moment on, we will refer to each of them using the counterpart name.

    \begin{table*}
	\centering
  
	\caption{Final sample of selected sources with information on selection parameters as extracted or derived from the 4LAC-DR2 catalog, on their redshift and emission lines. The columns are: source name, associated counterpart name, photon index reported in 4LAC-DR2 catalog, synchrotron peak frequency reported in 4LAC-DR2 catalog, luminosity derived from energy flux and redshift reported in 4LAC-DR2 catalog, redshift found in the literature, emission lines observed in the optical spectrum, luminosity of each line, references. The references are:  \citetalias{Paiano2018_TXS} = \citet{Paiano2018_TXS}, \citetalias{ZBLL} = \citet{ZBLL}, \citetalias{Shaw2012} = \citet{Shaw2012}}.
	\label{tab:info_lines}
    \scriptsize
	\begin{tabular}{lccccccccc}
		\hline
		\multirow{2}{*}{$\gamma-$ray Source} & \multirow{2}{*}{Counterpart} & \multirow{2}{*}{$\Gamma_\mathrm{\gamma}$} & $\nu_\mathrm{S}$ & $\mathrm{L}_\gamma$ &	\multirow{2}{*}{$z_{\mathrm{ref}}$}  & Emission  & $\mathrm{L}_{\mathrm{lines}}$ & \multirow{2}{*}{Refs}  \\
		& &  & [$10^{14}$ Hz] & [$10^{46}$ erg s$^{-1}$] & & lines & [10$^{41}$ erg s$^{-1}$] &   \\
		\hline
\multirow{3}{*}{4FGL J0509.4+0542} & \multirow{3}{*}{TXS~0506+056} & \multirow{3}{*}{$2.079 \pm 0.014$} & \multirow{3}{*}{3.55} & \multirow{3}{*}{$2.00 \pm 0.05$} & \multirow{3}{*}{0.3365} & [O II] & 1.0 & \multirow{3}{*}{\citetalias{Paiano2018_TXS, ZBLL}} \\
& & & & &  & [O III] & 0.05 &  \\
& & & & &  & [N II] & 0.05 &   \\
		\hline
        \hline
\multirow{3}{*}{4FGL J0050.7-0929} & \multirow{3}{*}{PKS~0048-09} & \multirow{3}{*}{$2.016 \pm 0.018 $} & \multirow{3}{*}{1.32} & \multirow{3}{*}{$4.77 \pm 0.15$}  & \multirow{3}{*}{0.635} & [O II] & 1.0 & \multirow{3}{*}{\citetalias{ZBLL}} \\
& & & & &  & [O III] &  0.94 & \\
& & & & &  & H$\alpha$ & 1.6 & \\
		\hline
\multirow{2}{*}{4FGL J0114.8+1326} & \multirow{2}{*}{GB6~J0114+1325}& \multirow{2}{*}{$2.11 \pm 0.04$} & \multirow{2}{*}{6.31} & \multirow{2}{*}{$1.02 \pm 0.06$} & \multirow{2}{*}{0.685} & H$\beta$  & 124 & \multirow{2}{*}{\citetalias{Shaw2012}} \\
& & & & & & [O III] & 245 & \\
		\hline		
  \multirow{2}{*}{4FGL J1309.4+4305} & \multirow{2}{*}{B3~1307+433} & \multirow{2}{*}{$1.90 \pm 0.03$} & \multirow{2}{*}{5.19} & \multirow{2}{*}{$2.68 \pm 0.13$} & \multirow{2}{*}{0.693} & [O II] &  20.0 & \multirow{2}{*}{\citetalias{ZBLL}} \\
& & & & & & [O III] & 5.5 & \\
		\hline
	\end{tabular}
	\end{table*}

\subsection{Estimation of the accretion rate}
\label{sec:acc_rate}
From the optical/UV spectral observations, we are interested in the luminosity of the emission lines, which can be exploited to estimate different useful quantities such as the photon energy density within the BLR and the dusty torus (see Section \ref{sec:model}), or the accretion rate.
In this section we give an estimate of the accretion rate of the candidates in order to understand if they could share the same accretion mechanism and the same nature as 
TXS~0506+056.

Starting from the luminosity of the lines we extrapolated the luminosity of the BLR, $\mathrm{L_{BLR}}$, of the source through the use of the lines ratios reported in \citet[][Table 1]{Francis1991}, 
where the authors estimate the strength of the emission lines with respect to the Ly$\alpha$ line (to which a reference value of 100 is assigned) in a composite spectrum of $>700$ quasars.
These ratios are used to estimate $\mathrm{L_{BLR}}$ through the relation 
\begin{equation}
\mathrm{L}_{\mathrm{BLR}} = \sum_i \mathrm{L}_{\mathrm{i,obs}} \frac{\langle \mathrm{L}_{\mathrm{BLR}}^*\rangle}{\sum_i \mathrm{L}_{\mathrm{i,est}}}, 
\end{equation}
reported in \citet[][Eq. 1]{Celotti1997}, where $\mathrm{L}_{\mathrm{i,obs}}$ is the observed luminosity, $\mathrm{L}_{\mathrm{i,est}}$ is the line ratio and $\langle \mathrm{L}_{\mathrm{BLR}}^*\rangle$ is the sum of the ratios.
Since one of our sources shows H$\alpha$ emission, we added this contribution \citep[not included in the list of][]{Francis1991} to the line ratios, as it was done in \citet{Celotti1997}.
$\mathrm{L_{BLR}}$ is a reliable tracer of the disk emission, since the broad emission lines observed in the spectrum arise from the plasma directly ionized by the disk radiation.
Thus, we estimated the luminosity of the accretion disk assuming that the BLR reprocesses about 10\% of the disk emission: $\mathrm{L}_{\mathrm{disk}}\simeq10\,\mathrm{L}_{\mathrm{BLR}}$.
Finally, from the disk luminosity, we recovered the accretion rate, defined as the quantity of matter accreted by the black hole by unit of time and given by $\dot{\mathrm{M}}=\mathrm{L}_{\mathrm{disk}}/(\eta c^2)$, with $\eta$ the accretion efficiency.
In terms of the Eddington luminosity, the accretion rate becomes: 
\begin{equation}
\dot{m} = \frac{1}{\eta} \frac{\mathrm{L}_{\mathrm{disk}}}{\mathrm{L}_{\mathrm{Edd}}}, 
\end{equation}
where $\mathrm{L}_{\mathrm{Edd}}$ can be written as $\mathrm{L}_{\mathrm{Edd}}=1.3\times10^{38}\frac{M}{M_{\odot}}\frac{\mathrm{erg}}{\mathrm{s}}$, and $\eta$ is the accretion efficiency, which we assume to be $\eta=0.1$ for black holes \citep{Thorne1974Jul, Frank2002Jan}.
The exact value of the black hole mass of the candidates is not known.
Therefore, in order not to provide a result depending too strongly on given assumptions, we decided to select two reasonable values for the black hole mass, i.~e. $M_{BH}= 10^8, \, 10^9 \, M_{\odot}$, defining a possible range in which the accretion rate can lie.
This choice is in accordance with the black hole mass estimations from \cite{Labita2006Dec}. In addition
the chosen mass range is consistent with the mass value estimated by \citet{Paliya2021} for PKS~0048-09.
The results we found are summarized in Table \ref{tab:acc_rate}.
\begin{table*}
	\centering
	\caption{Estimated values of BLR luminosity ($\mathrm{L}_{\mathrm{BLR}}$), disk luminosity ($\mathrm{L}_{\mathrm{disk}}$) and accretion rate ($\dot{m}$).}
	\label{tab:acc_rate}
	\begin{tabular}{lcccc}
		\hline
		\multirow{2}{*}{Source} & $\mathrm{L}_{\mathrm{BLR}}$ & $\mathrm{L}_{\mathrm{disk}}$ &	\multirow{2}{*}{$\dot{m}_{low}$} & \multirow{2}{*}{$\dot{m}_{high}$} \\
		& [erg s$^{-1}$] & [erg s$^{-1}$] & &  \\
		\hline
TXS~0506+056 & 2.55$\times 10^{43}$ & 2.55$\times 10^{44}$ & 1.96$\times 10^{-2}$ & 1.96$\times 10^{-1}$ \\
\hline\hline
PKS~0048-09 & 2.42$\times 10^{42}$ & 2.42$\times 10^{43}$ & 1.86$\times 10^{-3}$ & 1.86$\times 10^{-2}$ \\
GB6~J0114+1325 & 3.66$\times 10^{44}$ & 3.66$\times 10^{45}$ & 2.82$\times 10^{-1}$ & 2.82 \\
B3~1307+433 &  $3.39\times10^{44}$ & $3.39\times 10^{45}$ & 2.61$\times10^{-1}$ & 2.61 \\
		\hline		
	\end{tabular}
\end{table*}
Taking into account the division proposed by \citet{division_bll-fsrq}, we suppose a dividing value between FSRQs and BL Lacs of 
$\dot{m}_{\mathrm{d}}\sim5\times10^{-2}$, 
assuming FSRQs to have a higher $\dot{m}$ value while BL Lacs a lower one.
Despite the expectations, we can see that the estimated accretion rates are quite different from source to source.
In particular, B3 1307+433 and GB6 J0114+1325 show high values, lying above the dividing line, while PKS 0048-09 show values below the threshold, although quite close to it. 
This may suggest a different internal environment for the selected objects.
Regarding TXS 0506+056, the estimated range is consistent with the value found in \citet{Padovani2019_TXS} ($\mathrm{L_{BLR}}/\mathrm{L_{Edd}} \sim 10^{-3}$, which in our case translates into $\dot{m}\sim10^{-1}$), who proposed the source to be a FSRQ.
However, the situation seems more delicate, since, with the assumed mass values, the estimated accretion rate range lies around $\dot{m}_d$. 
If we assume the source host galaxy to be a giant elliptical with absolute magnitude in the R-band of $M_R=-22.9$ \citep[as done in e.~g.][]{Falomo2014, TeVbll, Paiano2018_TXS}, then the black hole mass can be estimated to be $\mathrm{M_{BH}\sim3\times10^8}$ \citep{McLure2002}.
By taking into account the redshift lower limit of $z>0.3$ by \citet{Paiano2018_TXS}, this mass value results to be an upper limit. 
With these additional assumptions, therefore the low edge of the accretion rate range can be estimated as $\dot{m}_{low}=6.55\times10^{-2}$, lying this time above the threshold values, although very close to it.
The results obtained for PKS~0048-09 and TXS~0506+056 are very interesting themselves, as they suggest that close to the critical accretion rate value the assumptions usually made for the BLR may break down. Indeed, The H$\alpha$ line observed for PKS~0048-09 is usually associated with the BLR. Possible explanations of the results obtained for the accretion rate can be related to the possibility that these objects host a non-standard BLR, where factors such as a reduced covering factor, lower gas density, or an atypical geometry (e.g., non-spherical or incomplete coverage of 4$\pi$) could lead to discrepancies between the expected and estimated values. 
This makes these sources interesting for further studies.

\section{Data analysis}
\label{sec:analysis}
For each source, we analyzed public data from \textit{Fermi}-LAT, \emph{Swift/XRT} and \emph{Swift/UVOT}.

\subsection{\textit{Fermi}-LAT}
\label{sec:fermi}

The LAT is the main instrument onboard the \textit{Fermi} satellite \citep{2009ApJ...697.1071A}. The LAT detects photons by conversion into electron-positron pairs and has an operational energy range from 20\,MeV to more than 300 GeV.

We performed a dedicated analysis of the \textit{Fermi}-LAT data of each selected source (namely B3\,1307+433, GB6\,J0114+1325, PKS\,0048-09 as well as TXS\,0506+056) using 12 years of LAT observations taken between 2008, August 4 and 2020, August 4. 
We then repeated the analysis on shorter time intervals based on the $\gamma-$ray activity of the selected sources as well as on the availability of quasi-simultaneous multi-wavelength observations (see second and third columns of Table \ref{tab:fermi_results} for further details).

For the data selection, we used P8R3 (v3) Source class events \citep{2018arXiv181011394B}, in the energy range between 100\,MeV and 1\,TeV, in a region of interest (ROI) of $15^{\circ}\times 15^{\circ}$ centered on the source position. 
The low energy threshold is motivated by the large uncertainties in the arrival directions of the photons below 100 MeV, leading to a possible confusion between point-like sources and the Galactic diffuse component (see \citet{2018A&A...618A..22P} for issues at low energies with \textit{Fermi}-LAT and a possible alternative analysis method).

The analysis consists of the following steps: model optimization, and source localization, spectrum and emission variability studies.
It was performed with \texttt{Fermipy}\footnote{http://fermipy.readthedocs.io/en/latest/} \citep[v1.0.1,][]{FermiPy}, a python library making use of the Fermi Tools, of which the version 2-0-18 was used. 
We created counts maps using a pixel size of $0.1^{\circ}$. All $\gamma$-rays with zenith angle larger than 105$^{\circ}$ were excluded in order to limit the contamination from secondary $\gamma$-rays from the Earth’s limb \citep{2009PhRvD..80l2004A}. 
We applied different cuts for the data selections at low energies and selected event types with the best point spread function\footnote{A measure of the quality of the direction reconstruction is used to assign events to four quartiles. In Pass 8 data, $\gamma-$rays can be separated into 4 PSF event types: 0, 1, 2, 3, where PSF0 has the largest point spread function and PSF3 has the best one.} (PSF). In particular, for energies below 300 MeV we excluded events with zenith angle larger than 85$^{\circ}$, and photons from PSF0 and PSF1 event types. Between 300 MeV and 1 GeV we excluded events with zenith angle larger than 95$^{\circ}$, as well as photons from the PSF0 event type. Above 1 GeV we use all events with zenith angles less than 105$^{\circ}$.

We used the P8R3\_SOURCE\_V3 instrument response functions. 
The model used to describe the sky includes all point-like and extended sources located at a distance $<15^{\circ}$ from the source position and listed in the 4FGL-DR2 \citep{4FGL}, as well as the Galactic diffuse and isotropic emission. For the two latter contributions, we adopted the same templates\footnote{\url{https://fermi.gsfc.nasa.gov/ssc/data/access/lat/BackgroundModels.html}} adopted to derive the 4FGL-DR2 catalog. The spectral analysis was performed leaving free to vary the diffuse background template normalization and the spectral parameters of the sources within 3$^\circ$ from our targets.
For the sources in a radius between 3$^{\circ}$--5$^{\circ}$ and all variable sources only the normalization was fit, while we fixed the parameters of all the remaining sources within the ROI at larger angular distances from our target.
We modeled the spectrum of the sources using the same spectral function adopted in the 4FGL-DR2 catalog. 
A log-parabola  
function 
\begin{equation*}
   \dfrac{dN}{dE} = N_{0}\times \left(\frac{E}{E_{b}}\right)^{- [\alpha+\beta \textrm{ln} (E/E_b)]} 
\end{equation*} 
was used for TXS\,0506+056, B3~1307+433 and PKS\,0048-09, while a power-law 
function 
\begin{equation*}
   \dfrac{dN}{dE} = N_{0} \times \left(\frac{E}{E_{b}}\right)^{- \Gamma}
\end{equation*} 
was used for GB6\,J0114+1325. 
Finally we extracted a light curve using time bins with a duration of two months. 

\begin{table*}
	\centering
    \scriptsize
	\caption{\textit{Fermi}-LAT results in the 0.1--1000 GeV energy range. For TXS~0506+056, PKS~0048-09 and B3~1307+433 a log-parabola function was used for fitting their spectra using a $E_0$ of 1.07 GeV, 924 MeV and 1.39 GeV, respectively. A power-law function with $E_0=$ 1.4 GeV was used for GB6~J0114+1325. The reported flux values are in unit of $10^{-8}$ ph cm$^{-2}$ s$^{-1}$, while normalisation values are in unit of $10^{-12}$ MeV$^{-1}$ cm$^{-2}$ s$^{-1}$.  }
	\label{tab:fermi_results}
	\begin{tabular}{cccccccc}
		\hline
        \multirow{2}{*}{Source} &  \multirow{2}{*}{Period} & \multirow{2}{*}{MJD} & \multirow{2}{*}{TS} & \multirow{2}{*}{Flux} & \multicolumn{3}{c}{Spectral values}  \\
        & & & &  & $N_0$ & $\alpha$ ($\Gamma$) & $\beta$ \\
        & [days] & & & [$10^{-8} \, \mathrm{ph \,cm^{-2}\, s^{-1} }$] & [$10^{-12}$ MeV$^{-1}$ cm$^{-2}$ s$^{-1}$] & & \\
		\hline
TXS~0506+056 & 12 years & $54682 - 59065$ & 15400 & 7.97$\pm$0.28 & $7.62\pm0.13$ & $2.03\pm0.02$ & $0.051\pm0.008$ \\
\hline
\hline
\multirow{3}{*}{PKS~0048-09} &  12 years & $54682 - 59065$ & 7500 & 3.31$\pm$0.15 & $4.97\pm0.12$ & $1.91\pm0.02$ & $0.062\pm0.010$ \\
 & low & $55100 - 55600$ & 610 & 2.04$\pm$0.34 &$8.11\pm1.14$ & $1.97\pm0.11$ & $0.044\pm0.005$\\
 &  high & $55880 - 55950$ & 230 & 6.00$\pm$1.22 & $3.30\pm0.26$ & $1.92\pm0.09$ & $0.089\pm0.040$ \\
\hline
\multirow{2}{*}{B3~1307+433}  &  12 years &$54682 - 59065$  & 2930 &  0.76$\pm$0.07 & $0.76\pm0.03$ & $1.74\pm0.04$ & $0.093\pm0.017$\\
& average 
& $59488 - 59759$ & 80 & 0.65$\pm$0.30 & $0.76\pm0.19$ & $1.59\pm0.23$ & $0.090\pm0.085$\\
\hline
\multirow{2}{*}{GB6~J0114+1325} &  12 years & $54682 - 59065$ & 1260 & 1.39$\pm$0.10 & $0.56\pm0.03$ & $2.09\pm0.03$\\
 & average & $55450 - 55600$ & 95 & 2.54$\pm$0.66 & $0.86\pm0.15$ & $2.18\pm0.13$\\
		\hline		
	\end{tabular}
\end{table*}

\begin{table*}
	\centering
    \caption{Values of the parameters obtained from the fit and total exposure time for each source.}
	\small
	\begin{tabular}{cccccc}
		\hline
		\multirow{2}{*}{Source} & State of & $n_H$ & $\Gamma_X$ & $K$ & Exp. time\\
        & activity & [cm$^{-2}$] &  & [keV$^{-1}$ cm$^{-2}$ s$^{-1}$] &  [ks] \\
		\hline
        \multirow{2}{*}{PKS~0048-09} & high & \multirow{2}{*}{$3.10\times 10^{20}$} & $2.16\pm0.17 $ & $ (7.0\pm0.5) \times 10^{-4}$ & 2.94 \\
         & low &  & $2.12\pm0.07$  & $(8.7\pm0.4)\times 10^{-4}$  & 6.69 \\
         GB6 J0114+1325 & average & $3.27\times 10^{20}$  & $4.0\pm0.7$  & $(7.7\pm3.6)\times 10^{-5}$  & 1.18  \\
         B3 1307+433 & average & $1.85\times 10^{20}$  & $2.38\pm0.18$ & $(6.9\pm0.6)\times 10^{-5}$ & 21.29  \\
		\hline
	\end{tabular}
	
	\label{tab:xrt-results}
\end{table*}
\begin{table*}
	\centering
    \scriptsize
	\caption{Recovered magnitudes in the Vega system for each source in the selection. For PKS~0048-09 we report both the magnitudes obtained in the chosen low state (upper line) and high state (lower line).}
	\label{tab:uvot_mag}
	\begin{tabular}{lccccccc}
		\hline
		\multirow{2}{*}{Source} & State of & $v$ &	$b$ & $u$ & $w1$ & $m2$ & $w2$ \\
		& activity & [mag] & [mag] & [mag] & [mag] & [mag] & [mag]  \\
		\hline
\multirow{2}{*}{PKS~0048-09} & high & $16.4\pm0.04 $ & $16.81\pm0.03$ & $15.99\pm0.02$ & $16.02\pm0.03 $ & $16.05\pm0.03$ & $16.18\pm0.02$ \\ 
 & low & $14.85\pm0.03 $ & $15.22\pm0.02 $ & $14.39\pm0.02 $ & $14.41\pm0.02$ & $14.41\pm0.03 $ & $14.69\pm0.02$ \\ 
 GB6~J0114+1325 & average & $16.85\pm0.1 $ & $17.21\pm0.07 $ & $16.29\pm0.05 $ & $16.29\pm0.05$ & $16.27\pm0.06$ & $16.28\pm0.04$ \\
 B3~1307+433 & average & $17.31\pm0.04 $ & $17.72\pm0.03$ & $16.86\pm0.03$ & $16.86\pm0.03$ & $16.9\pm0.03 $ & $17.01\pm0.02 $ \\
		\hline		
	\end{tabular}
\end{table*}

\subsection{\emph{Swift}}
\label{Swift}
The {\em Neil Gehrels Swift observatory} satellite \citep{Swift} is a multiwavelength satellite observatory equipped with 3 instruments: the X-ray Telescope \citep[XRT;][$0.2-10$ keV]{Burrows2005_xrt}, the Ultraviolet/Optical Telescope \citep[UVOT;][170–600 nm]{Roming2005_uvot}, and the Burst Alert Telescope \citep[BAT;][$15-150$ keV]{Barthelmy2005_bat}.
For our sources, we took data from the \emph{XRT} and \emph{UVOT}.
Similarly to what we did for the \emph{Fermi}-LAT analysis, we firstly performed a dedicated analysis of each selected source using all available observations.
Then, we repeated the analysis on smaller time intervals chosen based on the activity of the source (see Sect.~\ref{sec:analysis_results} for more details on the selection of the period).
In the first case the single observations were analyzed, while in the second one all observations belonging to the selected period were summed.

All \emph{XRT} observations were performed in photon counting mode \citep{Hill2004_readout-modes}, providing photon counts within specific energy ranges defined by the instrument channels.
The spectrum data of the source were taken from the UK Swift Science Data Centre\footnote{\url{https://www.swift.ac.uk/user_objects/index.php}} and analyzed by using the \textsc{heasoft} v. 6.29 and \texttt{XSPEC v.12.12.0g} software packages.
Each spectrum was rebinned in energy using the \texttt{grppha} tool of \textsc{heasoft} 
in order to have at least $25$ 
counts per bin to apply the $\chi^2$ test.
When counts were not sufficient, as in the observations of GB6~J0114+1325, the Cash statistic was used \citep{Cash1979_Cstat}.
All \emph{XRT} spectra were analyzed in the $0.3 - 10$ keV energy range and modeled with an absorbed power-law with a fixed Galactic column density $N_H$, measured according to \cite{Kalberla2005}.
For all sources, the spectral parameters obtained from the fit, i. e. the normalization $K$ and the photon index $\Gamma_X$, as well as the value used for $N_H$ and the total exposure time are reported in Table \ref{tab:xrt-results}.

Simultaneous \emph{UVOT} observations were performed during the \emph{XRT} pointings.
The \emph{UVOT} is composed of 6 broad-band filters, three of which in the visible band ($v$, $b$, $u$) and three in the ultraviolet ($w1$, $m2$, $w2$).
All of them were used in the observation of the selected sources. 
Photometric data were downloaded from the High Energy Astrophysics Science Archive Research Center (HEASARC\footnote{\url{https://heasarc.gsfc.nasa.gov/cgi-bin/W3Browse/w3browse.pl}}).
Data were analyzed with the \texttt{uvotimsum} and \texttt{uvotsource} tasks and the 20201215 CALDB-UVOTA release of the calibration database.
We used \texttt{uvotimsum} to integrate the available  images and \texttt{uvotsource} to recover the sources magnitudes through differential photometry techniques.
They were applied for each filter to photometric images belonging to the same observation date.
For each candidate, the source counts were extracted from a circular region centered on the source, with a radius of $5$ 
arcsec; while background counts were obtained from a circular region with radii between of $40$ and $60$ arcsec located in a source-free area.
All the computed magnitudes were then corrected for Galactic extinction, given by the relation $A_{\lambda} = E\left( B-V \right) \left[a_\lambda R_v+b_\lambda\right]$ \citep{Roming2009_extinction}, where $E(B-V)$ is taken from the NASA/IPAC Infrared Science
Archive\footnote{\url{https://irsa.ipac.caltech.edu/applications/DUST/}} according to \citet{S&F2011}, while the other coefficients are reported in \citet{Roming2009_extinction}.
The corrected magnitudes were finally converted into energy fluxes by using the parameters in \cite{UVOT_parameters}.
For each source the detection was obtained in all photometric bands.

\section{Analysis results}
\label{sec:analysis_results}

In the following we report the results from our \emph{Swift} and \emph{Fermi} analyses.
For each source we firstly recovered the multiwavelength emission over time in order to identify low and high or flaring states of activity. {As a criterion to define a gamma-ray emission state as low or high, we used the condition that the flux in the bi-monthly time bins of the \emph{Fermi}-LAT light curve differed by at least two sigma from the flux estimated over the 12-year period, as shown by the values reported in Table \ref{tab:fermi_results}}.
Then, once identified, we chose a period for both low and high states for which we have contemporaneous multiwavelength data.
Quasi-simultaneous observations in AGN modeling are crucial for accurately capturing the variability of the active galactic nucleus across different wavelengths, ensuring that changes in emission are correctly correlated, and therefore related to same emission mechanism, and not misinterpreted due to time lags between observations.
Among our sources, B3~1307+433 and GB6~J0114+1325 do not show significant variability in their lightcurves, so for these sources we just created a SED in their average state.
The source PKS~0048-09 is the only one for which we could define both a low and a high state period, given that it is the only source in our sample showing significant variability.
For all sources, the spectral results from \textit{Fermi}-LAT and \textit{Swift/XRT} analyses have been collected in Tables \ref{tab:fermi_results} and \ref{tab:xrt-results} respectively, while the magnitudes obtained from \emph{UVOT} analysis are presented in Table~\ref{tab:uvot_mag}.
The multi-wavelength SEDs containing the contemporaneous data in the selected periods are shown in Fig. \ref{fig:sed_all} for each source.

\begin{figure*}
	\includegraphics[width=\linewidth]{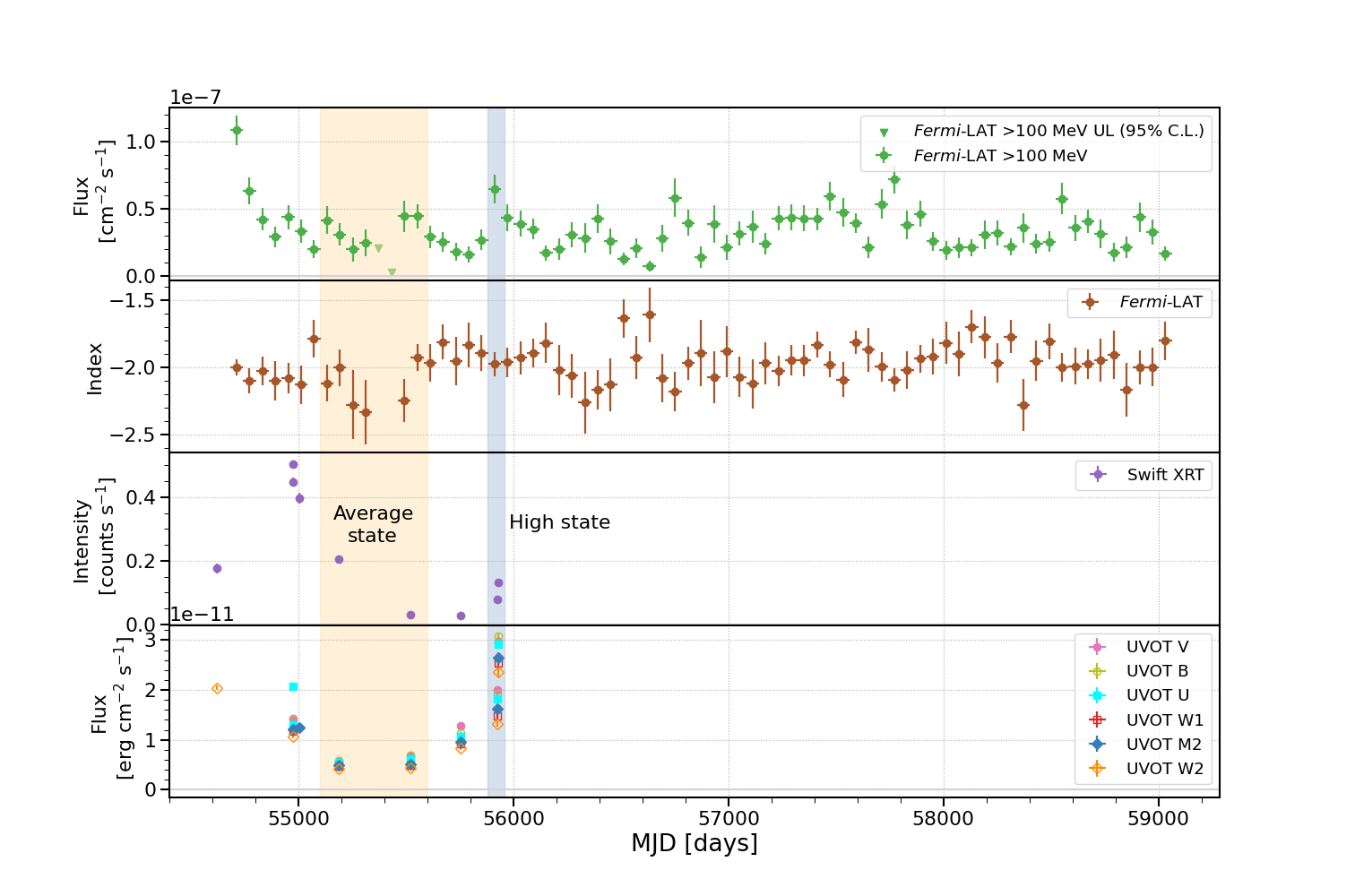} 
	\caption{Multiwavelength lightcurve of the source PKS~0048-09. From top to bottom, the panels show: \textit{Fermi}-LAT flux in 60-days bins, values assumed by the $\gamma-$ray spectral index in the same bins, \emph{Swift/XRT} flux in daily bins, \emph{Swift/UVOT} flux in daily bins. The shaded areas identify the boundaries of the low (orange) and high (blue) state periods chosen to build the SED.}
	\label{fig:pks_lc}
\end{figure*}

\subsection{PKS~0048-09}
\label{sec:results-pks}

The source was observed by \emph{Swift} 9 times between 2008, June 4 (MJD 54621) and 2012, January 13 (MJD 55939) with both \emph{XRT} and \emph{UVOT}.
The multiwavelength lightcurve of the source is quite peculiar, since it shows three flares, one in each of the energy ranges analyzed (\textit{Fermi}-LAT, \emph{Swift/XRT}, \emph{Swift/UVOT}) which are not coincident in time.
Indeed they happened in different periods and do not correspond to the brightest activity in the other bands, 
so it is not easy to define a unique high state of activity for the source.
The \textit{Fermi}-LAT flare has no coincident \emph{Swift} simultaneous observations, so we cannot say anything about the source activity in the other bands in that period and cannot consider it to build the simultaneous high-state SED of the source.
Regarding the brightest \emph{XRT} flare, it happened in May$-$June 2009 (MJD $\sim54952-55012$) and 
is not coincident with the brightest activity in \emph{UVOT} bands.
The same happens for the optical-UV flare observed by \emph{UVOT} between December 2011 and January 2012 (MJD $\sim55896-55939$): it corresponds to a 
relatively high activity of the source in the X-ray band, 
but not to the brightest X-ray flux.
We finally decided to consider as a high state a period of few months coincident with the \emph{UVOT} flare because of a simultaneous slightly-enhanced activity observed in \textit{Fermi}-LAT data at the same time.
The full period we considered for \textit{Fermi}-LAT analysis in the high state ranges from 2011 November 15 to 2012 January 24 (MJD $55880-55950$, see the blue shaded area in Fig.~\ref{fig:pks_lc}).
Instead, given that a clearly visible low state cannot be identified from the source LC, we concentrated on an average state of activity. We considered the period between 2009 September 26 and 2011 February 8, containing two \emph{Swift} pointings (MJD $55100-55600$, see the yellow shaded area in Fig.~\ref{fig:pks_lc}).
The full full multi-wavelength lightcurve of this source is shown in Fig.~\ref{fig:pks_lc}.
PKS 0048-09 is significantly detected with \emph{Fermi}-LAT in the 12-years period
(TS = 7500, $>85\sigma$) and presents flux and spectral parameters 
in agreement with the results for this source in the 4FGL-DR3 catalog \citep{2022ApJS..260...53A}, similarly based on 12 years of \textit{Fermi}-LAT data. 
During the high-state period the source presents a slightly higher flux (about twice the one derived in our 12-years analysis) while the log-parabola indexes are compatible within the statistical uncertainties.
The spectral parameters recovered from the analysis are reported in Table~\ref{tab:fermi_results}, while the values reported in the 4FGL-DR3 catalog are $\alpha=1.92\pm0.03$, $\beta=0.06\pm0.01$.

\subsection{GB6~J0114+1325}
\begin{figure*}
	\includegraphics[width=\linewidth]{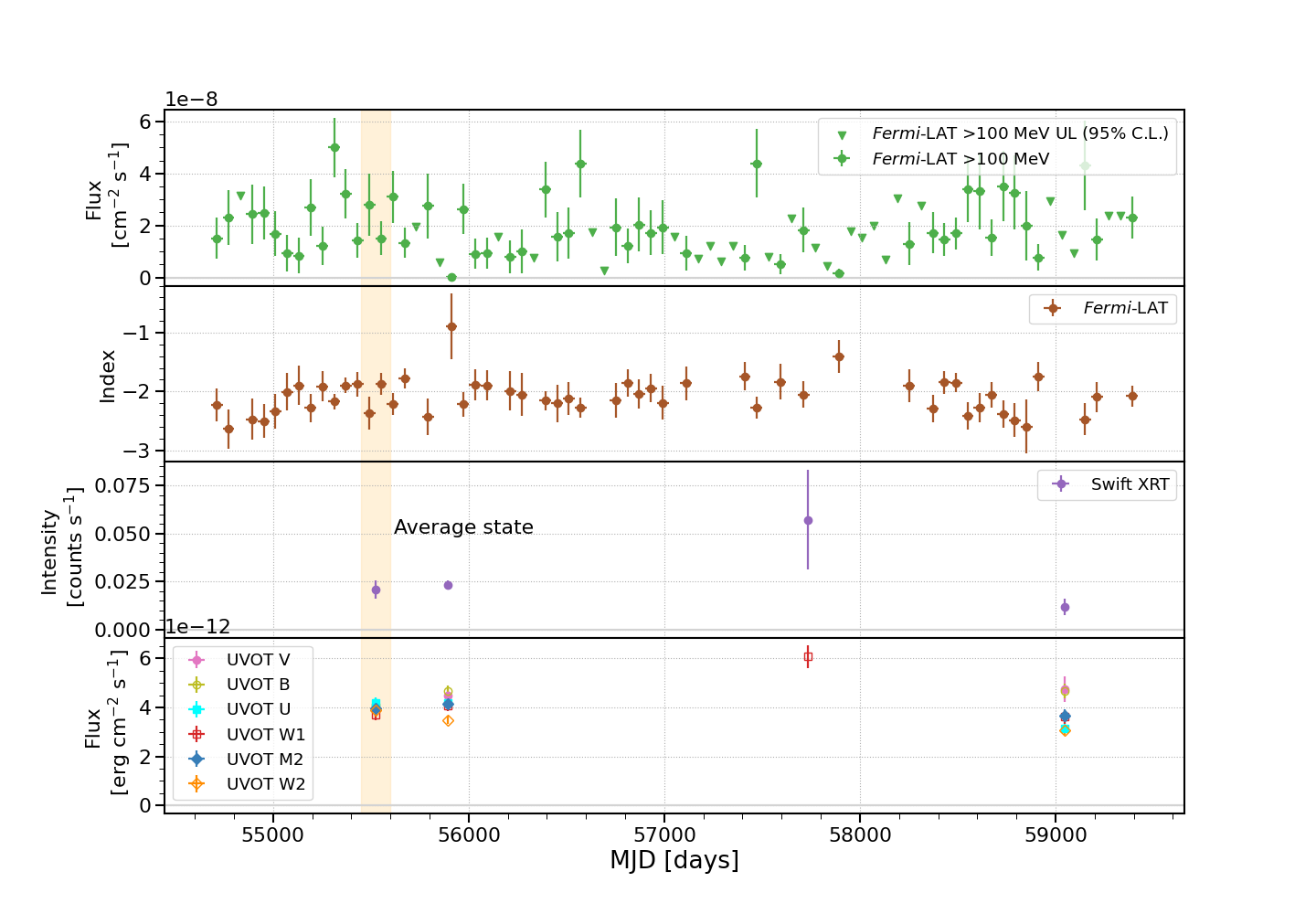} 
	\caption{Multiwavelength lightcurve of the source GB6~J0114+1325. 
    The reported information are the same as in Fig.~\ref{fig:pks_lc}.
    }
	\label{fig:gb_lc}
\end{figure*}

The source was observed by \emph{Swift} only 4 times between 2010, November 22 (MJD 55522) and 2020, July 17 (MJD 59047), with both the \emph{XRT} and \emph{UVOT}.
Like for the other sources, we first built the lightcurve of the candidate in order to identify suitable low and high states of activity, finding no a clear variation in the source flux (see Fig.~\ref{fig:gb_lc}).
However, since the data are so sparse in time, it is not possible to select a time interval with a large number of \emph{Swift} pointings 
for the construction of the SED, and thus we consider the average state of the SED.
We decided to consider an interval ranging from 2010, September 11 and 2011, February 8 (MJD $55450-55600$).
The chosen interval lies around the first \emph{Swift/XRT} observation (on 2010, November 22), which is one of the most significant ones.
The \textit{Fermi}-LAT analysis is then performed over the selected period.
The \emph{XRT} observations show a low total exposure (see Table \ref{tab:xrt-results}), because of which the number of counts obtained in each energy bin created with \texttt{grppha} is not enough to apply the $\chi^2$ statistic, so in this case the Cash statistic is used.
With \textit{Fermi}-LAT, GB6\,J0114+1325 is detected with a $\mathrm{TS}= 1260$, corresponding to a significance of $>35\sigma$. 
The flux and spectral parameters (see Table \ref{tab:fermi_results}) are in agreement with the 4FGL-DR3 results ($\Gamma=2.06\pm0.04$) for this source. During the selected period the source presents a slightly higher flux (less than twice the one derived in the 12-years analysis) and a power-law index compatible within the statistical uncertainties. 




\subsection{B3~1307+433}
\begin{figure*}
	\includegraphics[width=\linewidth]{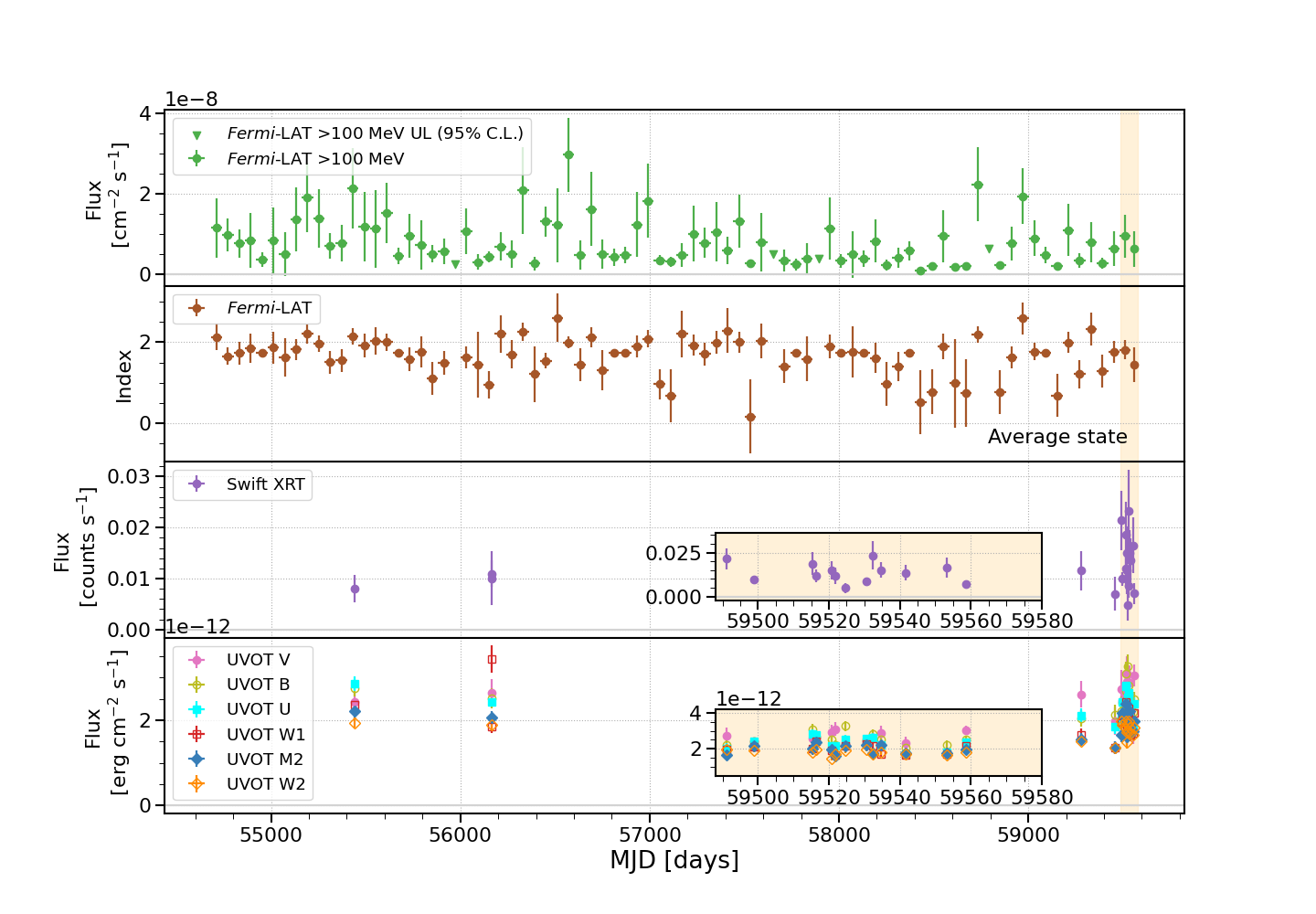} 
	\caption{Multiwavelength lightcurve of the source B3~1307+433. 
    The reported information are the same as in Fig.~\ref{fig:pks_lc}.
    The insets show the trend of the source in X-ray and optical/UV bands during the selected period.
    }
	\label{fig:b3_lc}
\end{figure*}

This source was observed by \emph{Swift} 18 times between 2010, September 2 (MJD 55441) and 2021, December 10 (MJD 59558), with the majority of the pointings in 2021, thanks to a dedicated observation proposal submitted by the authors.
In particular, we have 13 observations in the period from 2021, October 4 to 2021, December 10 (MJD $59491-59558 $).
We firstly generated the daily lightcurve of the source for both \emph{XRT} and \emph{UVOT}.
In the first case the lightcurve was generated with the flux in units of counts s$^{-1}$ with the data product generator tool available at the UK Swift Science Data Centre, while in the second case we performed the analysis of the photometric images for each observation date. 
The lightcurves in each band are shown in Fig.~\ref{fig:b3_lc}.
Since neither the \textit{Fermi}-LAT nor the \emph{Swift} light curves show significant variations of the source flux in the period from  October to December 2021 and given the large number of observations in this period, we chose these data to build the SED of the source in its average state. 
\textit{Fermi}-LAT analysis was performed in the whole period from 2021, October 1 to 2021, December 31 (MJD $59488-59579$).
Analyzing 12 years of \emph{Fermi}-LAT data (see Table 4 for spectral results)
the source is detected in $\gamma$ rays with a TS = 2931, corresponding to a significance $>50\sigma$.
The spectral parameters of this source are in agreement with the 4FGL-DR3 results, which are $\alpha=1.76\pm0.04$ and $\beta=0.09\pm0.02$. For the selected average-state period the source presents a similar emission with flux and spectral parameters compatible within the uncertainties with the one that we obtained using 12 years.




\section{Modeling}
\label{sec:model}

\begin{figure*}
	\includegraphics[width=\linewidth]{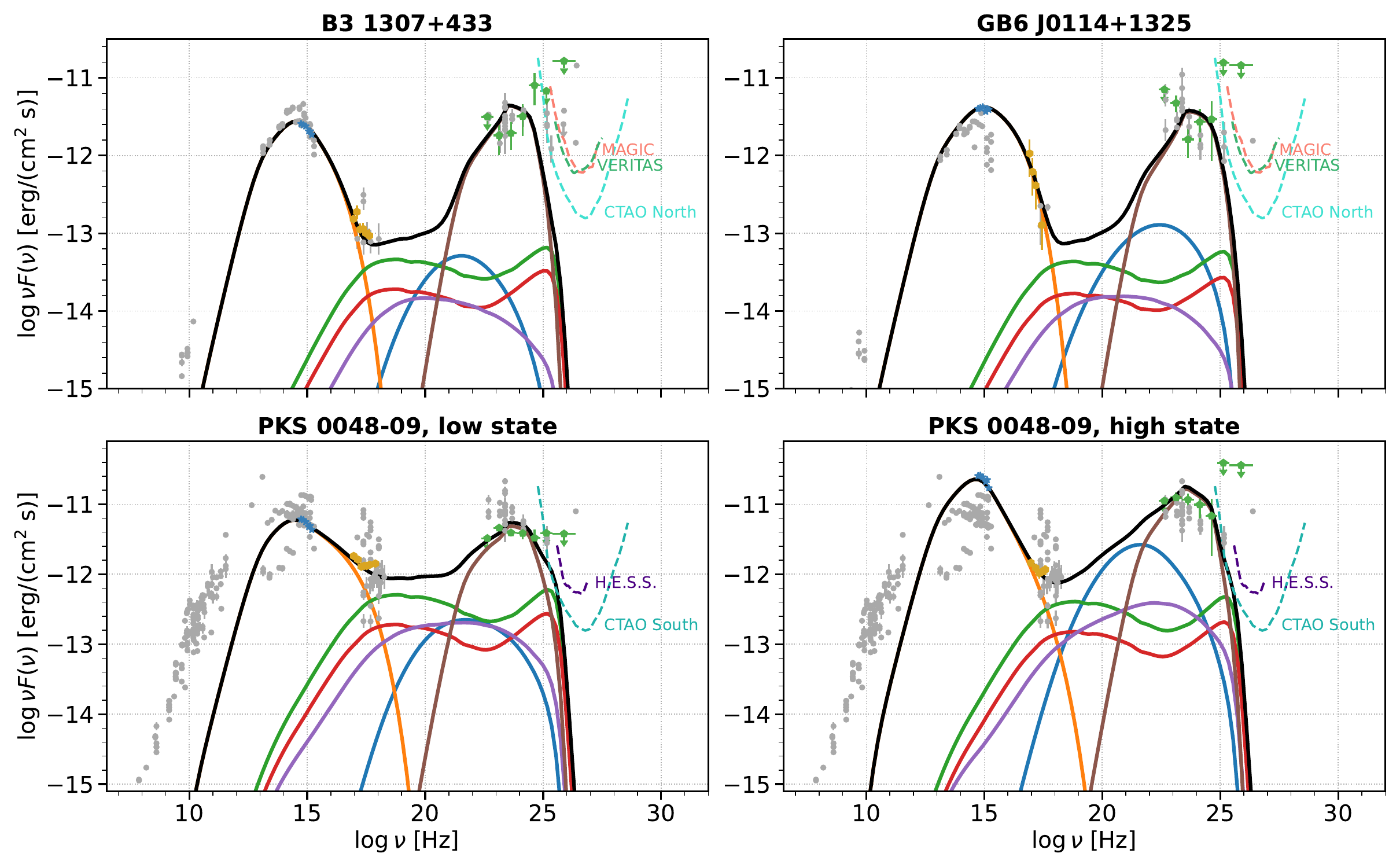} 
	\caption{Multi-wavelength SED of the candidates in the selected periods and modeling in lepto-hadronic context. Grey points show the archival data from ASI ASDC \citep{asi-asdc}, while light-blue stars, gold hexagons, and green pentagons show \emph{UVOT}, \emph{XRT} and \textit{Fermi}-LAT data in the selected periods, respectively. The model lines show: electron synchrotron emission (orange), synchrotron self Compton emission (blue), external Compton emission (brown), Bethe-Heitler cascades (violet), cascades from $\pi^0$ (green), and $\pi^\pm$ (red) decay. For the two sources in the northern hemisphere (GB6~J0114+1325, B3~1307+433) a comparison with the sensitivity of MAGIC \citep[salmon dashed line,][]{Aleksic2016}, VERITAS \citep[light-green dashed line,][]{Park2016}, and the CTAO Northern array \citep[cyan dashed line,][]{ctao_performance} 
    for 50 hours of observation, are also shown, while for the only source in the southern emisphere (i. e. PKS~0048-09), the comparison is done with H.E.S.S. \citep[purple dashed line][]{hess-sensitivity} and CTAO Southern array (seagreen dashed line).}
	\label{fig:sed_all}
\end{figure*}
\begin{figure*}
	\includegraphics[width=\linewidth]{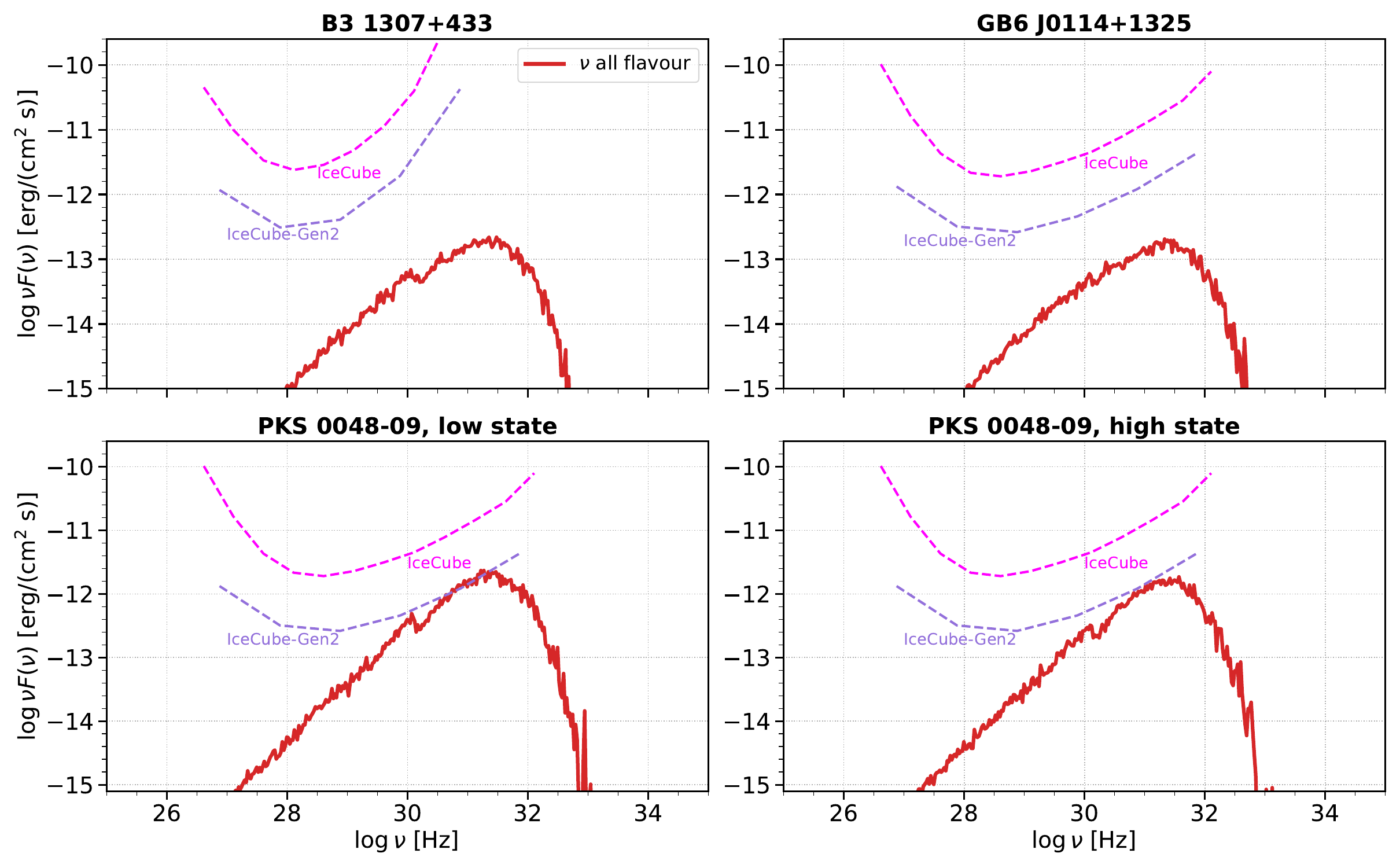}
	\caption{Neutrino spectrum expected from the model in Fig. \ref{fig:sed_all} of the sources under consideration, compared with the IceCube 
    sensitivity for point-source searches for 8 years of exposure \citep{Aartsen2019}. The sensitivity is shown for a declination of $30^\circ$ in the upper left panel and of $0^\circ$ in the other panels, based on the proximity to the source declination.}
	\label{fig:sed_all_nu}
\end{figure*}

The modeling of the TXS~0506+056 multi-messenger campaign in 2017 indicates that the most likely emission scenario is a mixed lepto-hadronic one, in which the origin of the $\gamma$-ray SED component is predominantly leptonic (inverse Compton scattering by primary electrons), with subdominant hadronic emission by ${\rm e}^\pm$ cascades triggered by pion decay and Bethe-Heitler pair production. 
Due to energetic constraints, the target photon field for p-$\gamma$ interactions is more likely to be external to the jet, and relativistically boosted in the reference frame of the emitting plasmoid in the jet. 
In this section we model the SED of the three targets 
that share similar observational properties (see Sect. \ref{sec:selection}), within the same framework.
The numerical code we use here is \textit{LeHa} \citep{Cerruti15}, that was already applied in the study of TXS~0506+056, albeit assuming a single-zone scenario \citep[i.e. in absence of external photon field,][]{Cerruti19}, and for a spine-layer structured jet \citep{MAGIC0506_paper3}. The code calculates the photon and neutrino emission from a spherical plasmoid in the jet, parametrized by its Doppler factor $\delta$, its radius $R^\prime$, and average magnetic field $B^\prime$. 
The energy distributions of primary electrons and protons are both described by a broken power law with exponential cut-off at the maximum energy. 
p-$\gamma$ interactions include photo-meson \citep[computed using SOPHIA,][]{Sophia}, and Bethe-Heitler pair-production. \textit{LeHa} is a stationary code, that computes SEDs assuming that all particles have the time to reach equilibrium between injection and cooling. Arbitrary external photon fields can be provided as input for both p-$\gamma$ and $\gamma$-$\gamma$ interactions.

For the purposes of this work, the important parameter is the radiation energy density of the broad emission line photons.
In order to compute this quantity we need information on the luminosity and radius of the BLR.
We gave an estimation for the former in Sect. \ref{sec:acc_rate} thanks to our study on the spectral lines, while the latter can be recovered from the disk luminosity by assuming a simple scaling relation: $\mathrm{R_{BLR}} = 10^{17} \mathrm{L}^{1/2}_{\mathrm{disk,45}} \, \mathrm{cm}$ \citep{GhiselliniTavecchio2008}, where $\mathrm{L_{disk,45}}$ is the luminosity of the disk in units of $10^{45} \, \mathrm{erg \, s^{-1}}$.
The recovered values for each source are the following: $1.84\times 10^{17}\, \mathrm{cm}$ for B3~1307+433, $1.91\times10^{17}\, \mathrm{cm}$ for GB6~J0114+1325 and $1.56\times10^{16}\, \mathrm{cm}$ for PKS~0048-09.
This relation implies that, within the BLR, the energy density, $u_{\mathrm{BLR}}$, of the line photons is constant and equal for all sources.
The value obtained in the observer's frame is
\begin{equation}
\label{eq:uBLR}
u_{\mathrm{BLR}} = \frac{\mathrm{L}_{\mathrm{BLR}}}{4\pi c \mathrm{R_{BLR}}^2}\times2.24 = 5.95\times10^{-2} \, \mathrm{erg \, cm^{-3}}.
\end{equation}
In order to recover the value of the photon energy density in the jet frame ($u_{\mathrm{ph,BLR}}^\prime$) one needs to multiply the above result by the square of the bulk Lorentz factor, $\Gamma^2$.
A similar procedure can be followed in order to obtain the luminosity and photon energy density within the torus.
In this case we assume that the accretion luminosity reprocessed by the torus is about 40\%, so $\mathrm{L}_{\mathrm{torus}}=0.4\,\mathrm{L}_{\mathrm{disk}}$.
As in the case of the BLR, we assume a scaling relation between the luminosity of the disk and the one of the torus: $R_{\mathrm{torus}}=2.5\times10^{18}\,\mathrm{L_{disk,45}}^{1/2}\, \mathrm{cm}$ \citep{GhiselliniTavecchio2008}.
For B3~1307+433 we obtain $\mathrm{L_{torus}}=1.36\times10^{45}\, \mathrm{erg\, s^{-1}}$ and $\mathrm{R_{torus}}=4.60\times10^{18}\, \mathrm{cm}$,
for GB6~J0114+1325 we obtain $\mathrm{L_{torus}}=1.46\times10^{45} \, \mathrm{erg\, s^{-1}}$ and $\mathrm{R_{torus}}=4.78\times10^{18}\, \mathrm{cm}$,
for PKS~0048-09 we obtain $\mathrm{L_{torus}}=9.68\times10^{42} \, \mathrm{erg\, s^{-1}}$ and $\mathrm{R_{torus}}=3.89\times10^{17}\, \mathrm{cm}$.

For the purpose of the numerical modeling, the free parameters are the energy densities of photons from the BLR and the torus in the reference frame of the plasmoid. 
We compare them to the predicted energy densities as a function of the location of the emitting region $r$, the bulk Lorentz factor $\Gamma$, and $\mathrm{R_{BLR}, L_{BLR}, R_{torus}, L_{torus}}$ as provided above. 
The energy distributions of the external fields are assumed to be a delta function for the BLR (we consider for simplicity that the whole BLR is dominated by the Ly$\alpha$ line), and a black body with temperature $T=420$ K for the torus \citep{Nenkova2008_torus}. 
The energy densities as a function of $r$ are computed following a modified version of \citet{Nalewajko14} approach, where the BLR is considered as a shell with an opening angle of $\pi/4$ along the jet direction. Here, we consider a BLR full shell with internal and external radii of $0.9\ \mathrm{R_{BLR}}$ and $1.1\ \mathrm{R_{BLR}}$ respectively \citep[see][for details]{HESS0736}, as well as a torus reprocessing fraction of $0.4$.

The modeling of the SEDs is provided in Figure \ref{fig:sed_all}, the expected neutrino spectrum is shown in Fig. \ref{fig:sed_all_nu}, and the detailed list of model parameters in Table \ref{tab:model_parameters}. The modeling for TXS~0506+056 can be found in \citet{Acciari2022Mar}.
For all sources, we start by performing a leptonic modeling with a fixed value of $\delta=30$. 
The energy density of the BLR photon field in the reference frame of the emitting region ($u^\prime_{\mathrm{BLR}}$) is a free parameter of the model, as it depends on the location of the emitting region and the Lorentz factor; on the other hand, we assume that the emitting region is always within $\mathrm{R_{torus}}$, which gives thus $u^\prime_{\mathrm{torus}} = 0.028$ erg cm$^{-3}$. Once the leptonic component is able to reproduce the data, we then add a hadronic contribution, maximized in order not to overshoot the constraints in the X-rays. 
In all cases, the modeling results are in line with the observed data, and the discrepancy between data and model at most $2\sigma$ at HE.  
This applies also in the case of B3 1307+433, which shows a spectral hardening at HE, that, however, cannot be fully reproduced by the Inverse Compton emission from a single-zone model such as the one adopted here.

We stress here that the solutions computed for all our sources are sample models. We show that a lepto-hadronic modeling inspired by the ones developed for TXS~0506+056, and assuming as target photon field the radiation from the BLR and the torus, can reproduce the SED, and provide 
the estimated
neutrino spectrum 
from these objects. 
We do not perform an extensive study of the parameter space, nor perform a minimization. In particular, the hadronic component here is not constrained by the data, and the underlying proton distribution is assumed for all models to be a power-law with index $\alpha_p = 2$, and high-energy exponential cut-off at the Lorentz factor $\gamma_{p, Max} = 10^8$.
    \begin{table*}[htbp]
	\centering
	\caption{Parameters of the lepto-hadronic models. The quantities flagged with a star ($^\star$) are derived quantities, and not model parameters. The luminosity of the emitting region was calculated as \mbox{$L=2 \pi R^{\prime 2}c\Gamma^2(u^\prime_{\rm B}+u^\prime_{\rm e}+u^\prime_{\rm p})$}, where $u^\prime_{\rm B}$, $u^\prime_{\rm e}$, and $u^\prime_{\rm p}$ are the energy densities of the magnetic field, the electrons, and the protons, respectively. }
	\label{tab:model_parameters}
	\small
	\begin{tabular}{lcccc}
  		\hline
        Source & PKS 0048-09 low & PKS 0048-09 high & GB6 J0114+1325 & B3 1307+433 \\
        \hline
        $z$ & $0.635$ & $0.635$ & $0.685$ & $0.693$\\
 		$\delta$ & $30$ & $30$ & $30$ & $30$\\
        $\Gamma$ & 20 & 20 & 20 & 20 \\
 $R^\prime$ [10$^{16}$ cm] & $1.25$ & $1.25$ & $1.25$ & $1.25$\\
 $^\star \tau_{\rm var}$ [hours] & $6.3$ & $6.3$ & $6.5$ & $6.5$\\
 		\hline
 		 $B^\prime$ [G] & $1$ & $1$ & $1$ & $1$  \\
 		$^\star u^\prime_B$ [erg cm$^{-3}$] & $0.04$ & $0.04$ & $0.04$ & $0.04$ \\
 		\hline
 		$\gamma_{\rm e,min}^\prime $& $300$ &  $300$ & $300$ & $300$  \\
 		$\gamma_{\rm e,break}^\prime $& $1.6\times10^3$ & $2.6\times10^3$ & $3.0\times10^3$ & $2.2\times10^3$   \\
 		$\gamma_{\rm e,max}^\prime$&    $5\times10^4$ & $7\times10^4$ & $1.6\times10^4$ & $1.5\times10^4$\\ 
 		$\alpha_{\rm e,1}=\alpha_{p,1}$ & $2.0$ & $2.0$ & $2.0$ & $2.0$\\
 		$\alpha_{\rm e,2}$ & $3.35$ & $4.1$ & $3.0$ & $3.5$  \\
 		$K^\prime_{\rm e}$ [cm$^{-3}$] & $1.6\times10^3$ & $4.5\times10^3$ & $8.0\times10^2$ & $8.0\times10^2$\\
 		$^\star u^\prime_{\rm e}$ [erg$\,$cm$^{-3}$] & $3.0\times10^{-3}$ & $9.5\times10^{-3}$ & $1.8\times10^{-3}$ & $1.5\times10^{-3}$\\
 		\hline
 		$\gamma_{\rm p,min}^\prime$& $1$ & $1$ & $1$ & $1$ \\
 		$\gamma_{\rm p,max}^\prime [10^8]$& $1$ & $1$ & $1$ & $1$\\
        $K_{\rm p}^\prime$ [cm$^{-3}$]  & $800$ & $630$ & $80$ & $56$\\
 		$^\star u^\prime_{\rm p}$ [erg$\,$cm$^{-3}$] & $21.4$ &  $16.9$ & $2.1$ & $1.5$\\
 		\hline
 		$u^\prime_{\rm BLR}$ [erg cm$^{-3}$] &$7.0\times10^{-2}$ & $7.0\times10^{-2}$ & $0.12$ & $0.20$ \\
        $u^\prime_{\rm torus}$ [erg cm$^{-3}$] &$2.8\times10^{-2}$ & $2.8\times10^{-2}$ & $2.8\times10^{-2}$ & $2.8\times10^{-2}$  \\
        $^\star r [R_{BLR}]$ &$2.0$ & $2.0$ & $1.9$ &  $1.7$ \\
 		\hline
 		$^\star u^\prime_{\rm e}/u^\prime_{\rm B}$ & $0.077$ & $0.24$ & $0.045$ & $0.039$\\
    	$^\star u^\prime_{\rm p}/u^\prime_{\rm B}$ & $540$ & $420$ & $54$ & $38$\\
 		$^\star L$ [10$^{46}$ erg s$^{-1}$] & $16.6$ & $13.1$ & $1.7$  & $1.2$\\
 		\hline
	\end{tabular}
	\end{table*}

\section{Discussion and Conclusions}
\label{sec:discussion}

In this work we searched for blazar sources with similar observational properties to TXS~0506+056, with the goal of identifying potential similarities in their accretion mechanism and non-thermal processes at work in their jet emission.

We selected the sources from the 4LAC \emph{Fermi}-LAT catalog of AGNs, based on their $\gamma-$ray luminosity, $\gamma-$ray photon index and synchrotron peak frequency. 
These selection criteria excluded a priori all sources without an established value of their redshift, which represent $\sim63\%$ of the BL Lac population in the catalog.
The initial sample we obtained consisted of 27 objects.
However, the need for precise information on the spectral emission lines of the selected sources for the determination of their accretion rate brought us to 
only three blazars, in addition to TXS~0506+056, all classified as BL Lacs.
Indeed, the great majority of the 
27 objects in the initial sample lack a detailed spectral analysis in the literature, and almost half of them lacks any measurement of their optical/UV spectra.
Therefore, we strongly support detailed spectroscopic measurements of the blazars which we had to rule out from the sample, as well as campaigns aimed at measuring the redshift of \emph{Fermi/LAT} blazars.
These measurements will be essential for the knowledge of the luminosity of the sources in the catalog and for a deeper investigation of the selected objects, which will allow us to enlarge the number of sources in the sample and to perform statistical studies.
New spectroscopic measurements of blazars can have an important role also for blazar studies with the upcoming Cherenkov Telescope Array Observatory (CTAO; see e.g. \cite{2024A&A...683A.222D}).

For each of the selected sources we provided a simple estimation of their accretion rate starting from the information on their optical emission lines.
For two out of four sources in the selection (i.~e. B3~1307+433, and GB6 J0114+1325) the values we found for the accretion rate lie above the dividing line set by \citet{division_bll-fsrq}, which may point to a more efficient accretion process and be more consistent with FSRQs, 
while for one source, PKS~0048-09, values below the threshold are found, appearing more compatible with an inefficient accretion flow, suggestive of a BL Lac.
The accretion rate range derived for all selected sources is based on the assumption of two typical mass values for AGNs and is consistent with typical accretion rate estimates in AGNs.
However, for TXS~0506+056 the situation is more delicate, based also on the assumptions made, so a precise measurement of its mass would be needed for a better understanding of its nature.

Public multiwavelength data from the selected sources were analyzed, searching for contemporaneous high and low emission activity.
A flaring state could be identified only for PKS~0048-09, while for the other sources simultaneous periods of average activity were selected.
The SED of the candidates in the selected periods was built with the available multiwavelength data and modeled in the same framework used for TXS~0506+056, assuming a lepto-hadronic emission.
Under the hypothesis that the emitting region is located either within the BLR or beyond it but still within the torus, and thus assuming that the external photon fields from these structures contribute non-negligibly to the emitting region itself, the resulting SED modeling is compatible with a dominant contribution from the EC process.
For PKS~0048-09, the only source that shows a flaring behaviour and for which we modeled two different flux states, a successful modeling of the high state is achieved by changing, with respect to the low state, only the energy distribution of primary electrons in the emitting region, without modifying the parameters of the jet, nor of the external photon field. This explains the higher SSC contribution in the SED of the flaring state of PKS~0048-09 compared to the other sources. 


The hadronic component in the models maximizes the neutrino emission without having a high impact on the electromagnetic emission in most of the cases.
This component emerges mainly in X-rays as the contribution of pion decay and Bethe-Heitler pair production, coming from photo-meson interactions, but the obtained SEDs can be fitted also with a purely leptonic model.
For the source PKS 0048-09, the hadronic component is visible also at VHE, where it arises as the result of photo-meson interactions mostly. A detection in that band would therefore be crucial in constraining this component (see below for a more detailed discussion).
Regarding the X-ray band,
for B3~1307+433 the observations we took thanks to our \emph{Swift} monitoring program play a relevant role in constraining the contribution of the hadronic component, thanks to their high photon statistics and the slope resulting from the analysis.
Therefore, additional measurements in this band are essential for a more precise characterization of the X-ray emission of these sources and to rule out the possibility of a purely leptonic emission.
In particular, additional data from \emph{Swift/XRT} for the objects with few observations would provide a better characterization of the $0.3-10$ keV band, while data from e.~g. \emph{NuSTAR} are essential to characterize the higher energies, since no data in this band are available at the moment for the selected sources.

The models were computed assuming both the BLR and the dusty torus as target photon fields, the photon energy densities of which were 
free model parameters.
The target photon field energy density in the reference frame of the emitting region depends on its position with respect to the accretion disk, the BLR, and the torus. It can thus be used to infer the position of the emitting region in the jet. The values used in the models constrain the emitting region to be just outside the BLR, at a few $R_{BLR}$, since the fit value (see Table~\ref{tab:model_parameters}) is lower than the one estimated thanks to our study on the emission lines, which assumes a region located inside the BLR. 
The value obtained from this estimation is: $u\prime_{BLR}= u_{BLR}\,\Gamma^2 =23.8 \, \mathrm{erg \, cm^{-3}}$, where $u_{BLR}$ is given by Eq.~\ref{eq:uBLR} and $\Gamma$ is the same as is used in the fit.

The SEDs and models obtained for the sources were compared with the sensitivities, for 50 hours of observation, of the imaging air Cherenkov telescopes (IACT) H.E.S.S., MAGIC, VERITAS, and the upcoming Cherenkov Telescope Array Observatory (CTAO), in order to provide perspectives for VHE observations (see Fig.~\ref{fig:sed_all}).
Two of the
selected sources (i. e. B3 1307+433 and GB6 J0114+1325) appear to be 
out of reach for IACTs in their average state. 
B3~1307+433, was observed by VERITAS with an exposure time of 9.4 hours, obtaining a flux upper limit of $17.8\times10^{-12} \, \mathrm{cm^{-2}\, s^{-1}\, TeV^{-1}}$ at $256$ GeV \citep{Archambault2016}.
This is in line with a potential nature as masquerading BL Lacs, for which a stronger absorption at VHE from the source environment is expected with respect to standard, bare, BL Lacs.
Differently, the source PKS 0048-09 exhibit a more pronounced VHE emission, due to the hadronic component. It appears out for reach for the current generation of IACTs, but at the edges of CTAO sensitivity, being therefore detectable by CTAO with about 50 hours of observations in both its low and high states.
The same comparison is also made for the obtained neutrino spectra with the sensitivity of the IceCube Neutrino Observatory for 8 years of data-taking 
\citep{Aartsen2019} and for the future IceCube-Gen2 for 10 years of datataking \citep{icecube-gen2}. For both we considered
a declination of $0^\circ$ or $30^\circ$ (we chose the closest to the source declination).
This comparison is reported in Fig.~\ref{fig:sed_all_nu}. 
Similarly to the VHE detection, the neutrino fluxes expected from B3 1307+433 and GB6 J0114+1325 appear not detectable by the current IceCube Neutrino Observatory and the future IceCube-Gen2, while PKS 0048-09 shows the highest neutrino flux, placing at the edge of IceCube-Gen2 sensitivity and being therefore potentially detectable.
Taking into account the ongoing effort in neutrino astronomy, this source culd be a good target for future experiments, as e.g. HUNT \citep{Huang:2023R8_hunt}, expected to have a volume of about 30 km$^3$.


It is worth noting that this source is also the only one showing a sub-threshold value of the accretion rate, which may suggest the presence of a non-standard BLR in its environment, as discussed in Sec.~\ref{sec:acc_rate}.
This, together with its detectability prospects at VHE, makes it an interesting target for future studies.

Although in this work we decided to concentrate on photo-meson interaction models, which are the most common for blazars, it is worth mentioning that also p-p models can well describe the neutrino emission from blazars. It was pointed out e.g. in \citet{Sahakyan2018Oct, Liu2019Mar, Banik2020Mar}, where the authors considered a general target with typical density and radius values, the BLR clouds, and cold protons in the jet respectively as target material. This kind of models require a lower proton power of the jet with respect to p-$\gamma$ ones, allowing the production of lower-energy neutrinos.
Therefore, even if dense target matter is usually not abundant in blazars environment, it may be interesting investigating also this kind of processes for these objects in future studies.



Finally, a comparison with the blazar PKS~0735+17 is worth mentioning, since this object was proposed as the counterpart of the neutrino event IceCube-211208A\footnote{See \url{https://www.astronomerstelegram.org/?read=15099}.}, which was coincident with flaring activity of the source in several wavebands, and was shown to have a SED very similar to the one of TXS~0506+056 in the work of \citet[][see Fig.~5 of the paper]{pks0735+17_padovani}.
After an inspection of the source parameters reported in the 4LAC-DR2 catalog, we found that the $\gamma-$ray luminosity and spectral index lie in the ranges we defined in Sect.~\ref{sec:selection}.
Regarding the synchrotron peak frequency $\nu_S$, it was found to be quite close to the TXS~0506+056 value during the period of the flare ($\log \nu_S=15.17$) and shortly after ($\log \nu_S=14.13$; \citep{pks0735+17_padovani}).
However, in the catalog this blazar is classified as a BL Lac of the LSP type and the reported value of $\nu_S$ is too low to include the source in the selection ($\log \nu_S=13.43$).
The value in the 4LAC-DR2 refers to an average emission state of the source.
Therefore, the similarities observed in its flaring state are due to the variability of $\nu_S$ with the source activity and the SED in its average state results to be not as similar as suggested.
A further investigation of this object is needed for a better understanding of its similarities and differences with TXS~0506+056.
Indeed, while this study focuses on the averaged emission, where quasi-simultaneous data is more readily available, future research will compare the characteristics of the high state of TXS~0506+065—during which the neutrino event was detected—with those of other gamma-ray AGNs. This comparison aims to shed light on the mechanisms potentially responsible for neutrino emission during their flaring episodes.

\section*{Acknowldgements}
GP. This work is partially supported by ICSC – Centro Nazionale di Ricerca in High Performance Computing, Big Data and Quantum Computing, funded by European Union – NextGenerationEU.
The Fermi LAT Collaboration acknowledges generous ongoing support from a number of agencies and institutes that have supported both the development and the operation of the LAT as well as scientific data analysis. These include the National Aeronautics and Space Administration and the Department of Energy in the United States, the Commissariat à l'Energie Atomique and the Centre National de la Recherche Scientifique / Institut National de Physique Nucléaire et de Physique des Particules in France, the Agenzia Spaziale Italiana and the Istituto Nazionale di Fisica Nucleare in Italy, the Ministry of Education, Culture, Sports, Science and Technology (MEXT), High Energy Accelerator Research Organization (KEK) and Japan Aerospace Exploration Agency (JAXA) in Japan, and the K. A. Wallenberg Foundation, the Swedish Research Council and the Swedish National Space Board in Sweden. 

Additional support for science analysis during the operations phase is gratefully acknowledged from the Istituto Nazionale di Astrofisica in Italy and the Centre National d'Etudes Spatiales in France. This work is performed in part under DOE Contract DE-AC02-76SF00515.

IV thanks Caterina Boscolo Meneguolo and Massimo Persic for useful discussions.


\bibliographystyle{elsarticle-harv} 
\bibliography{elsarticle_jheap}

@article{foschini2022,
	author = {Foschini, Luigi and Lister, Matthew L. and Andernach, Heinz and Ciroi, Stefano and Marziani, Paola and Ant{\ifmmode\acute{o}\else\'{o}\fi}n, Sonia and Berton, Marco and Dalla Bont{\ifmmode\grave{a}\else\`{a}\fi}, Elena and J{\ifmmode\ddot{a}\else\"{a}\fi}rvel{\ifmmode\ddot{a}\else\"{a}\fi}, Emilia and March{\ifmmode\tilde{a}\else\~{a}\fi}, Maria J. M. and Romano, Patrizia and Tornikoski, Merja and Vercellone, Stefano and Vietri, Amelia},
	title = {{A New Sample of Gamma-Ray Emitting Jetted Active Galactic Nuclei}},
	journal = {Universe},
	volume = {8},
	number = {11},
	pages = {587},
	year = {2022},
	month = nov,
	issn = {2218-1997},
	publisher = {Multidisciplinary Digital Publishing Institute},
	doi = {10.3390/universe8110587}
}

@article{Labita2006Dec,
	author = {Labita, M. and Treves, A. and Falomo, R. and Uslenghi, M.},
	title = {{The BH mass of nearby QSOs: a comparison of the bulge luminosity and virial methods}},
	journal = {Mon. Not. R. Astron. Soc.},
	volume = {373},
	number = {2},
	pages = {551--560},
	year = {2006},
	month = dec,
	issn = {0035-8711},
	publisher = {Oxford Academic},
	doi = {10.1111/j.1365-2966.2006.10878.x}
}

@ARTICLE{2024A&A...683A.222D,
       author = {{D'Ammando}, F. and {Goldoni}, P. and {Max-Moerbeck}, W. and {Becerra Gonz{\'a}lez}, J. and {Kasai}, E. and {Williams}, D.~A. and {Alvarez-Crespo}, N. and {Backes}, M. and {Barres de Almeida}, U. and {Boisson}, C. and {Cotter}, G. and {Fallah Ramazani}, V. and {Hervet}, O. and {Lindfors}, E. and {Mukhi-Nilo}, D. and {Pita}, S. and {Splettstoesser}, M. and {van Soelen}, B.},
        title = "{Optical spectroscopy of blazars for the Cherenkov Telescope Array - III}",
      journal = {A\&A},
         year = 2024,
        month = mar,
       volume = {683},
          eid = {A222},
        pages = {A222},
          doi = {10.1051/0004-6361/202348507},
archivePrefix = {arXiv},
       eprint = {2401.07911},
 primaryClass = {astro-ph.HE},
       adsurl = {https://ui.adsabs.harvard.edu/abs/2024A&A...683A.222D}
}

@ARTICLE{2022ApJS..263...24A,
       author = {{Ajello}, M. and {Baldini}, L. and {Ballet}, J. and {Bastieri}, D. and {Becerra Gonzalez}, J. and {Bellazzini}, R. and {Berretta}, A. and {Bissaldi}, E. and {Bonino}, R. and {Brill}, A. and {Bruel}, P. and {Buson}, S. and {Caputo}, R. and {Caraveo}, P.~A. and {Cheung}, C.~C. and {Chiaro}, G. and {Cibrario}, N. and {Ciprini}, S. and {Crnogorcevic}, M. and {Cutini}, S. and {D'Ammando}, F. and {De Gaetano}, S. and {Di Lalla}, N. and {Di Venere}, L. and {Dom{\'\i}nguez}, A. and {Ramazani}, V. Fallah and {Ferrara}, E.~C. and {Fiori}, A. and {Fukazawa}, Y. and {Funk}, S. and {Fusco}, P. and {Gammaldi}, V. and {Gargano}, F. and {Garrappa}, S. and {Gasparrini}, D. and {Giglietto}, N. and {Giordano}, F. and {Giroletti}, M. and {Green}, D. and {Grenier}, I.~A. and {Guiriec}, S. and {Horan}, D. and {Hou}, X. and {Kayanoki}, T. and {Kuss}, M. and {Larsson}, S. and {Latronico}, L. and {Lewis}, T. and {Li}, J. and {Liodakis}, I. and {Longo}, F. and {Loparco}, F. and {Lott}, B. and {Lovellette}, M.~N. and {Lubrano}, P. and {Madejski}, G.~M. and {Maldera}, S. and {Manfreda}, A. and {Mart{\'\i}-Devesa}, G. and {Mazziotta}, M.~N. and {Mereu}, I. and {Michelson}, P.~F. and {Mirabal}, N. and {Mitthumsiri}, W. and {Mizuno}, T. and {Monzani}, M.~E. and {Morselli}, A. and {Moskalenko}, I.~V. and {Negro}, M. and {Ojha}, R. and {Orienti}, M. and {Orlando}, E. and {Ormes}, J.~F. and {Pei}, Z. and {Pe{\~n}a-Herazo}, H. and {Persic}, M. and {Pesce-Rollins}, M. and {Petrosian}, V. and {Pillera}, R. and {Poon}, H. and {Porter}, T.~A. and {Principe}, G. and {Rain{\`o}}, S. and {Rando}, R. and {Rani}, B. and {Razzano}, M. and {Razzaque}, S. and {Reimer}, A. and {Reimer}, O. and {Scotton}, L. and {Serini}, D. and {Sgr{\`o}}, C. and {Siskind}, E.~J. and {Spandre}, G. and {Spinelli}, P. and {Suson}, D.~J. and {Tajima}, H. and {Torres}, D.~F. and {Valverde}, J. and {Yassin}, H. and {Zaharijas}, G.},
        title = "{The Fourth Catalog of Active Galactic Nuclei Detected by the Fermi Large Area Telescope: Data Release 3}",
      journal = {ApJS},
         year = 2022,
        month = dec,
       volume = {263},
       number = {2},
          eid = {24},
        pages = {24},
          doi = {10.3847/1538-4365/ac9523},
archivePrefix = {arXiv},
       eprint = {2209.12070},
 primaryClass = {astro-ph.HE},
       adsurl = {https://ui.adsabs.harvard.edu/abs/2022ApJS..263...24A}
}

@ARTICLE{2022ApJS..260...53A,
       author = {{Abdollahi}, S. and {Acero}, F. and {Baldini}, L. and {Ballet}, J. and {Bastieri}, D. and {Bellazzini}, R. and {Berenji}, B. and {Berretta}, A. and {Bissaldi}, E. and {Blandford}, R.~D. and {Bloom}, E. and {Bonino}, R. and {Brill}, A. and {Britto}, R.~J. and {Bruel}, P. and {Burnett}, T.~H. and {Buson}, S. and {Cameron}, R.~A. and {Caputo}, R. and {Caraveo}, P.~A. and {Castro}, D. and {Chaty}, S. and {Cheung}, C.~C. and {Chiaro}, G. and {Cibrario}, N. and {Ciprini}, S. and {Coronado-Bl{\'a}zquez}, J. and {Crnogorcevic}, M. and {Cutini}, S. and {D'Ammando}, F. and {De Gaetano}, S. and {Digel}, S.~W. and {Di Lalla}, N. and {Dirirsa}, F. and {Di Venere}, L. and {Dom{\'\i}nguez}, A. and {Fallah Ramazani}, V. and {Fegan}, S.~J. and {Ferrara}, E.~C. and {Fiori}, A. and {Fleischhack}, H. and {Franckowiak}, A. and {Fukazawa}, Y. and {Funk}, S. and {Fusco}, P. and {Galanti}, G. and {Gammaldi}, V. and {Gargano}, F. and {Garrappa}, S. and {Gasparrini}, D. and {Giacchino}, F. and {Giglietto}, N. and {Giordano}, F. and {Giroletti}, M. and {Glanzman}, T. and {Green}, D. and {Grenier}, I.~A. and {Grondin}, M. -H. and {Guillemot}, L. and {Guiriec}, S. and {Gustafsson}, M. and {Harding}, A.~K. and {Hays}, E. and {Hewitt}, J.~W. and {Horan}, D. and {Hou}, X. and {J{\'o}hannesson}, G. and {Karwin}, C. and {Kayanoki}, T. and {Kerr}, M. and {Kuss}, M. and {Landriu}, D. and {Larsson}, S. and {Latronico}, L. and {Lemoine-Goumard}, M. and {Li}, J. and {Liodakis}, I. and {Longo}, F. and {Loparco}, F. and {Lott}, B. and {Lubrano}, P. and {Maldera}, S. and {Malyshev}, D. and {Manfreda}, A. and {Mart{\'\i}-Devesa}, G. and {Mazziotta}, M.~N. and {Mereu}, I. and {Meyer}, M. and {Michelson}, P.~F. and {Mirabal}, N. and {Mitthumsiri}, W. and {Mizuno}, T. and {Moiseev}, A.~A. and {Monzani}, M.~E. and {Morselli}, A. and {Moskalenko}, I.~V. and {Negro}, M. and {Nuss}, E. and {Omodei}, N. and {Orienti}, M. and {Orlando}, E. and {Paneque}, D. and {Pei}, Z. and {Perkins}, J.~S. and {Persic}, M. and {Pesce-Rollins}, M. and {Petrosian}, V. and {Pillera}, R. and {Poon}, H. and {Porter}, T.~A. and {Principe}, G. and {Rain{\`o}}, S. and {Rando}, R. and {Rani}, B. and {Razzano}, M. and {Razzaque}, S. and {Reimer}, A. and {Reimer}, O. and {Reposeur}, T. and {S{\'a}nchez-Conde}, M. and {Saz Parkinson}, P.~M. and {Scotton}, L. and {Serini}, D. and {Sgr{\`o}}, C. and {Siskind}, E.~J. and {Smith}, D.~A. and {Spandre}, G. and {Spinelli}, P. and {Sueoka}, K. and {Suson}, D.~J. and {Tajima}, H. and {Tak}, D. and {Thayer}, J.~B. and {Thompson}, D.~J. and {Torres}, D.~F. and {Troja}, E. and {Valverde}, J. and {Wood}, K. and {Zaharijas}, G.},
        title = "{Incremental Fermi Large Area Telescope Fourth Source Catalog}",
      journal = {ApJS},
         year = 2022,
        month = jun,
       volume = {260},
       number = {2},
          eid = {53},
        pages = {53},
          doi = {10.3847/1538-4365/ac6751},
archivePrefix = {arXiv},
       eprint = {2201.11184},
 primaryClass = {astro-ph.HE},
       adsurl = {https://ui.adsabs.harvard.edu/abs/2022ApJS..260...53A}
}

@ARTICLE{IC2017_system,
       author = {{Aartsen}, M.~G. and {Ackermann}, M. and {Adams}, J. and {Aguilar}, J.~A. and {Ahlers}, M. and {Ahrens}, M. and {Altmann}, D. and {Andeen}, K. and {Anderson}, T. and {Ansseau}, I. and {Anton}, G. and {Archinger}, M. and {Arg{\"u}elles}, C. and {Auer}, R. and {Auffenberg}, J. and {Axani}, S. and {Baccus}, J. and {Bai}, X. and {Barnet}, S. and {Barwick}, S.~W. and {Baum}, V. and {Bay}, R. and {Beattie}, K. and {Beatty}, J.~J. and {Becker Tjus}, J. and {Becker}, K. -H. and {Bendfelt}, T. and {BenZvi}, S. and {Berley}, D. and {Bernardini}, E. and {Bernhard}, A. and {Besson}, D.~Z. and {Binder}, G. and {Bindig}, D. and {Bissok}, M. and {Blaufuss}, E. and {Blot}, S. and {Boersma}, D. and {Bohm}, C. and {B{\"o}rner}, M. and {Bos}, F. and {Bose}, D. and {B{\"o}ser}, S. and {Botner}, O. and {Bouchta}, A. and {Braun}, J. and {Brayeur}, L. and {Bretz}, H. -P. and {Bron}, S. and {Burgman}, A. and {Burreson}, C. and {Carver}, T. and {Casier}, M. and {Cheung}, E. and {Chirkin}, D. and {Christov}, A. and {Clark}, K. and {Classen}, L. and {Coenders}, S. and {Collin}, G.~H. and {Conrad}, J.~M. and {Cowen}, D.~F. and {Cross}, R. and {Day}, C. and {Day}, M. and {de Andr{\'e}}, J.~P.~A.~M. and {De Clercq}, C. and {del Pino Rosendo}, E. and {Dembinski}, H. and {De Ridder}, S. and {Descamps}, F. and {Desiati}, P. and {de Vries}, K.~D. and {de Wasseige}, G. and {de With}, M. and {DeYoung}, T. and {D{\'\i}az-V{\'e}lez}, J.~C. and {di Lorenzo}, V. and {Dujmovic}, H. and {Dumm}, J.~P. and {Dunkman}, M. and {Eberhardt}, B. and {Edwards}, W.~R. and {Ehrhardt}, T. and {Eichmann}, B. and {Eller}, P. and {Euler}, S. and {Evenson}, P.~A. and {Fahey}, S. and {Fazely}, A.~R. and {Feintzeig}, J. and {Felde}, J. and {Filimonov}, K. and {Finley}, C. and {Flis}, S. and {F{\"o}sig}, C. -C. and {Franckowiak}, A. and {Fr{\`e}re}, M. and {Friedman}, E. and {Fuchs}, T. and {Gaisser}, T.~K. and {Gallagher}, J. and {Gerhardt}, L. and {Ghorbani}, K. and {Giang}, W. and {Gladstone}, L. and {Glauch}, T. and {Glowacki}, D. and {Gl{\"u}senkamp}, T. and {Goldschmidt}, A. and {Gonzalez}, J.~G. and {Grant}, D. and {Griffith}, Z. and {Gustafsson}, L. and {Haack}, C. and {Hallgren}, A. and {Halzen}, F. and {Hansen}, E. and {Hansmann}, T. and {Hanson}, K. and {Haugen}, J. and {Hebecker}, D. and {Heereman}, D. and {Helbing}, K. and {Hellauer}, R. and {Heller}, R. and {Hickford}, S. and {Hignight}, J. and {Hill}, G.~C. and {Hoffman}, K.~D. and {Hoffmann}, R. and {Hoshina}, K. and {Huang}, F. and {Huber}, M. and {Hulth}, P.~O. and {Hultqvist}, K. and {In}, S. and {Inaba}, M. and {Ishihara}, A. and {Jacobi}, E. and {Jacobsen}, J. and {Japaridze}, G.~S. and {Jeong}, M. and {Jero}, K. and {Jones}, A. and {Jones}, B.~J.~P. and {Joseph}, J. and {Kang}, W. and {Kappes}, A. and {Karg}, T. and {Karle}, A. and {Katz}, U. and {Kauer}, M. and {Keivani}, A. and {Kelley}, J.~L. and {Kemp}, J. and {Kheirandish}, A. and {Kim}, J. and {Kim}, M. and {Kintscher}, T. and {Kiryluk}, J. and {Kitamura}, N. and {Kittler}, T. and {Klein}, S.~R. and {Kleinfelder}, S. and {Kleist}, M. and {Kohnen}, G. and {Koirala}, R. and {Kolanoski}, H. and {Konietz}, R. and {K{\"o}pke}, L. and {Kopper}, C. and {Kopper}, S. and {Koskinen}, D.~J. and {Kowalski}, M. and {Krasberg}, M. and {Krings}, K. and {Kroll}, M. and {Kr{\"u}ckl}, G. and {Kr{\"u}ger}, C. and {Kunnen}, J. and {Kunwar}, S. and {Kurahashi}, N. and {Kuwabara}, T. and {Labare}, M. and {Laihem}, K. and {Landsman}, H. and {Lanfranchi}, J.~L. and {Larson}, M.~J. and {Lauber}, F. and {Laundrie}, A. and {Lennarz}, D. and {Leich}, H. and {Lesiak-Bzdak}, M. and {Leuermann}, M. and {Lu}, L. and {Ludwig}, J. and {L{\"u}nemann}, J. and {Mackenzie}, C. and {Madsen}, J. and {Maggi}, G. and {Mahn}, K.~B.~M. and {Mancina}, S. and {Mandelartz}, M. and {Maruyama}, R. and {Mase}, K. and {Matis}, H. and {Maunu}, R. and {McNally}, F. and {McParland}, C.~P. and {Meade}, P. and {Meagher}, K. and {Medici}, M. and {Meier}, M. and {Meli}, A. and {Menne}, T. and {Merino}, G. and {Meures}, T. and {Miarecki}, S. and {Minor}, R.~H. and {Montaruli}, T. and {Moulai}, M. and {Murray}, T. and {Nahnhauer}, R. and {Naumann}, U. and {Neer}, G. and {Newcomb}, M. and {Niederhausen}, H. and {Nowicki}, S.~C. and {Nygren}, D.~R. and {Obertacke Pollmann}, A. and {Olivas}, A. and {O'Murchadha}, A. and {Palczewski}, T. and {Pandya}, H. and {Pankova}, D.~V. and {Patton}, S. and {Peiffer}, P. and {Penek}, {\"O}. and {Pepper}, J.~A. and {P{\'e}rez de los Heros}, C. and {Pettersen}, C. and {Pieloth}, D. and {Pinat}, E. and {Price}, P.~B. and {Przybylski}, G.~T. and {Quinnan}, M. and {Raab}, C. and {R{\"a}del}, L. and {Rameez}, M. and {Rawlins}, K. and {Reimann}, R. and {Relethford}, B. and {Relich}, M. and {Resconi}, E. and {Rhode}, W. and {Richman}, M. and {Riedel}, B. and {Robertson}, S. and {Rongen}, M. and {Roucelle}, C. and {Rott}, C. and {Ruhe}, T. and {Ryckbosch}, D. and {Rysewyk}, D. and {Sabbatini}, L. and {Sanchez Herrera}, S.~E. and {Sandrock}, A. and {Sandroos}, J. and {Sandstrom}, P. and {Sarkar}, S. and {Satalecka}, K. and {Schlunder}, P. and {Schmidt}, T. and {Schoenen}, S. and {Sch{\"o}neberg}, S. and {Schukraft}, A. and {Schumacher}, L. and {Seckel}, D. and {Seunarine}, S. and {Solarz}, M. and {Soldin}, D. and {Song}, M. and {Spiczak}, G.~M. and {Spiering}, C. and {Stanev}, T. and {Stasik}, A. and {Stettner}, J. and {Steuer}, A. and {Stezelberger}, T. and {Stokstad}, R.~G. and {St{\"o}{\ss}l}, A. and {Str{\"o}m}, R. and {Strotjohann}, N.~L. and {Sulanke}, K. -H. and {Sullivan}, G.~W. and {Sutherland}, M. and {Taavola}, H. and {Taboada}, I. and {Tatar}, J. and {Tenholt}, F. and {Ter-Antonyan}, S. and {Terliuk}, A. and {Te{\v{s}}i{\'c}}, G. and {Thollander}, L. and {Tilav}, S. and {Toale}, P.~A. and {Tobin}, M.~N. and {Toscano}, S. and {Tosi}, D. and {Tselengidou}, M. and {Turcati}, A. and {Unger}, E. and {Usner}, M. and {Vandenbroucke}, J. and {van Eijndhoven}, N. and {Vanheule}, S. and {van Rossem}, M. and {van Santen}, J. and {Vehring}, M. and {Voge}, M. and {Vogel}, E. and {Vraeghe}, M. and {Wahl}, D. and {Walck}, C. and {Wallace}, A. and {Wallraff}, M. and {Wandkowsky}, N. and {Weaver}, Ch. and {Weiss}, M.~J. and {Wendt}, C. and {Westerhoff}, S. and {Wharton}, D. and {Whelan}, B.~J. and {Wickmann}, S. and {Wiebe}, K. and {Wiebusch}, C.~H. and {Wille}, L. and {Williams}, D.~R. and {Wills}, L. and {Wisniewski}, P. and {Wolf}, M. and {Wood}, T.~R. and {Woolsey}, E. and {Woschnagg}, K. and {Xu}, D.~L. and {Xu}, X.~W. and {Xu}, Y. and {Yanez}, J.~P. and {Yodh}, G. and {Yoshida}, S. and {Zoll}, M.},
        title = "{The IceCube Neutrino Observatory: instrumentation and online systems}",
      journal = {Journal of Instrumentation},
         year = 2017,
        month = mar,
       volume = {12},
       number = {3},
        pages = {P03012},
          doi = {10.1088/1748-0221/12/03/P03012},
archivePrefix = {arXiv},
       eprint = {1612.05093},
 primaryClass = {astro-ph.IM},
       adsurl = {https://ui.adsabs.harvard.edu/abs/2017JInst..12P3012A}
}

@article{IceCube:2013low,
    author = "Aartsen, M. G. and others",
    collaboration = "IceCube",
    title = "{Evidence for High-Energy Extraterrestrial Neutrinos at the IceCube Detector}",
    eprint = "1311.5238",
    archivePrefix = "arXiv",
    primaryClass = "astro-ph.HE",
    doi = "10.1126/science.1242856",
    journal = "Science",
    volume = "342",
    pages = "1242856",
    year = "2013"
}

@article{TXS_2017,
author = {The IceCube Collaboration and others},
title = {Multimessenger observations of a flaring blazar coincident with high-energy neutrino IceCube-170922A},
journal = {Science},
volume = {361},
number = {6398},
pages = {eaat1378},
year = {2018},
doi = {10.1126/science.aat1378},
URL = {https://www.science.org/doi/abs/10.1126/science.aat1378},
eprint = {https://www.science.org/doi/pdf/10.1126/science.aat1378},
}

@article{Gao2019_txs,
	author = {Gao, Shan and Fedynitch, Anatoli and Winter, Walter and Pohl, Martin},
	title = {{Modelling the coincident observation of a high-energy neutrino and a bright blazar flare}},
	journal = {Nat. Astron.},
	volume = {3},
	pages = {88--92},
	year = {2019},
	month = jan,
	issn = {2397-3366},
	publisher = {Nature Publishing Group},
	doi = {10.1038/s41550-018-0610-1}
}

@article{Keivani2018_txs,
	author = {Keivani, A. and Murase, K. and Petropoulou, M. and Fox, D. B. and Cenko, S. B. and Chaty, S. and Coleiro, A. and DeLaunay, J. J. and Dimitrakoudis, S. and Evans, P. A. and Kennea, J. A. and Marshall, F. E. and Mastichiadis, A. and Osborne, J. P. and Santander, M. and Tohuvavohu, A. and Turley, C. F.},
	title = {{A Multimessenger Picture of the Flaring Blazar TXS 0506+056: Implications for High-energy Neutrino Emission and Cosmic-Ray Acceleration}},
	journal = {Astrophys. J.},
	volume = {864},
	number = {1},
	pages = {84},
	year = {2018},
	month = aug,
	issn = {0004-637X},
	publisher = {The American Astronomical Society},
	doi = {10.3847/1538-4357/aad59a}
}

@article{IceCube:2022der,
    author = "Abbasi, R. and others",
    collaboration = "IceCube",
    title = "{Evidence for neutrino emission from the nearby active galaxy NGC 1068}",
    eprint = "2211.09972",
    archivePrefix = "arXiv",
    primaryClass = "astro-ph.HE",
    doi = "10.1126/science.abg3395",
    journal = "Science",
    volume = "378",
    number = "6619",
    pages = "538--543",
    year = "2022"
}

@article{nu_from_GalPlane,
author = {{IceCube Collaboration} and R. Abbasi  and M. Ackermann  and J. Adams  and J. A. Aguilar  and M. Ahlers  and M. Ahrens  and J. M. Alameddine  and A. A. Alves  and N. M. Amin  and K. Andeen  and T. Anderson  and G. Anton  and C. Argüelles  and Y. Ashida  and S. Athanasiadou  and S. Axani  and X. Bai  and A. Balagopal V.  and S. W. Barwick  and V. Basu  and S. Baur  and R. Bay  and J. J. Beatty  and K.-H. Becker  and J. Becker Tjus  and J. Beise  and C. Bellenghi  and S. Benda  and S. BenZvi  and D. Berley  and E. Bernardini  and D. Z. Besson  and G. Binder  and D. Bindig  and E. Blaufuss  and S. Blot  and M. Boddenberg  and F. Bontempo  and J. Y. Book  and J. Borowka  and S. Böser  and O. Botner  and J. Böttcher  and E. Bourbeau  and F. Bradascio  and J. Braun  and B. Brinson  and S. Bron  and J. Brostean-Kaiser  and R. T. Burley  and R. S. Busse  and M. A. Campana  and E. G. Carnie-Bronca  and C. Chen  and Z. Chen  and D. Chirkin  and K. Choi  and B. A. Clark  and K. Clark  and L. Classen  and A. Coleman  and G. H. Collin  and A. Connolly  and J. M. Conrad  and P. Coppin  and P. Correa  and D. F. Cowen  and R. Cross  and C. Dappen  and P. Dave  and C. De Clercq  and J. J. DeLaunay  and D. Delgado López  and H. Dembinski  and K. Deoskar  and A. Desai  and P. Desiati  and K. D. de Vries  and G. de Wasseige  and T. DeYoung  and A. Diaz  and J. C. Díaz-Vélez  and M. Dittmer  and H. Dujmovic  and M. Dunkman  and M. A. DuVernois  and T. Ehrhardt  and P. Eller  and R. Engel  and H. Erpenbeck  and J. Evans  and P. A. Evenson  and K. L. Fan  and A. R. Fazely  and A. Fedynitch  and N. Feigl  and S. Fiedlschuster  and A. T. Fienberg  and C. Finley  and L. Fischer  and D. Fox  and A. Franckowiak  and E. Friedman  and A. Fritz  and P. Fürst  and T. K. Gaisser  and J. Gallagher  and E. Ganster  and A. Garcia  and S. Garrappa  and L. Gerhardt  and A. Ghadimi  and C. Glaser  and T. Glauch  and T. Glüsenkamp  and N. Goehlke  and A. Goldschmidt  and J. G. Gonzalez  and S. Goswami  and D. Grant  and T. Grégoire  and S. Griswold  and C. Günther  and P. Gutjahr  and C. Haack  and A. Hallgren  and R. Halliday  and L. Halve  and F. Halzen  and M. Ha Minh  and K. Hanson  and J. Hardin  and A. A. Harnisch  and A. Haungs  and K. Helbing  and F. Henningsen  and E. C. Hettinger  and S. Hickford  and J. Hignight  and C. Hill  and G. C. Hill  and K. D. Hoffman  and K. Hoshina  and W. Hou  and F. Huang  and M. Huber  and T. Huber  and K. Hultqvist  and M. Hünnefeld  and R. Hussain  and K. Hymon  and S. In  and N. Iovine  and A. Ishihara  and M. Jansson  and G. S. Japaridze  and M. Jeong  and M. Jin  and B. J. P. Jones  and D. Kang  and W. Kang  and X. Kang  and A. Kappes  and D. Kappesser  and L. Kardum  and T. Karg  and M. Karl  and A. Karle  and U. Katz  and M. Kauer  and M. Kellermann  and J. L. Kelley  and A. Kheirandish  and K. Kin  and J. Kiryluk  and S. R. Klein  and A. Kochocki  and R. Koirala  and H. Kolanoski  and T. Kontrimas  and L. Köpke  and C. Kopper  and S. Kopper  and D. J. Koskinen  and P. Koundal  and M. Kovacevich  and M. Kowalski  and T. Kozynets  and E. Krupczak  and E. Kun  and N. Kurahashi  and N. Lad  and C. Lagunas Gualda  and J. L. Lanfranchi  and M. J. Larson  and F. Lauber  and J. P. Lazar  and J. W. Lee  and K. Leonard  and A. Leszczyńska  and Y. Li  and M. Lincetto  and Q. R. Liu  and M. Liubarska  and E. Lohfink  and C. J. Lozano Mariscal  and L. Lu  and F. Lucarelli  and A. Ludwig  and W. Luszczak  and Y. Lyu  and W. Y. Ma  and J. Madsen  and K. B. M. Mahn  and Y. Makino  and S. Mancina  and I. C. Mariş  and I. Martinez-Soler  and R. Maruyama  and S. McCarthy  and T. McElroy  and F. McNally  and J. V. Mead  and K. Meagher  and S. Mechbal  and A. Medina  and M. Meier  and S. Meighen-Berger  and Y. Merckx  and J. Micallef  and D. Mockler  and T. Montaruli  and R. W. Moore  and K. Morik  and R. Morse  and M. Moulai  and T. Mukherjee  and R. Naab  and R. Nagai  and R. Nahnhauer  and U. Naumann  and J. Necker  and L. V. Nguyen  and H. Niederhausen  and M. U. Nisa  and S. C. Nowicki  and D. Nygren  and A. Obertacke Pollmann  and M. Oehler  and B. Oeyen  and A. Olivas  and E. O'Sullivan  and H. Pandya  and D. V. Pankova  and N. Park  and G. K. Parker  and E. N. Paudel  and L. Paul  and C. Pérez de los Heros  and L. Peters  and J. Peterson  and S. Philippen  and S. Pieper  and A. Pizzuto  and M. Plum  and Y. Popovych  and A. Porcelli  and M. Prado Rodriguez  and B. Pries  and G. T. Przybylski  and C. Raab  and J. Rack-Helleis  and A. Raissi  and M. Rameez  and K. Rawlins  and I. C. Rea  and Z. Rechav  and A. Rehman  and P. Reichherzer  and R. Reimann  and G. Renzi  and E. Resconi  and S. Reusch  and W. Rhode  and M. Richman  and B. Riedel  and E. J. Roberts  and S. Robertson  and G. Roellinghoff  and M. Rongen  and C. Rott  and T. Ruhe  and D. Ryckbosch  and D. Rysewyk Cantu  and I. Safa  and J. Saffer  and D. Salazar-Gallegos  and P. Sampathkumar  and S. E. Sanchez Herrera  and A. Sandrock  and M. Santander  and S. Sarkar  and S. Sarkar  and K. Satalecka  and M. Schaufel  and H. Schieler  and S. Schindler  and T. Schmidt  and A. Schneider  and J. Schneider  and F. G. Schröder  and L. Schumacher  and G. Schwefer  and S. Sclafani  and D. Seckel  and S. Seunarine  and A. Sharma  and S. Shefali  and N. Shimizu  and M. Silva  and B. Skrzypek  and B. Smithers  and R. Snihur  and J. Soedingrekso  and A. Sogaard  and D. Soldin  and C. Spannfellner  and G. M. Spiczak  and C. Spiering  and M. Stamatikos  and T. Stanev  and R. Stein  and J. Stettner  and T. Stezelberger  and B. Stokstad  and T. Stürwald  and T. Stuttard  and G. W. Sullivan  and I. Taboada  and S. Ter-Antonyan  and J. Thwaites  and S. Tilav  and F. Tischbein  and K. Tollefson  and C. Tönnis  and S. Toscano  and D. Tosi  and A. Trettin  and M. Tselengidou  and C. F. Tung  and A. Turcati  and R. Turcotte  and C. F. Turley  and J. P. Twagirayezu  and B. Ty  and M. A. Unland Elorrieta  and N. Valtonen-Mattila  and J. Vandenbroucke  and N. van Eijndhoven  and D. Vannerom  and J. van Santen  and J. Veitch-Michaelis  and S. Verpoest  and C. Walck  and W. Wang  and T. B. Watson  and C. Weaver  and P. Weigel  and A. Weindl  and M. J. Weiss  and J. Weldert  and C. Wendt  and J. Werthebach  and M. Weyrauch  and N. Whitehorn  and C. H. Wiebusch  and N. Willey  and D. R. Williams  and M. Wolf  and G. Wrede  and J. Wulff  and X. W. Xu  and J. P. Yanez  and E. Yildizci  and S. Yoshida  and S. Yu  and T. Yuan  and Z. Zhang  and P. Zhelnin },
title = {Observation of high-energy neutrinos from the Galactic plane},
journal = {Science},
volume = {380},
number = {6652},
pages = {1338-1343},
year = {2023},
doi = {10.1126/science.adc9818},
URL = {https://www.science.org/doi/abs/10.1126/science.adc9818},
eprint = {https://www.science.org/doi/pdf/10.1126/science.adc9818},
}

@ARTICLE{Ansoldi2018,
       author = {{Ansoldi}, S. and {Antonelli}, L.~A. and {Arcaro}, C. and {Baack}, D. and {Babi{\'c}}, A. and {Banerjee}, B. and {Bangale}, P. and {Barres de Almeida}, U. and {Barrio}, J.~A. and {Becerra Gonz{\'a}lez}, J. and {Bednarek}, W. and {Bernardini}, E. and {Berse}, R. Ch. and {Berti}, A. and {Besenrieder}, J. and {Bhattacharyya}, W. and {Bigongiari}, C. and {Biland}, A. and {Blanch}, O. and {Bonnoli}, G. and {Carosi}, R. and {Ceribella}, G. and {Chatterjee}, A. and {Colak}, S.~M. and {Colin}, P. and {Colombo}, E. and {Contreras}, J.~L. and {Cortina}, J. and {Covino}, S. and {Cumani}, P. and {D'Elia}, V. and {Da Vela}, P. and {Dazzi}, F. and {De Angelis}, A. and {De Lotto}, B. and {Delfino}, M. and {Delgado}, J. and {Di Pierro}, F. and {Dom{\'\i}nguez}, A. and {Dominis Prester}, D. and {Dorner}, D. and {Doro}, M. and {Einecke}, S. and {Elsaesser}, D. and {Fallah Ramazani}, V. and {Fattorini}, A. and {Fern{\'a}ndez-Barral}, A. and {Ferrara}, G. and {Fidalgo}, D. and {Foffano}, L. and {Fonseca}, M.~V. and {Font}, L. and {Fruck}, C. and {Gallozzi}, S. and {Garc{\'\i}a L{\'o}pez}, R.~J. and {Garczarczyk}, M. and {Gaug}, M. and {Giammaria}, P. and {Godinovi{\'c}}, N. and {Guberman}, D. and {Hadasch}, D. and {Hahn}, A. and {Hassan}, T. and {Hayashida}, M. and {Herrera}, J. and {Hoang}, J. and {Hrupec}, D. and {Inoue}, S. and {Ishio}, K. and {Iwamura}, Y. and {Konno}, Y. and {Kubo}, H. and {Kushida}, J. and {Lamastra}, A. and {Lelas}, D. and {Leone}, F. and {Lindfors}, E. and {Lombardi}, S. and {Longo}, F. and {L{\'o}pez}, M. and {Maggio}, C. and {Majumdar}, P. and {Makariev}, M. and {Maneva}, G. and {Manganaro}, M. and {Mannheim}, K. and {Maraschi}, L. and {Mariotti}, M. and {Mart{\'\i}nez}, M. and {Masuda}, S. and {Mazin}, D. and {Mielke}, K. and {Minev}, M. and {Miranda}, J.~M. and {Mirzoyan}, R. and {Moralejo}, A. and {Moreno}, V. and {Moretti}, E. and {Neustroev}, V. and {Niedzwiecki}, A. and {Nievas Rosillo}, M. and {Nigro}, C. and {Nilsson}, K. and {Ninci}, D. and {Nishijima}, K. and {Noda}, K. and {Nogu{\'e}s}, L. and {Paiano}, S. and {Palacio}, J. and {Paneque}, D. and {Paoletti}, R. and {Paredes}, J.~M. and {Pedaletti}, G. and {Pe{\~n}il}, P. and {Peresano}, M. and {Persic}, M. and {Pfrang}, K. and {Prada Moroni}, P.~G. and {Prandini}, E. and {Puljak}, I. and {Garcia}, J.~R. and {Rhode}, W. and {Rib{\'o}}, M. and {Rico}, J. and {Righi}, C. and {Rugliancich}, A. and {Saha}, L. and {Saito}, T. and {Satalecka}, K. and {Schweizer}, T. and {Sitarek}, J. and {{\v{S}}nidari{\'c}}, I. and {Sobczynska}, D. and {Stamerra}, A. and {Strzys}, M. and {Suri{\'c}}, T. and {Tavecchio}, F. and {Temnikov}, P. and {Terzi{\'c}}, T. and {Teshima}, M. and {Torres-Alb{\'a}}, N. and {Tsujimoto}, S. and {Vanzo}, G. and {Vazquez Acosta}, M. and {Vovk}, I. and {Ward}, J.~E. and {Will}, M. and {Zari{\'c}}, D. and {Cerruti}, Matteo},
        title = "{The Blazar TXS 0506+056 Associated with a High-energy Neutrino: Insights into Extragalactic Jets and Cosmic-Ray Acceleration}",
      journal = {ApJL},
         year = 2018,
        month = aug,
       volume = {863},
       number = {1},
          eid = {L10},
        pages = {L10},
          doi = {10.3847/2041-8213/aad083},
archivePrefix = {arXiv},
       eprint = {1807.04300},
 primaryClass = {astro-ph.HE},
       adsurl = {https://ui.adsabs.harvard.edu/abs/2018ApJ...863L..10A}
}

@article{Urry1995_bll-fsrq-ew,
	author = {Urry, C. Megan and Padovani, Paolo},
	title = {{UNIFIED SCHEMES FOR RADIO-LOUD ACTIVE GALACTIC NUCLEI}},
	journal = {Publ. Astron. Soc. Pac.},
	volume = {107},
	number = {715},
	pages = {803},
	year = {1995},
	month = sep,
	issn = {1538-3873},
	publisher = {The Astronomical Society of the Pacific},
	doi = {10.1086/133630}
}

@article{Ghisellini2009,
	author = {Ghisellini, G. and Maraschi, L. and Tavecchio, F.},
	title = {{The Fermi blazars' divide}},
	journal = {Mon. Not. R. Astron. Soc. Lett.},
	volume = {396},
	number = {1},
	pages = {L105--L109},
	year = {2009},
	month = jun,
	issn = {1745-3925},
	publisher = {Oxford Academic},
	doi = {10.1111/j.1745-3933.2009.00673.x}
}

@article{Fossati1998,
	author = {Fossati, G. and Maraschi, L. and Celotti, A. and Comastri, A. and Ghisellini, G.},
	title = "{A unifying view of the spectral energy distributions of blazars}",
	journal = {Monthly Notices of the Royal Astronomical Society},
	volume = {299},
	number = {2},
	pages = {433-448},
	year = {1998},
	month = {09},
	issn = {0035-8711},
	doi = {10.1046/j.1365-8711.1998.01828.x},
	url = {https://doi.org/10.1046/j.1365-8711.1998.01828.x},
	eprint = {https://academic.oup.com/mnras/article-pdf/299/2/433/3340292/299-2-433.pdf},
}

@article{GhiselliniTavecchio2008,
	author = {Ghisellini, G. and Tavecchio, F.},
	title = {{The blazar sequence: a new perspective}},
	journal = {Mon. Not. R. Astron. Soc.},
	volume = {387},
	number = {4},
	pages = {1669--1680},
	year = {2008},
	month = jul,
	issn = {0035-8711},
	publisher = {Oxford Academic},
	doi = {10.1111/j.1365-2966.2008.13360.x}
}

@ARTICLE{HESS0736,
       author = {{H.~E.~S.~S. Collaboration} and {Abdalla}, H. and {Adam}, R. and {Aharonian}, F. and {Ait Benkhali}, F. and {Ang{\"u}ner}, E.~O. and {Arakawa}, M. and {Arcaro}, C. and {Armand}, C. and {Ashkar}, H. and {Backes}, M. and {Barbosa Martins}, V. and {Barnard}, M. and {Becherini}, Y. and {Berge}, D. and {Bernl{\"o}hr}, K. and {Blackwell}, R. and {B{\"o}ttcher}, M. and {Boisson}, C. and {Bolmont}, J. and {Bonnefoy}, S. and {Bregeon}, J. and {Breuhaus}, M. and {Brun}, F. and {Brun}, P. and {Bryan}, M. and {B{\"u}chele}, M. and {Bulik}, T. and {Bylund}, T. and {Capasso}, M. and {Caroff}, S. and {Carosi}, A. and {Casanova}, S. and {Cerruti}, M. and {Chand}, T. and {Chandra}, S. and {Chen}, A. and {Colafrancesco}, S. and {Cury{\l}o}, M. and {Davids}, I.~D. and {Deil}, C. and {Devin}, J. and {deWilt}, P. and {Dirson}, L. and {Djannati-Ata{\"\i}}, A. and {Dmytriiev}, A. and {Donath}, A. and {Doroshenko}, V. and {Drury}, L.~O. 'C. and {Dyks}, J. and {Egberts}, K. and {Emery}, G. and {Ernenwein}, J. -P. and {Eschbach}, S. and {Feijen}, K. and {Fegan}, S. and {Fiasson}, A. and {Fontaine}, G. and {Funk}, S. and {F{\"u}{\ss}ling}, M. and {Gabici}, S. and {Gallant}, Y.~A. and {Gat{\'e}}, F. and {Giavitto}, G. and {Glawion}, D. and {Glicenstein}, J.~F. and {Gottschall}, D. and {Grondin}, M. -H. and {Hahn}, J. and {Haupt}, M. and {Heinzelmann}, G. and {Henri}, G. and {Hermann}, G. and {Hinton}, J.~A. and {Hofmann}, W. and {Hoischen}, C. and {Holch}, T.~L. and {Holler}, M. and {Horns}, D. and {Huber}, D. and {Iwasaki}, H. and {Jamrozy}, M. and {Jankowsky}, D. and {Jankowsky}, F. and {Jardin-Blicq}, A. and {Jung-Richardt}, I. and {Kastendieck}, M.~A. and {Katarzy{\'n}ski}, K. and {Katsuragawa}, M. and {Katz}, U. and {Khangulyan}, D. and {Kh{\'e}lifi}, B. and {King}, J. and {Klepser}, S. and {Klu{\'z}niak}, W. and {Komin}, Nu. and {Kosack}, K. and {Kostunin}, D. and {Kraus}, M. and {Lamanna}, G. and {Lau}, J. and {Lemi{\`e}re}, A. and {Lemoine-Goumard}, M. and {Lenain}, J. -P. and {Leser}, E. and {Levy}, C. and {Lohse}, T. and {Lypova}, I. and {Mackey}, J. and {Majumdar}, J. and {Malyshev}, D. and {Marandon}, V. and {Marcowith}, A. and {Mares}, A. and {Mariaud}, C. and {Mart{\'\i}-Devesa}, G. and {Marx}, R. and {Maurin}, G. and {Meintjes}, P.~J. and {Mitchell}, A.~M.~W. and {Moderski}, R. and {Mohamed}, M. and {Mohrmann}, L. and {Muller}, J. and {Moore}, C. and {Moulin}, E. and {Murach}, T. and {Nakashima}, S. and {de Naurois}, M. and {Ndiyavala}, H. and {Niederwanger}, F. and {Niemiec}, J. and {Oakes}, L. and {O'Brien}, P. and {Odaka}, H. and {Ohm}, S. and {de O{\~n}a Wilhelmi}, E. and {Ostrowski}, M. and {Oya}, I. and {Panter}, M. and {Parsons}, R.~D. and {Perennes}, C. and {Petrucci}, P. -O. and {Peyaud}, B. and {Piel}, Q. and {Pita}, S. and {Poireau}, V. and {Priyana Noel}, A. and {Prokhorov}, D.~A. and {Prokoph}, H. and {P{\"u}hlhofer}, G. and {Punch}, M. and {Quirrenbach}, A. and {Raab}, S. and {Rauth}, R. and {Reimer}, A. and {Reimer}, O. and {Remy}, Q. and {Renaud}, M. and {Rieger}, F. and {Rinchiuso}, L. and {Romoli}, C. and {Rowell}, G. and {Rudak}, B. and {Ruiz-Velasco}, E. and {Sahakian}, V. and {Saito}, S. and {Sanchez}, D.~A. and {Santangelo}, A. and {Sasaki}, M. and {Schlickeiser}, R. and {Sch{\"u}ssler}, F. and {Schulz}, A. and {Schutte}, H. and {Schwanke}, U. and {Schwemmer}, S. and {Seglar-Arroyo}, M. and {Senniappan}, M. and {Seyffert}, A.~S. and {Shafi}, N. and {Shiningayamwe}, K. and {Simoni}, R. and {Sinha}, A. and {Sol}, H. and {Specovius}, A. and {Spir-Jacob}, M. and {Stawarz}, {\L}. and {Steenkamp}, R. and {Stegmann}, C. and {Steppa}, C. and {Takahashi}, T. and {Tavernier}, T. and {Taylor}, A.~M. and {Terrier}, R. and {Tiziani}, D. and {Tluczykont}, M. and {Trichard}, C. and {Tsirou}, M. and {Tsuji}, N. and {Tuffs}, R. and {Uchiyama}, Y. and {van der Walt}, D.~J. and {van Eldik}, C. and {van Rensburg}, C. and {van Soelen}, B. and {Vasileiadis}, G. and {Veh}, J. and {Venter}, C. and {Vincent}, P. and {Vink}, J. and {Voisin}, F. and {V{\"o}lk}, H.~J. and {Vuillaume}, T. and {Wadiasingh}, Z. and {Wagner}, S.~J. and {White}, R. and {Wierzcholska}, A. and {Yang}, R. and {Yoneda}, H. and {Zacharias}, M. and {Zanin}, R. and {Zdziarski}, A.~A. and {Zech}, A. and {Ziegler}, A. and {Zorn}, J. and {{\.Z}ywucka}, N. and {Smith}, P.~S.},
        title = "{H.E.S.S. detection of very high-energy {\ensuremath{\gamma}}-ray emission from the quasar PKS 0736+017}",
      journal = {A\&A},
         year = 2020,
        month = jan,
       volume = {633},
          eid = {A162},
        pages = {A162},
          doi = {10.1051/0004-6361/201935906},
archivePrefix = {arXiv},
       eprint = {1911.04761},
 primaryClass = {astro-ph.HE},
       adsurl = {https://ui.adsabs.harvard.edu/abs/2020A&A...633A.162H}
}

@ARTICLE{Nalewajko14,
       author = {{Nalewajko}, Krzysztof and {Begelman}, Mitchell C. and {Sikora}, Marek},
        title = "{Constraining the Location of Gamma-Ray Flares in Luminous Blazars}",
      journal = {ApJ},
         year = 2014,
        month = jul,
       volume = {789},
       number = {2},
          eid = {161},
        pages = {161},
          doi = {10.1088/0004-637X/789/2/161},
archivePrefix = {arXiv},
       eprint = {1405.7694},
 primaryClass = {astro-ph.HE},
       adsurl = {https://ui.adsabs.harvard.edu/abs/2014ApJ...789..161N}
}

@ARTICLE{SOPHIA,
       author = {{M{\"u}cke}, A. and {Engel}, Ralph and {Rachen}, J.~P. and {Protheroe}, R.~J. and {Stanev}, Todor},
        title = "{Monte Carlo simulations of photohadronic processes in astrophysics}",
      journal = {Computer Physics Communications},
         year = 2000,
        month = feb,
       volume = {124},
       number = {2-3},
        pages = {290-314},
          doi = {10.1016/S0010-4655(99)00446-4},
archivePrefix = {arXiv},
       eprint = {astro-ph/9903478},
 primaryClass = {astro-ph},
       adsurl = {https://ui.adsabs.harvard.edu/abs/2000CoPhC.124..290M}
}

@ARTICLE{MAGIC0506_paper3,
       author = {{Acciari}, V.~A. and {Aniello}, T. and {Ansoldi}, S. and {Antonelli}, L.~A. and {Arbet Engels}, A. and {Artero}, M. and {Asano}, K. and {Baack}, D. and {Babi{\'c}}, A. and {Baquero}, A. and {Barres de Almeida}, U. and {Barrio}, J.~A. and {Batkovi{\'c}}, I. and {Becerra Gonz{\'a}lez}, J. and {Bednarek}, W. and {Bernardini}, E. and {Bernardos}, M. and {Berti}, A. and {Besenrieder}, J. and {Bhattacharyya}, W. and {Bigongiari}, C. and {Biland}, A. and {Blanch}, O. and {B{\"o}kenkamp}, H. and {Bonnoli}, G. and {Bo{\v{s}}njak}, {\v{Z}}. and {Busetto}, G. and {Carosi}, R. and {Ceribella}, G. and {Cerruti}, M. and {Chai}, Y. and {Chilingarian}, A. and {Cikota}, S. and {Colombo}, E. and {Contreras}, J.~L. and {Cortina}, J. and {Covino}, S. and {D'Amico}, G. and {D'Elia}, V. and {Vela}, P. Da and {Dazzi}, F. and {De Angelis}, A. and {De Lotto}, B. and {Del Popolo}, A. and {Delfino}, M. and {Delgado}, J. and {Mendez}, C. Delgado and {Depaoli}, D. and {Di Pierro}, F. and {Di Venere}, L. and {Do Souto Espi{\~n}eira}, E. and {Dominis Prester}, D. and {Donini}, A. and {Dorner}, D. and {Doro}, M. and {Elsaesser}, D. and {Fallah Ramazani}, V. and {Fari{\~n}a}, L. and {Fattorini}, A. and {Font}, L. and {Fruck}, C. and {Fukami}, S. and {Fukazawa}, Y. and {Garc{\'\i}a L{\'o}pez}, R.~J. and {Garczarczyk}, M. and {Gasparyan}, S. and {Gaug}, M. and {Giglietto}, N. and {Giordano}, F. and {Gliwny}, P. and {Godinovi{\'c}}, N. and {Green}, J.~G. and {Green}, D. and {Hadasch}, D. and {Hahn}, A. and {Hassan}, T. and {Heckmann}, L. and {Herrera}, J. and {Hoang}, J. and {Hrupec}, D. and {H{\"u}tten}, M. and {Inada}, T. and {Iotov}, R. and {Ishio}, K. and {Iwamura}, Y. and {Jim{\'e}nez Mart{\'\i}nez}, I. and {Jormanainen}, J. and {Jouvin}, L. and {Kerszberg}, D. and {Kobayashi}, Y. and {Kubo}, H. and {Kushida}, J. and {Lamastra}, A. and {Lelas}, D. and {Leone}, F. and {Lindfors}, E. and {Linhoff}, L. and {Lombardi}, S. and {Longo}, F. and {L{\'o}pez-Coto}, R. and {L{\'o}pez-Moya}, M. and {L{\'o}pez-Oramas}, A. and {Loporchio}, S. and {Machado de Oliveira Fraga}, B. and {Maggio}, C. and {Majumdar}, P. and {Makariev}, M. and {Mallamaci}, M. and {Maneva}, G. and {Manganaro}, M. and {Mannheim}, K. and {Mariotti}, M. and {Mart{\'\i}nez}, M. and {Mas Aguilar}, A. and {Mazin}, D. and {Menchiari}, S. and {Mender}, S. and {Mi{\'c}anovi{\'c}}, S. and {Miceli}, D. and {Miener}, T. and {Miranda}, J.~M. and {Mirzoyan}, R. and {Molina}, E. and {Moralejo}, A. and {Morcuende}, D. and {Moreno}, V. and {Moretti}, E. and {Nakamori}, T. and {Nava}, L. and {Neustroev}, V. and {Nievas Rosillo}, M. and {Nigro}, C. and {Nilsson}, K. and {Nishijima}, K. and {Noda}, K. and {Nozaki}, S. and {Ohtani}, Y. and {Oka}, T. and {Otero-Santos}, J. and {Paiano}, S. and {Palatiello}, M. and {Paneque}, D. and {Paoletti}, R. and {Paredes}, J.~M. and {Pavleti{\'c}}, L. and {Pe{\~n}il}, P. and {Persic}, M. and {Pihet}, M. and {Prada Moroni}, P.~G. and {Prandini}, E. and {Priyadarshi}, C. and {Puljak}, I. and {Rhode}, W. and {Rib{\'o}}, M. and {Rico}, J. and {Righi}, C. and {Rugliancich}, A. and {Sahakyan}, N. and {Saito}, T. and {Sakurai}, S. and {Satalecka}, K. and {Saturni}, F.~G. and {Schleicher}, B. and {Schmidt}, K. and {Schmuckermaier}, F. and {Schweizer}, T. and {Sitarek}, J. and {{\v{S}}nidari{\'c}}, I. and {Sobczynska}, D. and {Spolon}, A. and {Stamerra}, A. and {Stri{\v{s}}kovi{\'c}}, J. and {Strom}, D. and {Strzys}, M. and {Suda}, Y. and {Suri{\'c}}, T. and {Takahashi}, M. and {Takeishi}, R. and {Tavecchio}, F. and {Temnikov}, P. and {Terzi{\'c}}, T. and {Teshima}, M. and {Tosti}, L. and {Truzzi}, S. and {Tutone}, A. and {Ubach}, S. and {van Scherpenberg}, J. and {Vanzo}, G. and {Vazquez Acosta}, M. and {Ventura}, S. and {Verguilov}, V. and {Viale}, I. and {Vigorito}, C.~F. and {Vitale}, V. and {Vovk}, I. and {Will}, M. and {Wunderlich}, C. and {Yamamoto}, T. and {Zari{\'c}}, D. and {Hodges}, M. and {Hovatta}, T. and {Kiehlmann}, S. and {Liodakis}, I. and {Max-Moerbeck}, W. and {Pearson}, T.~J. and {Readhead}, A.~C.~S. and {Reeves}, R.~A. and {L{\"a}hteenm{\"a}ki}, A. and {Tornikoski}, M. and {Tammi}, J. and {D'Ammando}, F. and {Marchini}, A.},
        title = "{Investigating the Blazar TXS 0506+056 through Sharp Multiwavelength Eyes During 2017-2019}",
      journal = {ApJ},
         year = 2022,
        month = mar,
       volume = {927},
       number = {2},
          eid = {197},
        pages = {197},
          doi = {10.3847/1538-4357/ac531d},
archivePrefix = {arXiv},
       eprint = {2202.02600},
 primaryClass = {astro-ph.HE},
       adsurl = {https://ui.adsabs.harvard.edu/abs/2022ApJ...927..197A}
}

@ARTICLE{Cerruti19,
       author = {{Cerruti}, M. and {Zech}, A. and {Boisson}, C. and {Emery}, G. and {Inoue}, S. and {Lenain}, J. -P.},
        title = "{Leptohadronic single-zone models for the electromagnetic and neutrino emission of TXS 0506+056}",
      journal = {MNRAS},
         year = 2019,
        month = feb,
       volume = {483},
       number = {1},
        pages = {L12-L16},
          doi = {10.1093/mnrasl/sly210},
archivePrefix = {arXiv},
       eprint = {1807.04335},
 primaryClass = {astro-ph.HE},
       adsurl = {https://ui.adsabs.harvard.edu/abs/2019MNRAS.483L..12C}
}

@ARTICLE{Cerruti15,
       author = {{Cerruti}, M. and {Zech}, A. and {Boisson}, C. and {Inoue}, S.},
        title = "{A hadronic origin for ultra-high-frequency-peaked BL Lac objects}",
      journal = {MNRAS},
         year = 2015,
        month = mar,
       volume = {448},
       number = {1},
        pages = {910-927},
          doi = {10.1093/mnras/stu2691},
archivePrefix = {arXiv},
       eprint = {1411.5968},
 primaryClass = {astro-ph.HE},
       adsurl = {https://ui.adsabs.harvard.edu/abs/2015MNRAS.448..910C}
}

@INPROCEEDINGS{FermiPy,
       author = {{Wood}, M. and {Caputo}, R. and {Charles}, E. and {Di Mauro}, M. and {Magill}, J. and {Perkins}, J.~S. and {Fermi-LAT Collaboration}},
        title = "{Fermipy: An open-source Python package for analysis of Fermi-LAT Data}",
    booktitle = {35th International Cosmic Ray Conference (ICRC2017)},
         year = 2017,
       series = {International Cosmic Ray Conference},
       volume = {301},
        month = jul,
          eid = {824},
        pages = {824},
          doi = {10.22323/1.301.0824},
archivePrefix = {arXiv},
       eprint = {1707.09551},
 primaryClass = {astro-ph.IM},
       adsurl = {https://ui.adsabs.harvard.edu/abs/2017ICRC...35..824W}
}

@ARTICLE{2009ApJ...697.1071A,
       author = {{Atwood}, W.~B. and {Abdo}, A.~A. and {Ackermann}, M. and {Althouse}, W. and {Anderson}, B. and {Axelsson}, M. and {Baldini}, L. and {Ballet}, J. and {Band}, D.~L. and {Barbiellini}, G. and {Bartelt}, J. and {Bastieri}, D. and {Baughman}, B.~M. and {Bechtol}, K. and {B{\'e}d{\'e}r{\`e}de}, D. and {Bellardi}, F. and {Bellazzini}, R. and {Berenji}, B. and {Bignami}, G.~F. and {Bisello}, D. and {Bissaldi}, E. and {Blandford}, R.~D. and {Bloom}, E.~D. and {Bogart}, J.~R. and {Bonamente}, E. and {Bonnell}, J. and {Borgland}, A.~W. and {Bouvier}, A. and {Bregeon}, J. and {Brez}, A. and {Brigida}, M. and {Bruel}, P. and {Burnett}, T.~H. and {Busetto}, G. and {Caliandro}, G.~A. and {Cameron}, R.~A. and {Caraveo}, P.~A. and {Carius}, S. and {Carlson}, P. and {Casandjian}, J.~M. and {Cavazzuti}, E. and {Ceccanti}, M. and {Cecchi}, C. and {Charles}, E. and {Chekhtman}, A. and {Cheung}, C.~C. and {Chiang}, J. and {Chipaux}, R. and {Cillis}, A.~N. and {Ciprini}, S. and {Claus}, R. and {Cohen-Tanugi}, J. and {Condamoor}, S. and {Conrad}, J. and {Corbet}, R. and {Corucci}, L. and {Costamante}, L. and {Cutini}, S. and {Davis}, D.~S. and {Decotigny}, D. and {DeKlotz}, M. and {Dermer}, C.~D. and {de Angelis}, A. and {Digel}, S.~W. and {do Couto e Silva}, E. and {Drell}, P.~S. and {Dubois}, R. and {Dumora}, D. and {Edmonds}, Y. and {Fabiani}, D. and {Farnier}, C. and {Favuzzi}, C. and {Flath}, D.~L. and {Fleury}, P. and {Focke}, W.~B. and {Funk}, S. and {Fusco}, P. and {Gargano}, F. and {Gasparrini}, D. and {Gehrels}, N. and {Gentit}, F. -X. and {Germani}, S. and {Giebels}, B. and {Giglietto}, N. and {Giommi}, P. and {Giordano}, F. and {Glanzman}, T. and {Godfrey}, G. and {Grenier}, I.~A. and {Grondin}, M. -H. and {Grove}, J.~E. and {Guillemot}, L. and {Guiriec}, S. and {Haller}, G. and {Harding}, A.~K. and {Hart}, P.~A. and {Hays}, E. and {Healey}, S.~E. and {Hirayama}, M. and {Hjalmarsdotter}, L. and {Horn}, R. and {Hughes}, R.~E. and {J{\'o}hannesson}, G. and {Johansson}, G. and {Johnson}, A.~S. and {Johnson}, R.~P. and {Johnson}, T.~J. and {Johnson}, W.~N. and {Kamae}, T. and {Katagiri}, H. and {Kataoka}, J. and {Kavelaars}, A. and {Kawai}, N. and {Kelly}, H. and {Kerr}, M. and {Klamra}, W. and {Kn{\"o}dlseder}, J. and {Kocian}, M.~L. and {Komin}, N. and {Kuehn}, F. and {Kuss}, M. and {Landriu}, D. and {Latronico}, L. and {Lee}, B. and {Lee}, S. -H. and {Lemoine-Goumard}, M. and {Lionetto}, A.~M. and {Longo}, F. and {Loparco}, F. and {Lott}, B. and {Lovellette}, M.~N. and {Lubrano}, P. and {Madejski}, G.~M. and {Makeev}, A. and {Marangelli}, B. and {Massai}, M.~M. and {Mazziotta}, M.~N. and {McEnery}, J.~E. and {Menon}, N. and {Meurer}, C. and {Michelson}, P.~F. and {Minuti}, M. and {Mirizzi}, N. and {Mitthumsiri}, W. and {Mizuno}, T. and {Moiseev}, A.~A. and {Monte}, C. and {Monzani}, M.~E. and {Moretti}, E. and {Morselli}, A. and {Moskalenko}, I.~V. and {Murgia}, S. and {Nakamori}, T. and {Nishino}, S. and {Nolan}, P.~L. and {Norris}, J.~P. and {Nuss}, E. and {Ohno}, M. and {Ohsugi}, T. and {Omodei}, N. and {Orlando}, E. and {Ormes}, J.~F. and {Paccagnella}, A. and {Paneque}, D. and {Panetta}, J.~H. and {Parent}, D. and {Pearce}, M. and {Pepe}, M. and {Perazzo}, A. and {Pesce-Rollins}, M. and {Picozza}, P. and {Pieri}, L. and {Pinchera}, M. and {Piron}, F. and {Porter}, T.~A. and {Poupard}, L. and {Rain{\`o}}, S. and {Rando}, R. and {Rapposelli}, E. and {Razzano}, M. and {Reimer}, A. and {Reimer}, O. and {Reposeur}, T. and {Reyes}, L.~C. and {Ritz}, S. and {Rochester}, L.~S. and {Rodriguez}, A.~Y. and {Romani}, R.~W. and {Roth}, M. and {Russell}, J.~J. and {Ryde}, F. and {Sabatini}, S. and {Sadrozinski}, H.~F. -W. and {Sanchez}, D. and {Sander}, A. and {Sapozhnikov}, L. and {Parkinson}, P.~M. Saz and {Scargle}, J.~D. and {Schalk}, T.~L. and {Scolieri}, G. and {Sgr{\`o}}, C. and {Share}, G.~H. and {Shaw}, M. and {Shimokawabe}, T. and {Shrader}, C. and {Sierpowska-Bartosik}, A. and {Siskind}, E.~J. and {Smith}, D.~A. and {Smith}, P.~D. and {Spandre}, G. and {Spinelli}, P. and {Starck}, J. -L. and {Stephens}, T.~E. and {Strickman}, M.~S. and {Strong}, A.~W. and {Suson}, D.~J. and {Tajima}, H. and {Takahashi}, H. and {Takahashi}, T. and {Tanaka}, T. and {Tenze}, A. and {Tether}, S. and {Thayer}, J.~B. and {Thayer}, J.~G. and {Thompson}, D.~J. and {Tibaldo}, L. and {Tibolla}, O. and {Torres}, D.~F. and {Tosti}, G. and {Tramacere}, A. and {Turri}, M. and {Usher}, T.~L. and {Vilchez}, N. and {Vitale}, V. and {Wang}, P. and {Watters}, K. and {Winer}, B.~L. and {Wood}, K.~S. and {Ylinen}, T. and {Ziegler}, M.},
        title = "{The Large Area Telescope on the Fermi Gamma-Ray Space Telescope Mission}",
      journal = {ApJ},
         year = 2009,
        month = jun,
       volume = {697},
       number = {2},
        pages = {1071-1102},
          doi = {10.1088/0004-637X/697/2/1071},
archivePrefix = {arXiv},
       eprint = {0902.1089},
 primaryClass = {astro-ph.IM},
       adsurl = {https://ui.adsabs.harvard.edu/abs/2009ApJ...697.1071A}
}

@ARTICLE{2018arXiv181011394B,
       author = {{Bruel}, P. and {Burnett}, T.~H. and {Digel}, S.~W. and {Johannesson}, G. and {Omodei}, N. and {Wood}, M.},
        title = "{Fermi-LAT improved Pass\raisebox{-0.5ex}\textasciitilde8 event selection}",
      journal = {arXiv e-prints},
         year = 2018,
        month = oct,
          eid = {arXiv:1810.11394},
        pages = {arXiv:1810.11394},
          doi = {10.48550/arXiv.1810.11394},
archivePrefix = {arXiv},
       eprint = {1810.11394},
 primaryClass = {astro-ph.IM},
       adsurl = {https://ui.adsabs.harvard.edu/abs/2018arXiv181011394B}
}

@ARTICLE{2018A&A...618A..22P,
       author = {{Principe}, G. and {Malyshev}, D. and {Ballet}, J. and {Funk}, S.},
        title = "{The first catalog of Fermi-LAT sources below 100 MeV}",
      journal = {A\&A},
         year = 2018,
        month = oct,
       volume = {618},
          eid = {A22},
        pages = {A22},
          doi = {10.1051/0004-6361/201833116},
archivePrefix = {arXiv},
       eprint = {1806.10865},
 primaryClass = {astro-ph.HE},
       adsurl = {https://ui.adsabs.harvard.edu/abs/2018A&A...618A..22P}
}

@ARTICLE{2009PhRvD..80l2004A,
       author = {{Abdo}, A.~A. and {Ackermann}, M. and {Ajello}, M. and {Atwood}, W.~B. and {Baldini}, L. and {Ballet}, J. and {Barbiellini}, G. and {Bastieri}, D. and {Baughman}, B.~M. and {Bechtol}, K. and {Bellazzini}, R. and {Berenji}, B. and {Bloom}, E.~D. and {Bonamente}, E. and {Borgland}, A.~W. and {Bouvier}, A. and {Bregeon}, J. and {Brez}, A. and {Brigida}, M. and {Bruel}, P. and {Buehler}, R. and {Burnett}, T.~H. and {Buson}, S. and {Caliandro}, G.~A. and {Cameron}, R.~A. and {Caraveo}, P.~A. and {Casandjian}, J.~M. and {Cecchi}, C. and {{\c{C}}elik}, {\"O}. and {Charles}, E. and {Chekhtman}, A. and {Chiang}, J. and {Ciprini}, S. and {Claus}, R. and {Cohen-Tanugi}, J. and {Conrad}, J. and {de Palma}, F. and {Digel}, S.~W. and {Do Couto E Silva}, E. and {Drell}, P.~S. and {Dubois}, R. and {Dumora}, D. and {Farnier}, C. and {Favuzzi}, C. and {Fegan}, S.~J. and {Focke}, W.~B. and {Fortin}, P. and {Frailis}, M. and {Fukazawa}, Y. and {Funk}, S. and {Fusco}, P. and {Gargano}, F. and {Gehrels}, N. and {Germani}, S. and {Giebels}, B. and {Giglietto}, N. and {Giordano}, F. and {Glanzman}, T. and {Godfrey}, G. and {Grenier}, I.~A. and {Grondin}, M. -H. and {Grove}, J.~E. and {Guillemot}, L. and {Guiriec}, S. and {Hays}, E. and {Horan}, D. and {Hughes}, R.~E. and {J{\'o}hannesson}, G. and {Johnson}, A.~S. and {Johnson}, T.~J. and {Johnson}, W.~N. and {Kamae}, T. and {Katagiri}, H. and {Kataoka}, J. and {Kawai}, N. and {Kerr}, M. and {Kn{\"o}dlseder}, J. and {Kuss}, M. and {Lande}, J. and {Latronico}, L. and {Lemoine-Goumard}, M. and {Longo}, F. and {Loparco}, F. and {Lott}, B. and {Lovellette}, M.~N. and {Lubrano}, P. and {Makeev}, A. and {Mazziotta}, M.~N. and {McEnery}, J.~E. and {Meurer}, C. and {Michelson}, P.~F. and {Mitthumsiri}, W. and {Mizuno}, T. and {Monte}, C. and {Monzani}, M.~E. and {Morselli}, A. and {Moskalenko}, I.~V. and {Murgia}, S. and {Nolan}, P.~L. and {Norris}, J.~P. and {Nuss}, E. and {Ohsugi}, T. and {Okumura}, A. and {Omodei}, N. and {Orlando}, E. and {Ormes}, J.~F. and {Paneque}, D. and {Panetta}, J.~H. and {Parent}, D. and {Pelassa}, V. and {Pepe}, M. and {Pesce-Rollins}, M. and {Piron}, F. and {Porter}, T.~A. and {Rain{\`o}}, S. and {Rando}, R. and {Razzano}, M. and {Reimer}, A. and {Reimer}, O. and {Reposeur}, T. and {Rochester}, L.~S. and {Rodriguez}, A.~Y. and {Roth}, M. and {Sadrozinski}, H.~F. -W. and {Sander}, A. and {Saz Parkinson}, P.~M. and {Sgr{\`o}}, C. and {Share}, G.~H. and {Siskind}, E.~J. and {Smith}, D.~A. and {Smith}, P.~D. and {Spandre}, G. and {Spinelli}, P. and {Strickman}, M.~S. and {Suson}, D.~J. and {Takahashi}, H. and {Tanaka}, T. and {Thayer}, J.~B. and {Thayer}, J.~G. and {Thompson}, D.~J. and {Tibaldo}, L. and {Torres}, D.~F. and {Tosti}, G. and {Tramacere}, A. and {Uchiyama}, Y. and {Usher}, T.~L. and {Vasileiou}, V. and {Vilchez}, N. and {Vitale}, V. and {Waite}, A.~P. and {Wang}, P. and {Winer}, B.~L. and {Wood}, K.~S. and {Ylinen}, T. and {Ziegler}, M.},
        title = "{Fermi large area telescope observations of the cosmic-ray induced {\ensuremath{\gamma}}-ray emission of the Earth's atmosphere}",
      journal = {PRD},
         year = 2009,
        month = dec,
       volume = {80},
       number = {12},
          eid = {122004},
        pages = {122004},
          doi = {10.1103/PhysRevD.80.122004},
archivePrefix = {arXiv},
       eprint = {0912.1868},
 primaryClass = {astro-ph.HE},
       adsurl = {https://ui.adsabs.harvard.edu/abs/2009PhRvD..80l2004A}
}

@ARTICLE{4FGL,
       author = {{Abdollahi}, S. and {Acero}, F. and {Ackermann}, M. and {Ajello}, M. and {Atwood}, W.~B. and {Axelsson}, M. and {Baldini}, L. and {Ballet}, J. and {Barbiellini}, G. and {Bastieri}, D. and {Becerra Gonzalez}, J. and {Bellazzini}, R. and {Berretta}, A. and {Bissaldi}, E. and {Blandford}, R.~D. and {Bloom}, E.~D. and {Bonino}, R. and {Bottacini}, E. and {Brandt}, T.~J. and {Bregeon}, J. and {Bruel}, P. and {Buehler}, R. and {Burnett}, T.~H. and {Buson}, S. and {Cameron}, R.~A. and {Caputo}, R. and {Caraveo}, P.~A. and {Casandjian}, J.~M. and {Castro}, D. and {Cavazzuti}, E. and {Charles}, E. and {Chaty}, S. and {Chen}, S. and {Cheung}, C.~C. and {Chiaro}, G. and {Ciprini}, S. and {Cohen-Tanugi}, J. and {Cominsky}, L.~R. and {Coronado-Bl{\'a}zquez}, J. and {Costantin}, D. and {Cuoco}, A. and {Cutini}, S. and {D'Ammando}, F. and {DeKlotz}, M. and {de la Torre Luque}, P. and {de Palma}, F. and {Desai}, A. and {Digel}, S.~W. and {Di Lalla}, N. and {Di Mauro}, M. and {Di Venere}, L. and {Dom{\'\i}nguez}, A. and {Dumora}, D. and {Fana Dirirsa}, F. and {Fegan}, S.~J. and {Ferrara}, E.~C. and {Franckowiak}, A. and {Fukazawa}, Y. and {Funk}, S. and {Fusco}, P. and {Gargano}, F. and {Gasparrini}, D. and {Giglietto}, N. and {Giommi}, P. and {Giordano}, F. and {Giroletti}, M. and {Glanzman}, T. and {Green}, D. and {Grenier}, I.~A. and {Griffin}, S. and {Grondin}, M. -H. and {Grove}, J.~E. and {Guiriec}, S. and {Harding}, A.~K. and {Hayashi}, K. and {Hays}, E. and {Hewitt}, J.~W. and {Horan}, D. and {J{\'o}hannesson}, G. and {Johnson}, T.~J. and {Kamae}, T. and {Kerr}, M. and {Kocevski}, D. and {Kovac'evic'}, M. and {Kuss}, M. and {Landriu}, D. and {Larsson}, S. and {Latronico}, L. and {Lemoine-Goumard}, M. and {Li}, J. and {Liodakis}, I. and {Longo}, F. and {Loparco}, F. and {Lott}, B. and {Lovellette}, M.~N. and {Lubrano}, P. and {Madejski}, G.~M. and {Maldera}, S. and {Malyshev}, D. and {Manfreda}, A. and {Marchesini}, E.~J. and {Marcotulli}, L. and {Mart{\'\i}-Devesa}, G. and {Martin}, P. and {Massaro}, F. and {Mazziotta}, M.~N. and {McEnery}, J.~E. and {Mereu}, I. and {Meyer}, M. and {Michelson}, P.~F. and {Mirabal}, N. and {Mizuno}, T. and {Monzani}, M.~E. and {Morselli}, A. and {Moskalenko}, I.~V. and {Negro}, M. and {Nuss}, E. and {Ojha}, R. and {Omodei}, N. and {Orienti}, M. and {Orlando}, E. and {Ormes}, J.~F. and {Palatiello}, M. and {Paliya}, V.~S. and {Paneque}, D. and {Pei}, Z. and {Pe{\~n}a-Herazo}, H. and {Perkins}, J.~S. and {Persic}, M. and {Pesce-Rollins}, M. and {Petrosian}, V. and {Petrov}, L. and {Piron}, F. and {Poon}, H. and {Porter}, T.~A. and {Principe}, G. and {Rain{\`o}}, S. and {Rando}, R. and {Razzano}, M. and {Razzaque}, S. and {Reimer}, A. and {Reimer}, O. and {Remy}, Q. and {Reposeur}, T. and {Romani}, R.~W. and {Saz Parkinson}, P.~M. and {Schinzel}, F.~K. and {Serini}, D. and {Sgr{\`o}}, C. and {Siskind}, E.~J. and {Smith}, D.~A. and {Spandre}, G. and {Spinelli}, P. and {Strong}, A.~W. and {Suson}, D.~J. and {Tajima}, H. and {Takahashi}, M.~N. and {Tak}, D. and {Thayer}, J.~B. and {Thompson}, D.~J. and {Tibaldo}, L. and {Torres}, D.~F. and {Torresi}, E. and {Valverde}, J. and {Van Klaveren}, B. and {van Zyl}, P. and {Wood}, K. and {Yassine}, M. and {Zaharijas}, G.},
        title = "{Fermi Large Area Telescope Fourth Source Catalog}",
      journal = {ApJS},
         year = 2020,
        month = mar,
       volume = {247},
       number = {1},
          eid = {33},
        pages = {33},
          doi = {10.3847/1538-4365/ab6bcb},
archivePrefix = {arXiv},
       eprint = {1902.10045},
 primaryClass = {astro-ph.HE},
       adsurl = {https://ui.adsabs.harvard.edu/abs/2020ApJS..247...33A}
}

@article{icecube2017-txs,
	author = {{The IceCube Collaboration} and others},
	title = {{Multimessenger observations of a flaring blazar coincident with high-energy neutrino IceCube-170922A}},
	journal = {Science},
	volume = {361},
	number = {6398},
	year = {2018},
	month = jul,
	issn = {0036-8075},
	publisher = {American Association for the Advancement of Science},
	doi = {10.1126/science.aat1378}
}

@ARTICLE{4LAC,
       author = {{Ajello}, M. and {Angioni}, R. and {Axelsson}, M. and {Ballet}, J. and {Barbiellini}, G. and {Bastieri}, D. and {Becerra Gonzalez}, J. and {Bellazzini}, R. and {Bissaldi}, E. and {Bloom}, E.~D. and {Bonino}, R. and {Bottacini}, E. and {Bruel}, P. and {Buson}, S. and {Cafardo}, F. and {Cameron}, R.~A. and {Cavazzuti}, E. and {Chen}, S. and {Cheung}, C.~C. and {Ciprini}, S. and {Costantin}, D. and {Cutini}, S. and {D'Ammando}, F. and {de la Torre Luque}, P. and {de Menezes}, R. and {de Palma}, F. and {Desai}, A. and {Di Lalla}, N. and {Di Venere}, L. and {Dom{\'\i}nguez}, A. and {Dirirsa}, F. Fana and {Ferrara}, E.~C. and {Finke}, J. and {Franckowiak}, A. and {Fukazawa}, Y. and {Funk}, S. and {Fusco}, P. and {Gargano}, F. and {Garrappa}, S. and {Gasparrini}, D. and {Giglietto}, N. and {Giordano}, F. and {Giroletti}, M. and {Green}, D. and {Grenier}, I.~A. and {Guiriec}, S. and {Harita}, S. and {Hays}, E. and {Horan}, D. and {Itoh}, R. and {J{\'o}hannesson}, G. and {Kovac'evic'}, M. and {Krauss}, F. and {Kreter}, M. and {Kuss}, M. and {Larsson}, S. and {Leto}, C. and {Li}, J. and {Liodakis}, I. and {Longo}, F. and {Loparco}, F. and {Lott}, B. and {Lovellette}, M.~N. and {Lubrano}, P. and {Madejski}, G.~M. and {Maldera}, S. and {Manfreda}, A. and {Mart{\'\i}-Devesa}, G. and {Massaro}, F. and {Mazziotta}, M.~N. and {Mereu}, I. and {Meyer}, M. and {Migliori}, G. and {Mirabal}, N. and {Mizuno}, T. and {Monzani}, M.~E. and {Morselli}, A. and {Moskalenko}, I.~V. and {Negro}, M. and {Nemmen}, R. and {Nuss}, E. and {Ojha}, L.~S. and {Ojha}, R. and {Omodei}, N. and {Orienti}, M. and {Orlando}, E. and {Ormes}, J.~F. and {Paliya}, V.~S. and {Pei}, Z. and {Pe{\~n}a-Herazo}, H. and {Persic}, M. and {Pesce-Rollins}, M. and {Petrov}, L. and {Piron}, F. and {Poon}, H. and {Principe}, G. and {Rain{\`o}}, S. and {Rando}, R. and {Rani}, B. and {Razzano}, M. and {Razzaque}, S. and {Reimer}, A. and {Reimer}, O. and {Schinzel}, F.~K. and {Serini}, D. and {Sgr{\`o}}, C. and {Siskind}, E.~J. and {Spandre}, G. and {Spinelli}, P. and {Suson}, D.~J. and {Tachibana}, Y. and {Thompson}, D.~J. and {Torres}, D.~F. and {Torresi}, E. and {Troja}, E. and {Valverde}, J. and {van Zyl}, P. and {Yassine}, M.},
        title = "{The Fourth Catalog of Active Galactic Nuclei Detected by the Fermi Large Area Telescope}",
      journal = {ApJ},
         year = 2020,
        month = apr,
       volume = {892},
       number = {2},
          eid = {105},
        pages = {105},
          doi = {10.3847/1538-4357/ab791e},
archivePrefix = {arXiv},
       eprint = {1905.10771},
 primaryClass = {astro-ph.HE},
       adsurl = {https://ui.adsabs.harvard.edu/abs/2020ApJ...892..105A}
}

@ARTICLE{4LAC-DR2,
       author = {{Lott}, B. and {Gasparrini}, D. and {Ciprini}, S.},
        title = "{The Fourth Catalog of Active Galactic Nuclei Detected by the Fermi Large Area Telescope -- Data Release 2}",
      journal = {arXiv e-prints},
         year = 2020,
        month = oct,
          eid = {arXiv:2010.08406},
        pages = {arXiv:2010.08406},
          doi = {10.48550/arXiv.2010.08406},
archivePrefix = {arXiv},
       eprint = {2010.08406},
 primaryClass = {astro-ph.HE},
       adsurl = {https://ui.adsabs.harvard.edu/abs/2020arXiv201008406L}
}

@ARTICLE{4FGL-DR2,
       author = {{Ballet}, J. and {Burnett}, T.~H. and {Digel}, S.~W. and {Lott}, B.},
        title = "{Fermi Large Area Telescope Fourth Source Catalog Data Release 2}",
      journal = {arXiv e-prints},
         year = 2020,
        month = may,
          eid = {arXiv:2005.11208},
        pages = {arXiv:2005.11208},
          doi = {10.48550/arXiv.2005.11208},
archivePrefix = {arXiv},
       eprint = {2005.11208},
 primaryClass = {astro-ph.HE},
       adsurl = {https://ui.adsabs.harvard.edu/abs/2020arXiv200511208B}
}

@ARTICLE{sequence2.0,
       author = {{Ghisellini}, G. and {Righi}, C. and {Costamante}, L. and {Tavecchio}, F.},
        title = "{The Fermi blazar sequence}",
      journal = {MNRAS},
         year = 2017,
        month = jul,
       volume = {469},
       number = {1},
        pages = {255-266},
          doi = {10.1093/mnras/stx806},
archivePrefix = {arXiv},
       eprint = {1702.02571},
 primaryClass = {astro-ph.HE},
       adsurl = {https://ui.adsabs.harvard.edu/abs/2017MNRAS.469..255G}
}

@article{Padovani2019_TXS,
       author = {{Padovani}, P. and {Oikonomou}, F. and {Petropoulou}, M. and {Giommi}, P. and {Resconi}, E.},
        title = "{TXS 0506+056, the first cosmic neutrino source, is not a BL Lac}",
      journal = {MNRAS},
         year = 2019,
        month = mar,
       volume = {484},
       number = {1},
        pages = {L104-L108},
          doi = {10.1093/mnrasl/slz011},
archivePrefix = {arXiv},
       eprint = {1901.06998},
 primaryClass = {astro-ph.HE},
       adsurl = {https://ui.adsabs.harvard.edu/abs/2019MNRAS.484L.104P}
}

@ARTICLE{Shaw2013,
       author = {{Shaw}, Michael S. and {Romani}, Roger W. and {Cotter}, Garret and {Healey}, Stephen E. and {Michelson}, Peter F. and {Readhead}, Anthony C.~S. and {Richards}, Joseph L. and {Max-Moerbeck}, Walter and {King}, Oliver G. and {Potter}, William J.},
        title = "{Spectroscopy of the Largest Ever {\ensuremath{\gamma}}-Ray-selected BL Lac Sample}",
      journal = {ApJ},
         year = 2013,
        month = feb,
       volume = {764},
       number = {2},
          eid = {135},
        pages = {135},
          doi = {10.1088/0004-637X/764/2/135},
archivePrefix = {arXiv},
       eprint = {1301.0323},
 primaryClass = {astro-ph.HE},
       adsurl = {https://ui.adsabs.harvard.edu/abs/2013ApJ...764..135S}
}

@ARTICLE{Paiano2017,
       author = {{Paiano}, Simona and {Landoni}, Marco and {Falomo}, Renato and {Treves}, Aldo and {Scarpa}, Riccardo},
        title = "{Spectroscopy of 10 {\ensuremath{\gamma}}-Ray BL Lac Objects at High Redshift}",
      journal = {ApJ},
         year = 2017,
        month = aug,
       volume = {844},
       number = {2},
          eid = {120},
        pages = {120},
          doi = {10.3847/1538-4357/aa7aac},
archivePrefix = {arXiv},
       eprint = {1705.07454},
 primaryClass = {astro-ph.GA},
       adsurl = {https://ui.adsabs.harvard.edu/abs/2017ApJ...844..120P}
}

@ARTICLE{TeVbll,
       author = {{Paiano}, Simona and {Landoni}, Marco and {Falomo}, Renato and {Treves}, Aldo and {Scarpa}, Riccardo and {Righi}, Chiara},
        title = "{On the Redshift of TeV BL Lac Objects}",
      journal = {ApJ},
         year = 2017,
        month = mar,
       volume = {837},
       number = {2},
          eid = {144},
        pages = {144},
          doi = {10.3847/1538-4357/837/2/144},
archivePrefix = {arXiv},
       eprint = {1701.04305},
 primaryClass = {astro-ph.GA},
       adsurl = {https://ui.adsabs.harvard.edu/abs/2017ApJ...837..144P}
}

@ARTICLE{ZBLL,
       author = {{Landoni}, Marco and {Falomo}, R. and {Paiano}, S. and {Treves}, A.},
        title = "{ZBLLAC: A Spectroscopic Database of BL Lacertae Objects}",
      journal = {ApJS},
         year = 2020,
        month = oct,
       volume = {250},
       number = {2},
          eid = {37},
        pages = {37},
          doi = {10.3847/1538-4365/abb5ae},
archivePrefix = {arXiv},
       eprint = {2009.03439},
 primaryClass = {astro-ph.HE},
       adsurl = {https://ui.adsabs.harvard.edu/abs/2020ApJS..250...37L}
}

@ARTICLE{Paiano2018_TXS,
       author = {{Paiano}, Simona and {Falomo}, Renato and {Treves}, Aldo and {Scarpa}, Riccardo},
        title = "{The Redshift of the BL Lac Object TXS 0506+056}",
      journal = {ApJL},
         year = 2018,
        month = feb,
       volume = {854},
       number = {2},
          eid = {L32},
        pages = {L32},
          doi = {10.3847/2041-8213/aaad5e},
archivePrefix = {arXiv},
       eprint = {1802.01939},
 primaryClass = {astro-ph.GA},
       adsurl = {https://ui.adsabs.harvard.edu/abs/2018ApJ...854L..32P}
}

@ARTICLE{Shaw2012,
       author = {{Shaw}, Michael S. and {Romani}, Roger W. and {Cotter}, Garret and {Healey}, Stephen E. and {Michelson}, Peter F. and {Readhead}, Anthony C.~S. and {Richards}, Joseph L. and {Max-Moerbeck}, Walter and {King}, Oliver G. and {Potter}, William J.},
        title = "{Spectroscopy of Broad-line Blazars from 1LAC}",
      journal = {ApJ},
         year = 2012,
        month = mar,
       volume = {748},
       number = {1},
          eid = {49},
        pages = {49},
          doi = {10.1088/0004-637X/748/1/49},
archivePrefix = {arXiv},
       eprint = {1201.0999},
 primaryClass = {astro-ph.HE},
       adsurl = {https://ui.adsabs.harvard.edu/abs/2012ApJ...748...49S}
}

@ARTICLE{Francis1991,
       author = {{Francis}, Paul J. and {Hewett}, Paul C. and {Foltz}, Craig B. and {Chaffee}, Frederic H. and {Weymann}, Ray J. and {Morris}, Simon L.},
        title = "{A High Signal-to-Noise Ratio Composite Quasar Spectrum}",
      journal = {ApJ},
         year = 1991,
        month = jun,
       volume = {373},
        pages = {465},
          doi = {10.1086/170066},
       adsurl = {https://ui.adsabs.harvard.edu/abs/1991ApJ...373..465F}
}

@ARTICLE{Celotti1997,
       author = {{Celotti}, A. and {Padovani}, P. and {Ghisellini}, G.},
        title = "{Jets and accretion processes in active galactic nuclei: further clues}",
      journal = {MNRAS},
         year = 1997,
        month = apr,
       volume = {286},
       number = {2},
        pages = {415-424},
          doi = {10.1093/mnras/286.2.415},
archivePrefix = {arXiv},
       eprint = {astro-ph/9611111},
 primaryClass = {astro-ph},
       adsurl = {https://ui.adsabs.harvard.edu/abs/1997MNRAS.286..415C}
}

@ARTICLE{division_bll-fsrq,
       author = {{Ghisellini}, G. and {Tavecchio}, F. and {Foschini}, L. and {Ghirlanda}, G.},
        title = "{The transition between BL Lac objects and flat spectrum radio quasars}",
      journal = {MNRAS},
         year = 2011,
        month = jul,
       volume = {414},
       number = {3},
        pages = {2674-2689},
          doi = {10.1111/j.1365-2966.2011.18578.x},
archivePrefix = {arXiv},
       eprint = {1012.0308},
 primaryClass = {astro-ph.CO},
       adsurl = {https://ui.adsabs.harvard.edu/abs/2011MNRAS.414.2674G}
}

@article{swift,
	author = {Gehrels, N. and Chincarini, G. and Giommi, P. and Mason, K. O. and Nousek, J. A. and Wells, A. A. and White, N. E. and Barthelmy, S. D. and Burrows, D. N. and Cominsky, L. R. and Hurley, K. C. and Marshall, F. E. and M{\ifmmode\acute{e}\else\'{e}\fi}sz{\ifmmode\acute{a}\else\'{a}\fi}ros, P. and Roming, P. W. A. and Angelini, L. and Barbier, L. M. and Belloni, T. and Campana, S. and Caraveo, P. A. and Chester, M. M. and Citterio, O. and Cline, T. L. and Cropper, M. S. and Cummings, J. R. and Dean, A. J. and others},
	title = {{The Swift Gamma-Ray Burst Mission}},
	journal = {Astrophys. J.},
	volume = {611},
	number = {2},
	pages = {1005},
	year = {2004},
	month = aug,
	issn = {0004-637X},
	publisher = {IOP Publishing},
	doi = {10.1086/422091}
}

@ARTICLE{Burrows2005_xrt,
       author = {{Burrows}, David N. and {Hill}, J.~E. and {Nousek}, J.~A. and {Kennea}, J.~A. and {Wells}, A. and {Osborne}, J.~P. and {Abbey}, A.~F. and {Beardmore}, A. and {Mukerjee}, K. and {Short}, A.~D.~T. and {Chincarini}, G. and {Campana}, S. and {Citterio}, O. and {Moretti}, A. and {Pagani}, C. and {Tagliaferri}, G. and {Giommi}, P. and {Capalbi}, M. and {Tamburelli}, F. and {Angelini}, L. and {Cusumano}, G. and {Br{\"a}uninger}, H.~W. and {Burkert}, W. and {Hartner}, G.~D.},
        title = "{The Swift X-Ray Telescope}",
      journal = {Space Sci. Rev.},
         year = 2005,
        month = oct,
       volume = {120},
       number = {3-4},
        pages = {165-195},
          doi = {10.1007/s11214-005-5097-2},
archivePrefix = {arXiv},
       eprint = {astro-ph/0508071},
 primaryClass = {astro-ph},
       adsurl = {https://ui.adsabs.harvard.edu/abs/2005SSRv..120..165B}
}

@ARTICLE{Roming2005_uvot,
       author = {{Roming}, Peter W.~A. and {Kennedy}, Thomas E. and {Mason}, Keith O. and {Nousek}, John A. and {Ahr}, Lindy and {Bingham}, Richard E. and {Broos}, Patrick S. and {Carter}, Mary J. and {Hancock}, Barry K. and {Huckle}, Howard E. and {Hunsberger}, S.~D. and {Kawakami}, Hajime and {Killough}, Ronnie and {Koch}, T. Scott and {McLelland}, Michael K. and {Smith}, Kelly and {Smith}, Philip J. and {Soto}, Juan Carlos and {Boyd}, Patricia T. and {Breeveld}, Alice A. and {Holland}, Stephen T. and {Ivanushkina}, Mariya and {Pryzby}, Michael S. and {Still}, Martin D. and {Stock}, Joseph},
        title = "{The Swift Ultra-Violet/Optical Telescope}",
      journal = {Space Sci. Rev.},
         year = 2005,
        month = oct,
       volume = {120},
       number = {3-4},
        pages = {95-142},
          doi = {10.1007/s11214-005-5095-4},
archivePrefix = {arXiv},
       eprint = {astro-ph/0507413},
 primaryClass = {astro-ph},
       adsurl = {https://ui.adsabs.harvard.edu/abs/2005SSRv..120...95R}
}

@ARTICLE{Barthelmy2005_bat,
       author = {{Barthelmy}, Scott D. and {Barbier}, Louis M. and {Cummings}, Jay R. and {Fenimore}, Ed E. and {Gehrels}, Neil and {Hullinger}, Derek and {Krimm}, Hans A. and {Markwardt}, Craig B. and {Palmer}, David M. and {Parsons}, Ann and {Sato}, Goro and {Suzuki}, Masaya and {Takahashi}, Tadayuki and {Tashiro}, Makota and {Tueller}, Jack},
        title = "{The Burst Alert Telescope (BAT) on the SWIFT Midex Mission}",
      journal = {Space Sci. Rev.},
         year = 2005,
        month = oct,
       volume = {120},
       number = {3-4},
        pages = {143-164},
          doi = {10.1007/s11214-005-5096-3},
archivePrefix = {arXiv},
       eprint = {astro-ph/0507410},
 primaryClass = {astro-ph},
       adsurl = {https://ui.adsabs.harvard.edu/abs/2005SSRv..120..143B}
}

@INPROCEEDINGS{Hill2004_readout-modes,
       author = {{Hill}, Joanne E. and {Burrows}, David N. and {Nousek}, John A. and {Abbey}, Anthony F. and {Ambrosi}, Richard M. and {Br{\"a}uninger}, Heinrich W. and {Burkert}, Wolfgang and {Campana}, Sergio and {Cheruvu}, Chaitanya and {Cusumano}, Giancarlo and {Freyberg}, Michael J. and {Hartner}, Gisela D. and {Klar}, R. and {Mangels}, C. and {Moretti}, Alberto and {Mori}, Koji and {Morris}, Dave C. and {Short}, Alexander D.~T. and {Tagliaferri}, Gianpiero and {Watson}, D.~J. and {Wood}, P. and {Wells}, Alan A.},
        title = "{Readout modes and automated operation of the Swift X-ray Telescope}",
    booktitle = {X-Ray and Gamma-Ray Instrumentation for Astronomy XIII},
         year = 2004,
       editor = {{Flanagan}, Kathryn A. and {Siegmund}, Oswald H.~W.},
       series = {Society of Photo-Optical Instrumentation Engineers (SPIE) Conference Series},
       volume = {5165},
        month = feb,
        pages = {217-231},
          doi = {10.1117/12.505728},
       adsurl = {https://ui.adsabs.harvard.edu/abs/2004SPIE.5165..217H}
}

@ARTICLE{Cash1979_Cstat,
       author = {{Cash}, W.},
        title = "{Parameter estimation in astronomy through application of the likelihood ratio.}",
      journal = {ApJ},
         year = 1979,
        month = mar,
       volume = {228},
        pages = {939-947},
          doi = {10.1086/156922},
       adsurl = {https://ui.adsabs.harvard.edu/abs/1979ApJ...228..939C}
}

@ARTICLE{Kalberla2005,
       author = {{Kalberla}, P.~M.~W. and {Burton}, W.~B. and {Hartmann}, Dap and {Arnal}, E.~M. and {Bajaja}, E. and {Morras}, R. and {P{\"o}ppel}, W.~G.~L.},
        title = "{The Leiden/Argentine/Bonn (LAB) Survey of Galactic HI. Final data release of the combined LDS and IAR surveys with improved stray-radiation corrections}",
      journal = {A\&A},
         year = 2005,
        month = sep,
       volume = {440},
       number = {2},
        pages = {775-782},
          doi = {10.1051/0004-6361:20041864},
archivePrefix = {arXiv},
       eprint = {astro-ph/0504140},
 primaryClass = {astro-ph},
       adsurl = {https://ui.adsabs.harvard.edu/abs/2005A&A...440..775K}
}

@ARTICLE{Roming2009_extinction,
       author = {{Roming}, P.~W.~A. and {Koch}, T.~S. and {Oates}, S.~R. and {Porterfield}, B.~L. and {Vanden Berk}, D.~E. and {Boyd}, P.~T. and {Holland}, S.~T. and {Hoversten}, E.~A. and {Immler}, S. and {Marshall}, F.~E. and {Page}, M.~J. and {Racusin}, J.~L. and {Schneider}, D.~P. and {Breeveld}, A.~A. and {Brown}, P.~J. and {Chester}, M.~M. and {Cucchiara}, A. and {DePasquale}, M. and {Gronwall}, C. and {Hunsberger}, S.~D. and {Kuin}, N.~P.~M. and {Landsman}, W.~B. and {Schady}, P. and {Still}, M.},
        title = "{The First Swift Ultraviolet/Optical Telescope GRB Afterglow Catalog}",
      journal = {ApJ},
         year = 2009,
        month = jan,
       volume = {690},
       number = {1},
        pages = {163-188},
          doi = {10.1088/0004-637X/690/1/163},
archivePrefix = {arXiv},
       eprint = {0809.4193},
 primaryClass = {astro-ph},
       adsurl = {https://ui.adsabs.harvard.edu/abs/2009ApJ...690..163R}
}

@ARTICLE{S&F2011,
       author = {{Schlafly}, Edward F. and {Finkbeiner}, Douglas P.},
        title = "{Measuring Reddening with Sloan Digital Sky Survey Stellar Spectra and Recalibrating SFD}",
      journal = {ApJ},
         year = 2011,
        month = aug,
       volume = {737},
       number = {2},
          eid = {103},
        pages = {103},
          doi = {10.1088/0004-637X/737/2/103},
archivePrefix = {arXiv},
       eprint = {1012.4804},
 primaryClass = {astro-ph.GA},
       adsurl = {https://ui.adsabs.harvard.edu/abs/2011ApJ...737..103S}
}

@ARTICLE{UVOT_parameters,
       author = {{Poole}, T.~S. and {Breeveld}, A.~A. and {Page}, M.~J. and {Landsman}, W. and {Holland}, S.~T. and {Roming}, P. and {Kuin}, N.~P.~M. and {Brown}, P.~J. and {Gronwall}, C. and {Hunsberger}, S. and {Koch}, S. and {Mason}, K.~O. and {Schady}, P. and {vanden Berk}, D. and {Blustin}, A.~J. and {Boyd}, P. and {Broos}, P. and {Carter}, M. and {Chester}, M.~M. and {Cucchiara}, A. and {Hancock}, B. and {Huckle}, H. and {Immler}, S. and {Ivanushkina}, M. and {Kennedy}, T. and {Marshall}, F. and {Morgan}, A. and {Pandey}, S.~B. and {de Pasquale}, M. and {Smith}, P.~J. and {Still}, M.},
        title = "{Photometric calibration of the Swift ultraviolet/optical telescope}",
      journal = {MNRAS},
         year = 2008,
        month = jan,
       volume = {383},
       number = {2},
        pages = {627-645},
          doi = {10.1111/j.1365-2966.2007.12563.x},
archivePrefix = {arXiv},
       eprint = {0708.2259},
 primaryClass = {astro-ph},
       adsurl = {https://ui.adsabs.harvard.edu/abs/2008MNRAS.383..627P}
}

@article{asi-asdc,
	author = {Stratta, G. and Capalbi, M. and Giommi, P. and Primavera, R. and Cutini, S. and Gasparrini, D.},
	title = {{The ASDC SED Builder Tool description and Tutorial}},
	journal = {arXiv},
	year = {2011},
	month = mar,
	eprint = {1103.0749},
	doi = {10.48550/arXiv.1103.0749}
}

@article{Aleksic2016,
	author = {Aleksi{\ifmmode\acute{c}\else\'{c}\fi}, J. and Ansoldi, S. and Antonelli, L. A. and Antoranz, P. and Babic, A. and Bangale, P. and Barcel{\ifmmode\acute{o}\else\'{o}\fi}, M. and Barrio, J. A. and Becerra Gonz{\ifmmode\acute{a}\else\'{a}\fi}lez, J. and Bednarek, W. and Bernardini, E. and Biasuzzi, B. and Biland, A. and Bitossi, M. and Blanch, O. and Bonnefoy, S. and Bonnoli, G. and Borracci, F. and Bretz, T. and Carmona, E. and Carosi, A. and Cecchi, R. and Colin, P. and Colombo, E. and Contreras, J. L. and others},
	title = {{The major upgrade of the MAGIC telescopes, Part II: A performance study using observations of the Crab Nebula}},
	journal = {Astropart. Phys.},
	volume = {72},
	pages = {76--94},
	year = {2016},
	issn = {0927-6505},
	publisher = {North-Holland},
	doi = {10.1016/j.astropartphys.2015.02.005}
}

@inproceedings{Park2016,
  author = "Park, Nahee",
  title = "{Performance of the VERITAS experiment }",
  doi = "10.22323/1.236.0771",
  booktitle = "Proceedings of The 34th International Cosmic Ray Conference {\textemdash} PoS(ICRC2015)",
  year = 2016,
  volume = "236",
  pages = "771"
}

@article{Aartsen2019,
	author = {Aartsen, M. G. and Ackermann, M. and Adams, J. and Aguilar, J. A. and Ahlers, M. and Ahrens, M. and Altmann, D. and Andeen, K. and Anderson, T. and Ansseau, I. and Anton, G. and Arg{\ifmmode\ddot{u}\else\"{u}\fi}elles, C. and Auffenberg, J. and Axani, S. and Backes, P. and Bagherpour, H. and Bai, X. and Barbano, A. and Barron, J. P. and Barwick, S. W. and Baum, V. and Bay, R. and Beatty, J. J. and Becker Tjus, J. and Becker, K.-H. and others},
	title = {{Search for steady point-like sources in the astrophysical muon neutrino flux with 8 years of IceCube data}},
	journal = {Eur. Phys. J. C},
	volume = {79},
	number = {3},
	pages = {1--19},
	year = {2019},
	month = mar,
	issn = {1434-6052},
	publisher = {Springer Berlin Heidelberg},
	doi = {10.1140/epjc/s10052-019-6680-0}
}

@article{Archambault2016,
	author = {Archambault, S. and Archer, A. and Benbow, W. and Bird, R. and Biteau, J. and Buchovecky, M. and Buckley, J. H. and Bugaev, V. and Byrum, K. and Cerruti, M. and Chen, X. and Ciupik, L. and Connolly, M. P. and Cui, W. and Eisch, J. D. and Errando, M. and Falcone, A. and Feng, Q. and Finley, J. P. and Fleischhack, H. and Fortin, P. and Fortson, L. and Furniss, A. and Gillanders, G. H. and Griffin, S. and others},
	title = {{UPPER LIMITS FROM FIVE YEARS OF BLAZAR OBSERVATIONS WITH THE VERITAS CHERENKOV TELESCOPES}},
	journal = {Astron. J.},
	volume = {151},
	number = {6},
	pages = {142},
	year = {2016},
	month = may,
	issn = {1538-3881},
	publisher = {The American Astronomical Society},
	doi = {10.3847/0004-6256/151/6/142}
}

@inproceedings{hess-sensitivity,
  author = "Holler, Markus  and  Berge, David and van Eldik, Christopher  and  Zaborov, Dmitry  and  Lenain, Jean-Philippe  and  Marandon, Vincent  and  Murach, Thomas  and  Prokoph, Heike  and  de Naurois, Mathieu  and  Parsons, Robert Daniel",
  title = "{Observations of the Crab Nebula with H.E.S.S. phase II }",
  doi = "10.22323/1.236.0847",
  booktitle = "Proceedings of The 34th International Cosmic Ray Conference {\textemdash} PoS(ICRC2015)",
  year = 2016,
  volume = "236",
  pages = "847"
}

@article{pks0735+17_padovani,
	author = {Sahakyan, N. and Giommi, P. and Padovani, P. and Petropoulou, M. and B{\ifmmode\acute{e}\else\'{e}\fi}gu{\ifmmode\acute{e}\else\'{e}\fi}, D. and Boccardi, B. and Gasparyan, S.},
	title = {{A multimessenger study of the blazar PKS{\hspace{0.167em}}0735+178: a new major neutrino source candidate}},
	journal = {Mon. Not. R. Astron. Soc.},
	volume = {519},
	number = {1},
	pages = {1396--1408},
	year = {2023},
	month = feb,
	issn = {0035-8711},
	publisher = {Oxford Academic},
	doi = {10.1093/mnras/stac3607}
}

@article{Mannheim1993Feb,
	author = {Mannheim, Karl},
	title = {{The Proton Blazar}},
	journal = {arXiv},
	year = {1993},
	month = feb,
	eprint = {astro-ph/9302006},
	doi = {10.48550/arXiv.astro-ph/9302006}
}

@article{Atoyan2001Nov,
	author = {Atoyan, Armen and Dermer, Charles D.},
	title = {{High-Energy Neutrinos from Photomeson Processes in Blazars}},
	journal = {Phys. Rev. Lett.},
	volume = {87},
	number = {22},
	pages = {221102},
	year = {2001},
	month = nov,
	issn = {1079-7114},
	publisher = {American Physical Society},
	doi = {10.1103/PhysRevLett.87.221102}
}

@article{Bottcher2013Apr,
	author = {B{\ifmmode\ddot{o}\else\"{o}\fi}ttcher, M. and Reimer, A. and Sweeney, K. and Prakash, A.},
	title = {{LEPTONIC AND HADRONIC MODELING OF FERMI-DETECTED BLAZARS}},
	journal = {Astrophys. J.},
	volume = {768},
	number = {1},
	pages = {54},
	year = {2013},
	month = apr,
	issn = {0004-637X},
	publisher = {The American Astronomical Society},
	doi = {10.1088/0004-637X/768/1/54}
}






\end{document}